\documentclass[12pt]{article}

\usepackage{epsf,amsfonts,hyperref,amsmath}
\usepackage{url}
\usepackage{color}
\usepackage{cite}
\newcommand\encadremath[1]{\vbox{\hrule\hbox{\vrule\kern8pt 
\vbox{\kern8pt \hbox{$\displaystyle #1$}\kern8pt} 
\kern8pt\vrule}\hrule}}
\def\enca#1{\vbox{\hrule\hbox{
\vrule\kern8pt\vbox{\kern8pt \hbox{$\displaystyle #1$}
\kern8pt} \kern8pt\vrule}\hrule}}

\newcommand\figureframex[3]{
\begin{figure}[bth]
\hrule\hbox{\vrule\kern8pt 
\vbox{\kern8pt \vbox{
\begin{center}
{\mbox{\epsfxsize=#1.truecm\epsfbox{#2}}}
\end{center}
\caption{#3}
}\kern8pt} 
\kern8pt\vrule}\hrule
\end{figure}
}
\newcommand\figureframey[3]{
\begin{figure}[bth]
\hrule\hbox{\vrule\kern8pt 
\vbox{\kern8pt \vbox{
\begin{center}
{\mbox{\epsfysize=#1.truecm\epsfbox{#2}}}
\end{center}
\caption{#3}
}\kern8pt} 
\kern8pt\vrule}\hrule
\end{figure}
}

\makeatletter
\@addtoreset{equation}{section}
\makeatother
\newtheorem{theorem}{Theorem}[section]

\newtheorem{remark}{Remark}[section]
\newtheorem{proposition}{Proposition}[section]
\newtheorem{lemma}{Lemma}[section]
\newtheorem{corollary}{Corollary}[section]
\newtheorem{definition}{Definition}[section]
\def\br{\begin{remark}\rm\small}
\def\er{\end{remark}}
\def\bt{\begin{theorem}}
\def\et{\end{theorem}}
\def\bd{\begin{definition}}
\def\ed{\end{definition}}
\def\bp{\begin{proposition}}
\def\ep{\end{proposition}}
\def\bl{\begin{lemma}}
\def\el{\end{lemma}}
\def\bc{\begin{corollary}}
\def\ec{\end{corollary}}
\def\beaq{\begin{eqnarray}}
\def\eeaq{\end{eqnarray}}

\newcommand{\beq}{\begin{equation}}
\newcommand{\eeq}{\end{equation}}
\newcommand{\bea}{\begin{eqnarray}}
\newcommand{\eea}{\end{eqnarray}}

\renewcommand{\and}{{\qquad {\rm and} \qquad}}

\newcommand{\virg}{{\qquad , \qquad}}

 \newcommand{\Tr}{{\,\rm Tr}\:}
\newcommand{\tr}{{\,\rm tr}\:}

\newcommand{\Res}{\mathop{\,\rm Res\,}}

\newcommand{\td}[1]{{\tilde{#1}}}

\newcommand{\e}{{\,\rm e}\,}
\newcommand{\ee}[1]{{{\rm e}^{#1}}}

\renewcommand{\d}{{{\partial}}}

\newcommand{\Pint}{{\int\kern -1.em -\kern-.25em}}

\renewcommand{\L}{\Lambda}

\begin{document}
%=============================Page de titre===============
%\date{??}
%\author{Eynard}
%\title{Correlation functions for hermitian random matrices}
%\topmargin .5cm \textheight 21.5cm \textwidth 15.8cm 
%\oddsidemargin 0.54cm
%\evensidemargin 0.54cm 
\sloppy

%\maketitle

\pagestyle{empty}
\hfill IPT-11/145\\
\indent \hfill CERN-2011-121
\addtolength{\baselineskip}{0.20\baselineskip}
\baselineskip 16pt 
\begin{center}
\begin{Large}\fontfamily{cmss}
\fontsize{20pt}{30pt}
\selectfont
\medskip
\textbf{Loop equations and topological recursion for the arbitrary-$\beta$ two-matrix model}
\end{Large}\\
\bigskip
\bigskip
{\sl M.\ Berg\`{e}re}\hspace*{0.05cm}\footnote{ E-mail: michel.bergere@cea.fr }\\
Institut de physique th\'{e}orique\\
F-91191 Gif-sur-Yvette Cedex, France.\\
\vspace{6pt}
{\sl B.\ Eynard}\hspace*{0.05cm}\footnote{ E-mail: bertrand.eynard@cea.fr }\\
Institut de physique th\'{e}orique\\
F-91191 Gif-sur-Yvette Cedex, France.\\
CERN  PH-TH, Case C01600, CERN, CH-1211 Geneva 23.\\
\vspace{6pt}
{\sl O.\ Marchal}\hspace*{0.05cm}\footnote{ E-mail: olivier.marchal@polytechnique.org }\\
Department of Mathematics and Statistics\\
University of Alberta\\
Edmonton, Canada.\\
\vspace{6pt}
{\sl A.\ Prats--Ferrer}\hspace*{0.05cm}\footnote{ E-mail:pratsferrer@crm.umontreal.ca}\\
Department of mathematics\\
Concordia University\\
Montr\'eal, Canada.\\
\end{center}

\vspace{20pt}
\begin{center}
{\bf Abstract}

We write the loop equations for the $\beta$ two-matrix model, and we propose a topological recursion algorithm to solve them, order by order in a small parameter.
We find that to leading order, the spectral curve is a ``quantum" spectral curve, i.e. it is given by a differential operator (instead of an algebraic equation for the hermitian case).
Here, we study the case where that quantum spectral curve is completely degenerate, it satisfies a Bethe ansatz, and the spectral curve is the Baxter TQ relation.

\end{center}
%-----------------------------ABSTRACT--------------------------------------
%
%Abstract

%\begin{center}

%\end{center}

%\newpage
%\pagestyle{empty}

%\section*{}

%\newpage
\vspace{26pt}
\pagestyle{plain}
\setcounter{page}{1}

%*********************************************************************
%==================== ARTICLE ========================================
%*********************************************************************

\tableofcontents

\section{Introduction: Generalization of the two-matrix model \label{secloops}}

Random matrix models have played a very important role in physics and mathematics \cite{Wigner, Dyson, Dysondet, Dyson2, Guhr, Mehtabook, DGZ, BI}. The eigenvalue statistics of large random matrices tend to universal statistical laws, which can be observed in many systems in nature, ranging from nuclear physics to finance and biology. Moreover, random matrices, treated as formal series, provide generating functions for counting discretized surfaces \cite{BIPZ, DGZ, David, KazakovRMTcrit, PDFgraph, eynform}, and are a very useful tool for string theory \cite{betaensemble, MMbook, MMtor, DV, DV2, DV4, DW, KoKo, kostovhirota}. 

So far, the random matrices studied were mostly hermitian matrices, because it was the easiest case.
From the beginning, Wigner \cite{Wigner, Ginibre, Mehtabook} introduced 3 ensembles of matrices, classified by an exponent $2\beta$, Hermitian: $2\beta=2$, real--symmetric: $2\beta=1$, and quaternionic: $2\beta=4$. However, it is rather easy to analytically extend the joint eigenvalue distribution to arbitrary values of $\beta$ \cite{Dum, Bryc, desrosiers}.

\medskip
The case of one random matrix belonging to an arbitrary $\beta$ ensemble, although less studied than the hermitian case, has received lot of attention, and many works were done. In particular loop equations have been known for a long time, and their recursive solution was recently proposed in \cite{ChekEynbeta, MoiBertrand, MoiLeonidBertrand, MoiLeonidBertrand2}.

\medskip
For the hermitian case, it turned out that a 1-matrix model was not general enough, for instance the spectral curve for 1-matrix model is always quadratic, it can't have higher degree, the critical behaviors are related to only a small subset of all possible conformal field theories. A 2-matrix hermitian model turns out to be much more general, it allows to reach spectral curves of any degrees, and any conformal minimal model \cite{BookPDF, DKK, KazMar, DGZ}. Moreover, as applications to counting discretized surfaces, a 2-matrix model allows to count discrete surfaces carrying colors \cite{KazakovIsing, PDFgraph}.

Unfortunately, trying to generalize the 2-matrix model to arbitrary $\beta$ seemed very difficult, and was almost never studied.
One reason, is that the joint eigenvalue distribution was not known, because the integration over the angular parts of the matrix was not known (properties of angular integrals for hermitian matrix models can be found for example in \cite{Zuber}).

Recently, some progress was made \cite{Angular} in those angular integrals, and although we don't know yet how to compute angular integrals for arbitrary $\beta$, we know already that those angular integrals have to satisfy some differential equations, and this is sufficient to derive loop equations. This is what we do in this article, we derive the loop equations, and solve them perturbatively by expanding in some small parameter.

\subsection{The hermitian two-matrix model}

The hermitian two-matrix model is given by the partition function:
\beq \label{2HMM}
Z_{\textit{Herm}}=\ee{F}=
\int dM_1\, dM_2\,\ee{-\frac{N}{T}\tr\left[V_1(M_1)+V_2(M_2)-M_1 M_2\right]}
\eeq
where $M_1$ and $M_2$ are $N \times N$ hermitian matrices and the measure $dM_i$ is the corresponding Lebesgue measure associated with all independent entries of the matrices. 
$V_1$ and $V_2$ are called the potentials, we shall assume in this article that $V_1$ and $V_2$ are polynomials.
The parameter $T$, often called the temperature, is redundant, it can be absorbed by a change of variable $M_2\to T M_2$ and a redefinition of $V_2$, however, we prefer to keep it for convenience, and because of its future geometric interpretation.
The hermitian 2-matrix model was introduced in particular as a formal series to study the Ising model on a random surface, i.e. the Ising model coupled to 2D gravity \cite{KazakovIsing, DGZ, eynform}.

It is very well known \cite{Mehta1} now that this integral can be rewritten in term of its eigenvalues problem as:
\beq \label{Diagonalized2HMM} Z_{\textit{Herm}}=\int dX dY \Delta(X)^2 \Delta(Y)^2 e^{-\frac{N}{T}\left[\underset{i=1}{\overset{N}{\sum}}V_1(x_i)+\underset{j=1}{\overset{N}{\sum}}V_2(y_j)\right]}I(X,Y)\eeq
where $X={\rm diag}(x_1,\dots,x_N)$ and $Y={\rm diag}(y_1,\dots,y_N)$ are diagonal matrices representing the eigenvalues of $M_1$ and $M_2$, $\Delta(X)=\underset{i<j}{\prod} (x_j-x_i)$ is the Vandermonde determinant,  and $I(X,Y)$ is the Itzykson-Zuber integral \cite{IZ, HC} which corresponds to the integration about the angular variables:
\beq I(X,Y)=\int_{\mathcal{U}_N} dU e^{\frac{N}{T}\Tr(XUYU^{-1})}\eeq
where $\mathcal{U}_N$ is the unitary group equipped with the Haar measure.
Relatively to the Itzykson-Zuber integral, one can also define the quantities:
\beq M_{i,j}=\int_{\mathcal{U}_N} dU \| U_{i,j} \|^2 e^{\frac{N}{T}\Tr(XUYU^{-1})}\eeq
which can be used to determine the Itzykson-Zuber integral by the formula:
\beq \label{OO}I(X,Y)=\sum_{i=1}^N M_{i,j}=\sum_{j=1}^N M_{i,j}\eeq
Note in particular that the r.h.s. does not depend on the second index $i$ or $j$. Technically, the last formulas are obvious since $\underset{i=1}{\overset{N}{\sum}}\| U_{i,j} \|^2=1=\underset{j=1}{\overset{N}{\sum}}\| U_{i,j} \|^2$ in the unitary group. Eventually, in the hermitian case, it is also known that the $M_{i,j}$'s satisfy the Dunkl equation \cite{Calogero, Bryc, desrosiers}, namely:
\beq \label{CalogeroMoser} 
\forall\,  1\leq i,j\leq N\, :\, \frac{\partial}{\partial x_i} M_{i,j} + \sum_{k\neq i} \frac{M_{i,j}-M_{k,j}}{x_i-x_k}=y_j M_{i,j}
\eeq

We will see in the next sections that these properties can be extended uniquely in the case of the $\beta$-deformation of the hermitian two-matrix model case.

\subsection{Generalization to arbitrary $\beta$ two-matrix models}

Similarly to what happens in the one-matrix model, we would like to generalize the hermitian two-matrix model to other $\beta$--ensembles of matrices. As in the case of one-matrix models, the usual generalization is made from the diagonalized version of the problem \eqref{Diagonalized2HMM} by:
\beq \label{Diagonalized2betaMM} Z_{\beta}\overset{\text{def}}{=}\int dX dY \Delta(X)^{2\beta} \Delta(Y)^{2\beta} e^{-\frac{N\beta}{T}\left[\underset{i=1}{\overset{N}{\sum}}V_1(x_i)+\underset{j=1}{\overset{N}{\sum}}V_2(y_j)\right]}I_\beta(X,Y)\eeq
where $I_\beta(X,Y)$ is a generalized version of the Itzykson-Zuber integral that we will describe later. Note that with our convention, the hermitian case corresponds to $\beta=1$ (notation used by people working on AGT conjecture \cite{AGT}, and Laughlin wave function), whereas sometimes in the literature it is normalized to $\beta=2$ (Wigner's notation \cite{Wigner, Mehtabook}). We prefer to use the first notation in order to avoid unnecessary powers of 2 in all formulas. 

The diagonalized version \eqref{Diagonalized2betaMM} corresponds to the idea of the generalized matrix integral:
\beq 
Z_{\beta}``="
\int_{E_{N,\beta}} dM_1\, dM_2\,\ee{-\frac{N \beta}{T}\tr\left[V_1(M_1)+V_2(M_2)-M_1 M_2\right]}
\eeq
where $E_{N,\beta}$ would be an ensemble of matrices such that the diagonalization of $M_1$ and $M_2$ gives \eqref{Diagonalized2betaMM}. Note that this definition is only formal since apart from $\beta=\,1,\,\, 0,\,\, \frac{1}{2}$ or $2$ where $E_N$ corresponds to hermitian, diagonal, real-symmetric or quaternionic matrices, no other ensemble is known presently (it would be interesting to see if the $\beta$-matrix ensemble introduced by \cite{Dum} also reproduces the $I_\beta$ term).

\subsection{The angular integral}

We shall now define the $I_\beta(X,Y)$ angular integral, so that it coincides with the actual angular integral for the matrix cases $\beta=1,0,\frac{1}{2},2$, and in fact we shall use the angular matrix integral defined in \cite{Angular}.

Following \cite{Angular}, we first  define a good generalization of the $M_{i,j}$ and from \eqref{OO} we will define the generalized version of the Itzykson-Zuber integral.
In \cite{Angular}, the authors claim that the natural generalization of the $M_{i,j}$, noted here $M_{i,j}^{(\beta)}$ is the following:
\begin{enumerate}
 \item The $M_{i,j}^{(\beta)}$'s satisfy the generalized Dunkl-Calogero-Moser equations:
\beq \label{GeneralizedCM} \forall\,  1\leq i,j\leq N\, :\, \frac{\partial}{\partial x_i} M_{i,j}^{(\beta)}+ \beta\sum_{k\neq i} \frac{M_{i,j}^{(\beta)}-M_{k,j}^{(\beta)}}{x_i-x_k}=\frac{N\beta}{T} y_j M_{i,j}^{(\beta)}
\eeq (The factor $\frac{N\beta}{T}$ comes from the same factor in the exponential term of the partition function which can be absorbed by the change $Y \leftrightarrow \frac{N\beta}{T}Y$)
\item $M_{i,j}^{(\beta)}$ must be stochastic matrices, i.e. $I_\beta=\underset{i=1}{\overset{N}{\sum}} M_{i,j}^{(\beta)}$ must be independent from $j$ and $I_\beta=\underset{j=1}{\overset{N}{\sum}} M_{i,j}^{(\beta)}$ must be independent from $i$
\item They must have the symmetry $M_{i,j}^{(\beta)}(X,Y)=M_{j,i}^{(\beta)}(Y,X)$. In particular, this implies that $I_\beta(X,Y)=\underset{i=1}{\overset{N}{\sum}} M_{i,j}^{(\beta)}(X,Y)$ must be symmetric in the exchange $X \leftrightarrow Y$.
\item They must also have the symmetry $M_{\sigma(i),j}^{(\beta)}(X_\sigma,Y)=M_{i,j}^{(\beta)}(X,Y)$ where $\sigma\in \Sigma_N$ is a permutation. This implies that $I_\beta(X_\sigma,Y)=I_\beta(X,Y)$.
\item They are normalized so that $I_\beta({\rm Id}_{N\times N},Y) = \ee{\frac{N\beta}{T}\,\tr Y}$.
\end{enumerate}
These conditions define the $M_{i,j}^{(\beta)}$'s uniquely making the previous set of conditions a proper definition (in \cite{Angular}, an explicit solution for those $M_{i,j}^{(\beta)}$ is provided as an integral, or is given by a recursion on $N$). Moreover all these properties are standard results in the hermitian case $\beta=1$ and hold for $\beta=1, \frac{1}{2},2$. In particular note that $M_{i,j}^{(\beta)}(X,Y)=M_{j,i}^{(\beta)}(Y,X)$ gives another formulation of \eqref{GeneralizedCM} as:
\beq \label{GeneralizedCM2}\forall\,  1\leq i,j\leq N\, :\, \frac{\partial}{\partial y_j} M_{i,j}^{(\beta)}+ \beta\sum_{k\neq j} \frac{M_{i,j}^{(\beta)}-M_{i,k}^{(\beta)}}{y_j-y_k}=\frac{N\beta}{T} x_i  M_{i,j}^{(\beta)}
\eeq

Similarly to what happens in the hermitian case, it is logical to define the generalized Itzykson-Zuber integral by:
\beq \label{I} I_\beta(X,Y)\overset{\text{def}}{=} \sum_{i=1}^N M_{i,j}^{(\beta)}(X,Y)=\sum_{j=1}^N M_{i,j}^{(\beta)}(X,Y).\eeq
Note again that this definition recovers the known cases when $\beta=1, \frac{1}{2},2$. Eventually, from this definition we can prove (see \cite{Angular}) that the generalized Itzykson-Zuber integral $I_\beta(X,Y)$ satisfies the following Calogero-Moser equation \cite{Calogero}:
\beq \label{Hamilton} H_X^{(\beta)}I_\beta\mathop{=}^{def}\sum_{i=1}^N \frac{\partial^2 I_\beta}{\partial x_i^2} +\beta \sum_{i\neq j} \frac{ 1}{x_i-x_j} \left( \frac{ \partial I_\beta}{\partial x_i} - \frac{ \partial I_\beta}{\partial x_j}\right) =\left(\frac{N\beta}{T}\right)^2\left(\sum_{j=1}^N y_j^2\right) I_\beta \eeq
where $H_X^{(\beta)}$ is the Calogero-Moser Hamiltonian. Note again that $I_\beta(X,Y)$ is well known to satisfy \eqref{Hamilton} in the $\beta=1,\frac{1}{2},2$ cases.

\medskip
\medskip

Now that we have introduced a proper generalized two-matrix model, we can try to solve it in the large $N$ limit and in various regimes of the parameters. In particular, since no (bi)-orthogonal polynomials techniques are known in the case of an arbitrary exponent $\beta$ in the Vandermonde determinant (although some work has been done in \cite{ED2}), we will use in this article the method of the loop equations to determine the correlation functions of the models.

\section{Summary of the main results}

Our goal is to compute the asymptotic (possibly formal) expansion of the 2-matrix model integral and its correlation functions.
$$
Z_\beta "=" \int_{E_{N,\beta}} dM_1\,\,dM_2\,\, \ee{-\,\frac{N\beta}{T} \tr [ V_1(M_1)+V_2(M_2)-M_1 M_2]} 
$$
and the cumulants (subscript ${}_c$) of expectation values of products of traces of resolvents
$$
W_n(x_1,\dots,x_n) = \left< \Tr \frac{1}{x_1-M_1} \dots \frac{1}{x_n-M_1}\right>_c .
$$
We want to expand them in powers of a small parameter $\epsilon$ as
$$
\ln Z = \sum_g \epsilon^{2g-2}\,F_g
$$
$$
W_n = \sum_g \epsilon^{2g-2+n} W_n^{(g)}.
$$
The small parameter $\epsilon$ is a combination of $\frac{1}{N}$ and $\beta$.
There can be several asymptotic regimes of interest, which we discuss in detail in section \ref{sectoprecregimes}.

\medskip
{\bf Main results}
\smallskip

$\bullet$ In section \ref{secloopeqs} we derive the loop equations for the 2-matrix model with arbitrary $\beta$.

$\bullet$ In section \ref{secpsi} we solve the loop equation to leading order for the resolvent $W_1$, i.e. we compute $W_1^{(0)}$. We show that
$$
W_1^{(0)}(x) = \frac{T}{N}\,\frac{\psi'(x)}{\psi(x)}
$$
where $\psi(x)$ is solution of a linear differential equation with polynomial coefficients
$$
{\cal E}\left(x,\frac{T}{N}\,\frac{\d}{\d x}\right).\psi(x)=0
$$
where ${\cal E}(x,y)$ is a polynomial in 2 variables.

$\bullet$ This equation for $\psi$ can be thought of as a "quantum spectral curve". We remind that for $\beta=1$ one gets an algebraic equation ${\cal E}(x,W_1^{(0)}(x))=0$. Here, we replace this algebraic equation by a linear operator ${\cal E}(x,\hat y)$ where
$$ \hat y = \frac{T}{N}\,\frac{\d}{\d x}
\virg
[\hat y,x] =\frac{T}{N},$$
i.e. the non-commutative version of the algebraic equation ${\cal E}(x,y)=0$.

$\bullet$ This equation can also be seen as a Baxter T-Q equation where $\psi(x)$ is the $Q$ function. Its zeros $s_i$'s ($\psi(s_i)=0$) are thus the Bethe roots, and the equation ${\cal E}(x,\hat y).\psi=0$ can be translated into a Bethe Ansatz equation for the $s_i$'s. We do it in \ref{Bethe ansatz}.
The Bethe equation for $S={\rm diag}(s_1,\dots,s_N)$ can be written in the following way:
find a matrix $B$ such that:
$$
(V'_2(B)-S) e=0
\virg
e^t(V'_1(S)-B) = 0
\virg
[S,B]=\frac{T}{N}\,({\rm Id}- e\, e^t)
$$
where $ e = (1,1,\dots,1)^t$.

$\bullet$ In section \ref{secYangYang} we show that this Bethe ansatz equation can also be written as the extremization of a Yang-Yang action:
\bea\nonumber 
\frac{N}{T}\mathcal{S}(S,\td{S},A,\vec{u})&=& \tr V_1(S) +\tr V_2(\td{S})- \tr (SA\td{S}A^{-1}) - \frac{T}{N}\ln(\Delta(S))\cr
&&-\frac{T}{N}\ln(\Delta(\td{S}))+\frac{T}{N}\ln \det(A) -\frac{T}{N}\vec{u}^t \left(A\vec{e}-\vec{e}\right)
\eea
At the extremum of ${\cal S}$ we have $S= {\rm diag}(s_1,\dots,s_N)$.

The leading order $F_0$ of $\ln Z$ is the value of the Yang-Yang action at its extremum, and the first subleading term $F_1$ is half the log of the determinant of the Hessian of ${\cal S}$:
$$
\ln Z_N = \sum_{g=0}^\infty \epsilon^{2g-2} F_g
\virg
F_0 = -{\cal S}({\rm extremum})
\virg
F_1 = -\,\frac{1}{2}\ln{\det {\cal S}''}
$$

$\bullet$ In section \ref{secVar20} we compute the leading order of the 2-point function $W_2^{(0)}$ in terms of the Hessian ${\cal H} = {\cal S}''$ of the Yang-Yang action:
$$
W_2^{(0)}(x,x') = \sum_{i,j=1}^N \frac{\left({\cal H}^{-1}\right)_{i,j}}{(x-s_i)^2\,(x'-s_j)^2}.
$$
In a similar way, we compute the leading order of the 3-point function $W_3^{(0)}$ in terms of ${\cal H}={\cal S}''$ and ${\cal S}'''$ in \eqref{secVar30}.

$\bullet$ In section \ref{section 8} we provide a ``topological recursion" algorithm to compute every subleading correction to any $n$-point function, i.e. $W_n^{(g)}$, in the form:
$$
W_{n+1}^{(g)}(x_0,x_1,\dots,x_n) 
= \sum_{i=1}^N \Res_{x\to s_i} K(x_0,x)\,\Big({\rm combination\, of\,}W_{m}^{(h)}{\rm 's\,\,with\,}2h+m<2g+n\Big)
$$
However, in contrast with \cite{EO}, the recursion kernel $K(x_0,x)$ is here a matrix of dimensions $d_2=\deg V'_2$.
We derive the topological recursion in section \ref{sectoprec}, and then we describe the kernel $K(x_0,x)$ in more details in section \ref{secK} and in appendix.

\section{Loop equations for the $\beta$-deformed two-matrix model}\label{secloopeqs}

The loop equations, also called ``Schwinger--Dyson" equations, are a very powerful tool to study random matrices, perturbatively in the expansion in some small parameter (for the hermitian case, the small parameter is usually $1/N$ in the large $N$ limit), see \cite{Wadia, Virasoro, ACKM, Jevicki3, DGZ, eynloop1mat}. In short, Schwinger--Dyson equations exploit the fact that an integral is invariant by change of variable, or an alternative way to obtain the same equations is doing integration by parts.

\subsection{Notations for correlation functions}

For the hermitian two-matrix models, the loop equations are well known \cite{staudacher, eynm2m, eynm2mg1, eynmultimat, eynmetha, KazMar, CEO, EO} and they are very useful to compute the large $N$ expansion of correlations functions,  $W_n^{(g)}$ and symplectic invariants $F_g$ presented in a more general context in \cite{EO}. In the one-matrix model, it is also known \cite{MoiBertrand} that the loop equations method can be generalized for arbitrary $\beta$ quite directly from the hermitian case. Indeed, we remind the reader that we can see loop equations as integration by part or Schwinger-Dyson equations of the diagonalized problem \eqref{Diagonalized2betaMM}. In the present $\beta$-deformed two-matrix model case, we will see that we can get also some loop equations by following the same approach. Before writing down these equations, we need to introduce some notations (the same as in \cite{eynm2m, eynm2mg1, CEO}):
\begin{itemize}
 \item The potentials are assumed to be polynomials:
\beq V_1'(x)=\sum_{k=0}^{d_1} t_k x^k \virg V_2'(x)=\sum_{k=0}^{d_2} \tilde{t}_k x^k \eeq
\item The correlation functions are defined by:
\beq \label{Wn} W_n(z_1,\dots,z_n)=\left<\sum_{i_1,\dots,i_n=1}^N  \frac{1}{z_1-x_{i_1}}\dots \frac{1}{z_n-x_{i_n}}\right>_c \eeq
where the brackets $\left<\,\right>$ indicates that we take the expectation value relatively to the measure defined by \eqref{Diagonalized2betaMM} for the random variables $X={\rm diag}(x_1,\dots,x_n)$ and $Y={\rm diag}(y_1,\dots,y_N$, namely:
\beq \label{expected} 
\encadremath{
\left<A(X,Y)\right>\overset{\text{def}}{=}\frac{1}{Z_\beta} \int dX dY A(X,Y) e^{-\frac{N\beta}{T} \left(\tr V_1(X)+V_2(Y)\right)}\Delta(X)^{2\beta}\Delta(Y)^{2\beta} I_\beta(X,Y)}
\eeq
The index $_c$ stands for the connected component (also called cumulant) of the correlation function that is to say:
\bea <A_1> &=& <A_1>_c \cr
<A_1 A_2> &=& <A_1 A_2>_c + <A_1>_c <A_2>_c \cr
<A_1 A_2 A_3> &=& <A_1 A_2 A_3>_c + <A_1 A_2>_c <A_3> + <A_1 A_3>_c <A_2> \cr
&&+ <A_2 A_3>_c <A_3> + <A_1> <A_2> <A_3>\cr
<A_1\dots A_n>&=&<A_J>=\sum_{k=1}^n\sum_{I_1 \uplus I_2 \dots \uplus I_k=J}\prod_{i=1}^k <A_{I_i}>_c\cr 
\eea
In order to have more compact notations, we will sometimes simply denote $W(x)=W_1(x)$ for the first correlation function.
\item We can define similarly the second type of correlation functions:
\beq \td{W}_n(z_1,\dots,z_n)=\left<\sum_{i_1,\dots,i_n=1}^N  \frac{1}{z_1-y_{i_1}}\dots \frac{1}{z_n-y_{i_n}}\right>_c \eeq
and denote $\td{W}(y)=\td{W}_1(y)$ for the first correlation function.
\item To close the loop equations we will need to introduce the following functions:
\bea \label{Un} &&U_n(x,y;z_1,\dots,z_n)=\cr
&&\sum_{i,j,i_1,\dots,i_n=1}^N\left<  \frac{1}{x-x_i} \frac{M_{i,j}^{(\beta)}}{I_\beta}\frac{V'_2(y)-V'_2(y_j)}{y-y_j}\frac{1}{z_1-x_{i_1}} \dots  \frac{1}{z_n-x_{i_n}}\right>_c\cr
 \eea
Note that they are polynomials in the variable $y$, and remember that $M_{i,j}^{(\beta)}(X,Y)$ depends on $X$ and $Y$ and is also a random variable. 
In a similar way, we also introduce:
 \bea \label{Pn} &&P_n(x,y;z_1,\dots,z_n)=\cr
 &&\sum_{i,j,i_1,\dots,i_n=1}^N\left<  \frac{V'_1(x)-V'_2(x_i)}{x-x_i} \frac{M_{i,j}^{(\beta)}}{I_\beta}\frac{V'_2(y)-V'_2(y_j)}{y-y_j} \frac{1}{z_1-x_{i_1}} \dots \frac{1}{z_n-x_{i_n}}\right>_c
 \cr \eea
Note that this time $P_n(x,y;z_1,\dots,z_n)$ are polynomials in both $x$ and $y$.
\end{itemize}

The loop equations method is a powerful method of deriving an infinite set of equations connecting the correlation functions and the functions $P_n$ and $U_n$ introduced before. There are several ways to derive the loop equations and we choose here to use the integration by part way which is the most convenient in our setting. In the case of hermitian, real-symmetric or quaternionic matrix models, these loop equations are already known and can be derived directly from the matrix integral settings (before any diagonalization). They can be found in many different places in the literature \cite{DGZ, staudacher, KazakovIsing, DKK, Boulatov, eynm2m, eynm2mg1, eynmultimat, KazMar, CEO, EO}. In our arbitrary $\beta$ setting, we do not have the initial matrix integral model properly defined, thus we must adapt the derivation of the loop equations with our definitions \eqref{GeneralizedCM} and \eqref{I}. This is done in different successive steps that are fully detailed in the next sections.

\subsection{First step: deriving an auxiliary result}

%In order to lighten notations, we will note in this section $X={\rm diag}(x_1,\dots,x_N)$ and $Y={\rm diag}(y_1,\dots,y_n)$ as our random diagonal matrices. Also we will not write every time the dependence on $X$ and $Y$ of quantities like $M_{i,j}^{(\beta)}\equiv M_{i,j}^{(\beta)}(X,Y) $ and $I_\beta\equiv I_\beta(X,Y)$ which are assumed to be implicit. Let's now start the first step to derive the loop equations: 
We start by evaluating the following integral:
\beq 0=\sum_{i,j=1}^N \int dXdY \, \frac{\partial}{\partial y_j} \left(e^{-\frac{N\beta}{T} \tr (V_1(X)+V_2(Y) )} \Delta(X)^{2\beta} \Delta(Y)^{2\beta} \frac{1}{x-x_i} M_{i,j}^{(\beta)}\right)\eeq 
The result is clearly null if we choose an integration path which goes to $\infty$ in directions in which $\e^{-\frac N T V_1(X)}$ and $\e^{-\frac N T V_2(Y)}$ vanish exponentially, because we integrate a total derivative. 
%Now we can split the action of the derivative into each term of the product. Using the expected value notation $<>$ introduced before \eqref{expected}, we find 
The right hand side yields $3$ different contributions:
\begin{itemize}
 \item Acting on the exponential we find:
\beq -\frac{N\beta}{T}\sum_{i,j=1}^N\left< V_2'(y_j)\frac{1}{x-x_i}\frac{M_{i,j}^{(\beta)}}{I_\beta}\right>\eeq
\item Acting on the Vandermonde determinant, we find:
\beq \label{oo} 2\beta \left< \sum_{i,j=1}^N \sum_{k \neq j} \frac{1}{y_j-y_k} \frac{1}{x-x_i} \frac{M_{i,j}^{(\beta)}}{I_\beta}\right>\eeq
\item Eventually, acting on $M_{i,j}^{(\beta)}$ and using the Dunkl differential equation satisfied by the $M_{i,j}^{(\beta)}$'s  \eqref{GeneralizedCM2} we find:
\beq \label{ooo} \left< \sum_{i,j=1}^N\frac{1}{x-x_i} \left( \frac{N\beta}{T} x_iM_{i,j}^{(\beta)}-\beta \sum_{k\neq j} \frac{M_{i,j}^{(\beta)}-M_{i,k}^{(\beta)}}{y_j-y_k}\right) \frac{1}{I_\beta}\right> \eeq
\end{itemize}

Then we see that \eqref{oo} cancels with the last part of \eqref{ooo} so that we have our first equation:
\beq \sum_{i,j=1}^N\left< V_2'(y_j)\frac{1}{x-x_i}\frac{M_{i,j}^{(\beta)}}{I_\beta}\right>=\sum_{i,j=1}^N\left< \frac{x_i}{x-x_i}\frac{M_{i,j}^{(\beta)}}{I_\beta}\right> \eeq
Remembering now that $\underset{j=1}{\overset{N}{\sum}} M_{i,j}^{(\beta)}=I_\beta$ gives:
\beq \label{firstloop} \encadremath{\sum_{i,j=1}^N\left< V_2'(y_j)\frac{1}{x-x_i}\frac{M_{i,j}^{(\beta)}}{I_\beta}\right>=-N+xW(x)} \eeq

\subsection{Second step: finding the loop equations}

Then consider:
%The next step is to study the integral:
\beq 
0=\sum_{i,j=1}^N \int dXdY \, \frac{\partial}{\partial x_i} \left(e^{-\frac{N\beta}{T} \tr (V_1(X)+V_2(Y) )} \Delta(X)^{2\beta} \Delta(Y)^{2\beta} \frac{1}{x-x_i} M_{i,j}^{(\beta)}\frac{V'_2(y)-V'_2(y_j)}{y-y_j}\right)\eeq 
%which is again zero since there are no hard-edges and we deal with a total derivative. 
%We observe in particular that this integral is very similar to the definitions of the functions $U_0(x,y)$ and $P_0(x,y)$ defined by \eqref{Un} and \eqref{Pn}. Again we can make the derivation act on every term of the product. This time there are
whose r.h.s. yields $4$ contributions:
\begin{itemize}
 \item Acting on the exponential, we find:
\beq \label{i}-\frac{N\beta}{T} \sum_{i,j=1}^N \left< \frac{V'_1(x_i)}{x-x_i} \frac{M_{i,j}^{(\beta)}}{I_\beta} \frac{V'_2(y)-V'_2(y_j)}{y-y_j}\right> \, (i)\eeq 
\item Acting on the Vandermonde determinant we find:
\beq \label{ii}2\beta\sum_{i,j=1}^N \left<  \sum_{k \neq i} \frac{1}{x_i-x_k}\frac{1}{x-x_i} \frac{M_{i,j}^{(\beta)}}{I_\beta}\frac{V'_2(y)-V'_2(y_j)}{y-y_j}\right> \, (ii)\eeq
\item Acting on $\frac{1}{x-x_i}$ we find:
\beq \label{iii}\sum_{i,j=1}^N \left<  \frac{1}{(x-x_i)^2} \frac{M_{i,j}^{(\beta)}}{I_\beta}\frac{V'_2(y)-V'_2(y_j)}{y-y_j}\right> \, (iii)\eeq
\item Acting on $M_{i,j}^{(\beta)}$ and using \eqref{GeneralizedCM} we find:
\beq \label{iv} \sum_{i,j=1}^N \left< \frac{1}{x-x_i} \frac{1}{I_\beta}\left(\frac{N\beta}{T}y_j M_{i,j}^{(\beta)}-\beta\sum_{k \neq i}\frac{M_{i,j}^{(\beta)}-M_{k,j}^{(\beta)}}{x_i-x_k}\right) \frac{V'_2(y)-V'_2(y_j)}{y-y_j}\right> \, (iv)\eeq 
\end{itemize}
Now we observe the following identities: first in \eqref{i} we can perform $V'_1(x_i)=V'_1(x_i)-V'_1(x)+V'_1(x)$ so that we have (using the notations $U_0$ and $P_0$ introduced in \eqref{Un} and \eqref{Pn}):
\beq (i) \, \Leftrightarrow -\frac{N\beta}{T} \left( V_1'(x)U_0(x,y)-P_0(x,y)\right)\eeq
Secondly we can split $(ii)$ in the following form:
\bea
(ii)&=& \beta\sum_{i,j=1}^N \left<  \sum_{k \neq i} \frac{1}{x_i-x_k}\frac{1}{x-x_i} \frac{M_{i,j}^{(\beta)}}{I_\beta}\frac{V'_2(y)-V'_2(y_j)}{y-y_j}\right>\cr
&&+ \beta \sum_{i,j=1}^N \left<  \sum_{k \neq i} \frac{1}{x-x_i}\frac{1}{x-x_k} \frac{M_{i,j}^{(\beta)}}{I_\beta}\frac{V'_2(y)-V'_2(y_j)}{y-y_j}\right>\cr
&&+\beta \sum_{i,j=1}^N \left<  \sum_{k \neq i} \frac{1}{x-x_k}\frac{1}{x_i-x_k} \frac{M_{i,j}^{(\beta)}}{I_\beta}\frac{V'_2(y)-V'_2(y_j)}{y-y_j}\right>\cr
&=& \beta \sum_{i,j=1}^N \left<  \sum_{k \neq i} \frac{1}{x-x_i}\frac{1}{x-x_k} \frac{M_{i,j}^{(\beta)}}{I_\beta}\frac{V'_2(y)-V'_2(y_j)}{y-y_j}\right> \,(ii)'\cr
&&+\beta\sum_{i,j=1}^N \left<\sum_{k \neq i}\frac{M_{i,j}^{(\beta)}-M_{k,j}^{(\beta)}}{(x_i-x_k)(x-x_i)} \frac{V'_2(y)-V'_2(y_j)}{y-y_j} \right>\, (ii)''\cr
\eea
Note that $(ii)''$ is the same as the last terms of $(iv)$ so that it cancels out. We can also split $(ii)'$ into a sum over $i,k$ minus the case $i=k$ which is identical to $(iii)$ except as a factor $\beta$. Therefore we can regroup $(ii)$, $(iii)$ and $(iv)$ to get:
\bea  \label{v} (ii)+(iii)+(iv)&=&(1-\beta) \sum_{i,j=1}^N \left<  \frac{1}{(x-x_i)^2} \frac{M_{i,j}^{(\beta)}}{I_\beta}\frac{V'_2(y)-V'_2(y_j)}{y-y_j}\right> \, (1)\cr
&&+\beta \sum_{i,j,k=1}^N \left<\frac{1}{x-x_i}\frac{1}{x-x_k} \frac{M_{i,j}^{(\beta)}}{I_\beta}\frac{V'_2(y)-V'_2(y_j)}{y-y_j}\right> \, (2)\cr
&&+\frac{N\beta}{T}\sum_{i,j=1}^N \left< \frac{y_j}{x-x_i} \frac{M_{i,j}^{(\beta)}}{I_\beta}\frac{V'_2(y)-V'_2(y_j)}{y-y_j} \right> \, (3)
\eea
Observe now, that we have:
\beq (1)=(1-\beta)\sum_{i,j=1}^N \left<  \frac{1}{(x-x_i)^2} \frac{M_{i,j}^{(\beta)}}{I_\beta}\frac{V'_2(y)-V'_2(y_j)}{y-y_j}\right>=(\beta-1)\frac{\partial}{\partial x} U_0(x,y)\eeq
Then we have also:
\bea \frac{T}{N\beta}\left< \frac{\partial}{ \partial V_1(x)} U_0(x,y)\right>_c&=&\sum_{i,k,j=1}^N\left<\frac{1}{x-x_i}\frac{1}{x-x_k} \frac{M_{i,j}^{(\beta)}}{I_\beta}\frac{V'_2(y)-V'_2(y_j)}{y-y_j}\right>\cr
&&-\sum_{i,j=1}^N\left<\frac{1}{x-x_i}\frac{M_{i,j}^{(\beta)}}{I_\beta}\frac{V'_2(y)-V'_2(y_j)}{y-y_j}\right>\left< \frac{1}{x-x_k}\frac{M_{i,j}^{(\beta)}}{I_\beta}\right>\cr
&&=\sum_{i,k,j=1}^N\left<\frac{1}{x-x_i}\frac{1}{x-x_k} \frac{M_{i,j}^{(\beta)}}{I_\beta}\frac{V'_2(y)-V'_2(y_j)}{y-y_j}\right>\cr
&&-W(x)U_0(x,y)
\eea
We recognize here one the second term of \eqref{v}:
\bea (2) &\Leftrightarrow& \frac{T}{N}\left< \frac{\partial}{ \partial V_1(x)} U_0(x,y)\right>_c+\beta W(x)U_0(x,y)\cr
 &\Leftrightarrow& \beta U_1(x,y;x)+\beta W(x)U_0(x,y)
\eea
Eventually, we are left with $(3)$. We perform the change $y_j \leftrightarrow y_j-y+y$ and split it into two parts:
\bea
(3) &\Leftrightarrow& \frac{N\beta}{T}y U_0(x,y) -\frac{N\beta}{T}V'_2(y)\sum_{i,j=1}^N\left<\frac{1}{x-x_i}\frac{M_{i,j}^{(\beta)}}{I_\beta}\right>\cr
&&+ \frac{N\beta}{T}\sum_{i,j=1}^N\left<\frac{V'_2(y_j)}{x-x_i}\frac{M_{i,j}^{(\beta)}}{I_\beta}\right>
\eea

But remember that $\underset{j=1}{\overset{N}{\sum}} M_{i,j}^{(\beta)}=I_\beta$. and that from our first step we have \eqref{firstloop} so that eventually:
\beq (3) \Leftrightarrow \frac{N\beta}{T}y U_0(x,y)-\frac{N\beta}{T}V'_2(y)W(x)+\frac{N\beta}{T}(-N +x W(x))\eeq

Now we can put everything back together to get the loop equation:
\bea 0&=&-\frac{N\beta}{T} \left( V_1'(x)U_0(x,y)-P_0(x,y)\right)-(1-\beta)\frac{\partial}{\partial x} U_0(x,y)\cr
&&+\beta U_1(x,y;x)+\beta W(x)U_0(x,y)+\frac{N\beta}{T}y U_0(x,y)\cr
&&-\frac{N\beta}{T}V'_2(y)W(x)+\frac{N\beta}{T}(-N +x W(x))\cr
\eea
which can be rewritten (performing a multiplication by $-\frac{T}{N\beta}$) as \textbf{the master loop equation}:
\begin{equation} \label{loopequation}
\addtolength{\fboxsep}{10pt}
			\boxed{
				\begin{split}
				&\left(y-V'_1(x)+\frac{T}{N}W(x)+\hbar \partial_x\right)U_0(x,y)=(V'_2(y)-x)W(x)-P_0(x,y)\cr
			&+N-\frac{T}{N} U_1(x,y;x) \cr
				\end{split}
			}
\end{equation}

where we have defined $\hbar = \frac{T}{N}(1-\frac{1}{\beta})$ which we prefer to write
\beq
\encadremath{
\hbar = \frac{T}{N\sqrt\beta}\,\,\left( \sqrt\beta - \frac 1 {\sqrt\beta}\right).
}
\eeq

\subsection{Higher order loop equations}

With the same approach as before, one can deduce higher order loop equations for $U_n(x,y,\xi_1,\dots,,\xi_n)$. Indeed, if one look at the auxiliary integral:
\bea 0&=&\sum_{i,j=1}^N \int dXdY \,\frac{\partial}{\partial y_j}  \Big(e^{-\frac{N\beta}{T} \tr (V_1(X)+V_2(Y) )} \Big[\Delta(X)^{2\beta} \Delta(Y)^{2\beta} \frac{1}{x-x_i} M_{i,j}^{(\beta)}(X,Y)\,\,\, \cr
&&\left(\prod_{p=1}^n \tr \frac{1}{\xi_p-X}\right)_c - C_p(\xi_1,\dots,\xi_p) \Delta(X)^{2\beta} \Delta(Y)^{2\beta} \frac{1}{x-x_i} M_{i,j}^{(\beta)}\Big] \Big)\cr\eea
with
\beq C_p(\xi_1,\dots,\xi_p)=\int dXdZ  e^{-\frac{N\beta}{T} \tr (V_1(X)+V_2(Z) )} \Delta(X)^{2\beta} \Delta(Z)^{2\beta} I_\beta(X,Z) \left(\prod_{p=1}^n \tr \frac{1}{\xi_p-X}\right)_c\eeq

One finds:
\beq \sum_{i,j=1}^N \left< V_2'(y_j)\frac{1}{x-x_i}\frac{ M_{i,j}^{(\beta)}}{I_\beta}\prod_{p=1}^n \tr \frac{1}{\xi_p-X}\right>_c=x W_{n+1}(x,\xi_1,\dots,\xi_n)\eeq

Then the second step can be carried out using the integral:
\bea \label{newint}
0&=&\sum_{i,j=1}^N \int dXdY \, \frac{\partial}{\partial x_i} \Big(e^{-\frac{N\beta}{T} \tr (V_1(X)+V_2(Y) )} \Delta(X)^{2\beta} \Delta(Y)^{2\beta} \frac{1}{x-x_i} \cr
&&M_{i,j}^{(\beta)}\,\,\,\frac{V_2'(y)-V_2'(y_j)}{y-y_j} \prod_{p=1}^n \tr \frac{1}{\xi_p-X}\Big)\eea
First, we find the same terms as before (the new term $\underset{p=1}{\overset{n}{\prod}} \tr \frac{1}{\xi_p-X}$ being treated as a constant). This gives:
\bea \label{gen}&&-\frac{N\beta}{T} \Big( V_1'(x)U_n(x,y,\xi_1,\dots,\xi_n)-P_n(x,y,\xi_1,\dots,\xi_n)\Big)\cr
&&-(1-\beta)\frac{\partial}{\partial x} U_n(x,y,\xi_1,\dots,\xi_n)+\beta U_{n+1}(x,y;x,\xi_1,\dots,xi_n)\cr
&&+\beta W(x)U_n(x,y,\xi_1,\dots,\xi_n)+\frac{N\beta}{T}yU_n(x,y,\xi_1,\dots,\xi_n)\cr
&&+\beta \underset{ I\sqcup J= \{\xi_1,\dots,\xi_n\},\, I\neq \emptyset}{\sum} W_{1+|I|}(x,I)\,U_{|J|}(x,y;J) \cr
&&-\frac{N\beta}{T}V'_2(y)W_{n+1}(x,\xi_1,\dots,\xi_n) +x W_{n+1}(x,\xi_1,\dots,\xi_n)\Big)\cr
\eea 
However, the main difference is that now the derivative will also act on $\underset{p=1}{\overset{n}{\prod}} \tr \frac{1}{\xi_p-X}$. This gives an additional contribution:
\bea &&\sum_{i,j,k,i_K} \Big< \frac{1}{x-x_i}\frac{ M_{i,j}^{(\beta)}}{I_\beta}\frac{V_2'(y)-V_2'(y_j)}{y-y_j} \cr
&&\frac{1}{\xi_1-x_{i_1}}\dots \frac{1}{\xi_{k-1}-x_{i_{k-1}}} \frac{1}{(\xi_k-x_i)^2} \frac{1}{\xi_{k+1}-x_{i_{k+1}}}\dots  \frac{1}{\xi_n-x_{i_n}}\Big>\eea

Then observe the two identities:
\beq \frac{1}{(\xi_k-x_i)^2}=-\frac{\partial}{\partial \xi_k} \frac{1}{\xi_k-x_i}\eeq
and:
\beq \frac{1}{x-x_i}\frac{1}{\xi_k-x_i}= \frac{1}{x-\xi_k} \left( \frac{1}{\xi_k-x_i}-\frac{1}{x-x_i}\right)\eeq
Therefore the action of the derivative can be rewritten as:
\beq \label{deri}\sum_{k=1}^n\frac{\partial}{\partial \xi_k}\left( \frac{U_{n-1}(x,y,\xi_1,\dots,\xi_{k-1},\xi_{k+1},\dots,\xi_n)-U_{n-1}(\xi_k,y,\xi_1,\dots,\xi_{k-1},\xi_{k+1},\dots,\xi_n)}{x-\xi_k}\right)\eeq
 
Eventually, adding \eqref{gen} and \eqref{deri} to get \eqref{newint} we get the higher order loop equations:

\begin{equation} \label{loopequationn}
\addtolength{\fboxsep}{10pt}
			\boxed{
				\begin{split}
				&\left(y-V'_1(x)+\frac{T}{N}W(x)+\hbar \partial_x\right)U_n(x,y;\xi_1,\dots,\xi_n)+ \frac {T} {N} \,U_{n+1}(x,y;x,\xi_1,\dots,\xi_n)\cr
			& +  \frac {T}{ N} \underset{ I\sqcup J= \{\xi_1,\dots,\xi_n\},\, I\neq \emptyset}{\sum} W_{1+|I|}(x,I)\,U_{|J|}(x,y;J) + \frac{T}{N\beta}\underset{k=1}{\overset{n}{\sum}} \frac \partial {\partial \xi_k}\cr
& \left( \frac{U_{n-1}(x,y,\xi_1,\dots,\xi_{k-1},\xi_{k+1},\dots,\xi_n)-U_{n-1}(\xi_k,y,\xi_1,\dots,\xi_{k-1},\xi_{k+1},\dots,\xi_n)}{x-\xi_k}\right) \cr
& =(V'_2(y)-x)W_{n+1}(x,\xi_1,\dots,\xi_n)-P_n(x,y;\xi_1,\dots,\xi_n)\cr
				\end{split}
			}
	\end{equation}

These loop equations \eqref{loopequation}, \eqref{loopequationn} are exact equations satisfied by the correlation functions, they hold for every $N$ and every $\beta$.
Now, in order to solve them, we want to consider some limit and expansion around that limit, in which they can be solved order by order.

\subsection{Relationships between correlators}

The loop equations which we have written, involve the resolvent $W(x)$, as well as higher correlation functions $W_n$, and also some auxiliary functions $U_n$ and $P_n$.
However all of them are related.

First, notice that $P_n$ is the large $x$ polynomial part of $V'_1 U_{n}$:
\beq
P_n(x,y;x_1,\dots,x_n) = \left( V'_1(x)\,\, U_n(x,y;x_1,\dots,x_n) \right)_+
\eeq
Second, notice that the large $y$ behavior of $U_n$ is related to $W_{n+1}$:
\beq
U_n(x,y;x_1,\dots,x_n) \mathop{{\sim}}_{y\to\infty}  \td t_{d_2}\,\, y^{d_2-1}\,\,W_{n+1}(x,x_1,\dots,x_n) + O(y^{d_2-2})
\eeq
And then, the loop insertion operators allow to increase the index $n$ by 1.
Define:
\bea \label{insertionop}
 \frac{\d}{\d V_1(x)}&:=&-\sum_{k=1}^\infty \frac{1}{x^{k+1}}\,k\frac{\d}{\d t_{k-1}} \cr
\frac{\d}{\d V_2(x)}&:=&-\sum_{k=1}^\infty \frac{1}{y^{k+1}}\,k\frac{\d}{\d \td{t}_{k-1}}
\eea
These operators, called loop-insertion-operators, are formal and they have the property that:
\beq
\frac{\d V_j(x)}{\d V_l(x')}=\delta_{j,l}\,\frac{1}{x-x'}\virg \frac{\d V_j'(x)}{\d V_l(x')}=-\delta_{j,l}\,\frac{1}{(x-x')^2} 
\eeq
The main interest of these operators is that they can be used to connect the correlation functions $W_n(x_1,\dots,x_n)$ to the next correlation functions $W_{n+1}(x_1,\dots,x_{n+1})$:
\bea \frac{N\beta}{T}\, W(x)&=&\frac{\d F}{\d V_1(x)} %\,\, \virg \,\,  \frac{\d F_g}{\d V_1(x)}=W_1^{(g)}(x)
\cr
\frac{N\beta}{T}\,W_n(x_1,\dots,x_n) &=& \frac{\d W_{n-1}(x_1\dots,x_{n-1})}{\d V_1(x_n)}\cr
\frac{N\beta}{T}\,U_n(x,y;x_1,\dots,x_n) &=& \frac{\d U_{n-1}(x,y;x_1\dots,x_{n-1})}{\d V_1(x_n)}\cr
\eea
We also have
\bea
&&\frac{N\beta}{T}\,P_n(x,y;x_1,\dots,x_n) = \frac{\d P_{n-1}(x,y;x_1\dots,x_{n-1})}{\d V_1(x_n)} \cr
&&+  \frac{\partial}{\partial x_n} \,\,\frac{U_{n-1}(x,y;x_1\dots,x_{n-1})-U_{n-1}(x_n,y;x_1\dots,x_{n-1})}{x-x_n}\cr
\eea

\section{Topological expansion(s)}\label{sectoprecregimes}

One sees, that the loop equation \eqref{loopequation} involves $U_0$ and $U_1$, and similarly
 \eqref{loopequationn} involves $U_n$ and $U_{n+1}$.
This is why loop equations can only be solved perturbatively, order by order in some small parameter, in a regime where the $U_{n+1}$ term is subleading compared to the $U_n$ term.

We thus need to expand all our observables in powers of some parameter, and this expansion is usually called topological expansion. Most often for hermitian matrices, the small parameter is chosen to be $1/N$ where $N$ is the size of the matrices, i.e. the topological expansion is an asymptotic expansion for large random matrices.

Such a power series expansion does not always exist for convergent matrix models, but by definition it does for formal matrix models which is our context here (we do not interest ourselves in convergence here). For the subtle differences between convergent and formal matrix models in the hermitian case, we invite the interested reader to refer to \cite{Breakdown, eynconvergent}. 

\bigskip

We thus need a small parameter in which to compute an expansion.
For the hermitian case, the good choice is an expansion into powers of $1/N$ in the large $N$ limit. However, for $\beta\neq 1$, we can already see on loop equations, that the most interesting regime (in which loop equation are most conveniently solved) is a regime where $\hbar=O(1)$, and so is not simply $N\to\infty$.
In fact, depending on which application of random matrices one is interested in, several limit regimes can be interesting. Fortunately, all of them are related together, and knowing one expansion allows to recover the others. Let us present $3$ main regimes:

Before presenting those regimes, let us introduce some usual notations:
\bea
g_s = \frac{T}{N\sqrt{\beta}} & \qquad g_s={\rm string\,coupling\,constant}  \quad \cr
\epsilon_1 = g_s\,\sqrt\beta &\qquad 1^{\rm st}\, {\rm equivariant\, parameter}\cr
\epsilon_2 = -\,\frac{g_s}{\sqrt{\beta}} &\qquad 2^{\rm nd}\, {\rm equivariant\, parameter}\cr
\hbar = \epsilon_1+\epsilon_2  \cr
g_s^2=-\epsilon_1\epsilon_2\cr
\beta = -\, \frac{\epsilon_1}{\epsilon_2}
\eea
\medskip
Those notations are those used in different applications of matrix models, in particular string theory \cite{Dmodules, Fuji, HitchinSystems} and in relationship to the now famous AGT conjecture \cite{AGT}.

\subsection{Topological regime}

This is the regime where we expand in powers of $g_s$ at fixed $\hbar$.
This regime is the one interesting for applications to geometry and topological strings.
In this regime, we wish to expand:
\beq
F=\ln Z = \sum_{g=0}^\infty g_s^{2g-2} F_g(\hbar) = \sum_{g=0}^\infty \sum_{k=0}^\infty g_s^{2g-2}\,\,\hbar^k\,\, F_{g,k}
\eeq
Each coefficient $F_g(\hbar) = \underset{k=0}{\sum} \hbar^k F_{g,k}$ have a geometric interpretation. Notice that $F_{g,k}$ depends neither on $N$ nor on $\beta$ (but of course it depends on the potentials $V_1$, $V_2$ and on $T$).

In topological strings, $F_{g,k}$ is the generating function for counting equivariant Gromov-Witten numbers \cite{Bmodel}.

\subsection{Large $N$ regime}

This is an expansion in powers of $N$, at fixed $\beta$. This is the regime interesting for large random matrices. Most often, one is interested in specific values of $\beta$, like $\beta=1/2$ or $\beta=2$.
We shall write:
\beq
F=\ln Z = \sum_{l=0}^\infty (N/T)^{2-l}\,\, \hat F_l(\beta)
\eeq

This expansion is related to the topological expansion by
\beq \label{ruru}
\hat{F}_l(\beta)=\sum_{g=0}^{E(\frac{l}{2})} \beta^{1+g-l}(\beta-1)^{l-2g}F_{g,l-2g}.
\eeq
For example
\beq
\hat F_0(\beta) = \beta\,F_{0,0}
\virg
\hat F_1(\beta) =  (\beta-1)\,F_{0,1}
\eeq
\beq
\hat F_2(\beta) = F_{1,0} + \frac{(\beta-1)^2}\beta F_{0,2}
\virg
\hat F_3(\beta) = \frac{\beta-1}\beta F_{1,1} + \frac{(\beta-1)^3}{\beta^2} F_{0,3}
\eeq
and more generally by \eqref{ruru}

\subsection{WKB large $\beta$ regime}

This is an expansion in powers of $\beta$ at fixed $N$. This is the regime we shall consider below. It can be computed with standard WKB techniques, but we shall analyze it here with loop equations.

We expand in powers of $g_s=\frac{T}{N\sqrt\beta}$ at fixed $N$:
\beq
F=\ln Z = \sum_{h=0}^\infty g_s^{2h-2}\,f_h(N)
\eeq

This regime can be related to the topological regime as follows:
\bea\label{linktopoWkb}
f_h(N) 
&=& \sum_{g=0}^h\,\,\frac{(-1)^{g}}{g!}\, \sum_{k=0}^\infty \left(\frac{T}{N}\right)^{k-g}\,\, \frac{(k+g)!}{k!}\,\, F_{h-g,g+k}
\eea
For example
$$
f_0(N) = \sum_k \left(\frac{T}{N}\right)^k f_{0,k} = F_0(T/N)
\virg
f_1(N)  = F_1(\frac{T}{N}) - \frac{N}{T}\,F'_0(\frac{T}{N}).
$$

\subsection{Notations for the expansion}

In this article, we want to study perturbatively the BKW regime where $\beta \to \infty$, $N$ fixed, and we write
$g_s=\frac{T}{N\sqrt\beta}$.
% and $\hbar$ scales with $g_s$ like: $\hbar=\frac{T}{N}-\frac{N}{T}g_s^2$ and $\beta=\frac{T^2}{N^2g_s^2}\underset{g_s\to 0}\to\infty$. In this case, as mentioned in \ref{sectionNfixe} every function should be expanded like:
We thus expand all our correlation functions as:
\bea\label{topexpMod} W_n(x_1,\dots,x_n)&=&\left(\frac{N}{T}\right)^{n}\,\,\sum_{g=0}^\infty g_s^{2g-2+2n}\,\, W_n^{(g)}(x_1,\dots,x_n)\cr
U_0(x,y)&=&\frac{N}{T}\left(U_0^{(0)}(x,y)-x+V_2'(y)\right)+\frac{N}{T}\sum_{g=1}^\infty g_s^{2g}\, U_0^{(g)}(x,y)\cr
U_n(x,y;x_1,\dots,x_n)&=&\left(\frac{N}{T}\right)^{n+1}\,\,\sum_{g=0}^\infty g_s^{2g+2n}\, U_n^{(g)}(x,y;x_1,\dots,x_n)\cr
%P_0(x,y)&=&\frac{N}{T}\,P_0^{(0)}(x,y)+\frac{N}{T}\sum_{g=1}^\infty g_s^{2g}\, P_0^{(g)}(x,y)\cr
P_n(x,y;x_1,\dots,x_n)&=&\left(\frac{N}{T}\right)^{n+1}\,\,\sum_{g=0}^\infty  g_s^{2g+2n}\,\, P_n^{(g)}(x,y;x_1,\dots,x_n)\cr
\eea

\textbf{Note that the functions $W_n^{(g)}$, $U_n^{(g)}$, and $P_n^{(g)}$ are only functions of  $N$ (and of course of their corresponding $x_i$, of $T$ and of the potentials $V_1$, $V_2$) but not of $g_s$, $\beta$ or $\hbar$.}

Eventually, we define the numbers $f_g$ by the expansion of the logarithm of the partition function itself:
\beq Z=e^F \, \, \virg \,\, F=\sum_{g=0}^\infty g_s^{2g-2}\, f_g(N).
\eeq
%We recall that the parameters $g_s$ and $\hbar$ are given by \eqref{parameter}.
Again, $f_g(N)$ is also implicitly a function of $T$, $V_1$ and $V_2$, and we recall that $f_g(N)$ is related to $F_g(\hbar)$ which is the usual topological expansion.

\medskip
As we will see in the next sections, most of the formulas get nicer if we shift the first correlation function by the derivative of the potentials. Therefore, for convenience, we introduce the shifted functions:
\beq
Y(x):=(V'_1(x)-W_1^{(0)}(x))
\virg
X(y):=(V'_2(y)-\td{W}_1^{(0)}(y))
\eeq
It is also useful to redefine a shifted version of the polynomial $P^{(0)}(x,y)$ as:
\beq \label{spectr}
E(x,y):=(V'_1(x)-y)(V'_2(y)-x)-P_0^{(0)}(x,y) + T +\frac T N,
\eeq
which is a polynomial in both $x$ and $y$ (and it depends on $T$, $N$, $V_1$, $V_2$ but not on $\beta$).

%
%Moreover, as for the hermitian case, a central notion that will be developed in this article is the notion of {\bf spectral curve}, which is the equation obeyed by $W_1^{(0)}(x)$, i.e. by $Y(x)$. In our case, we will show later that it is given in terms of the following polynomial:

\section{Spectral curve and Loop equations as an ODE \label{secpsi}}

In this section, we shall see that the loop equation satisfied by the leading order $W_1^{(0)}$ of the resolvent  $W_1(x) = \left<\tr \frac{1}{x-M_1}\right>$ can be written as an ODE
\beq
{\cal E}\left(x,\frac{T}{N}\,\frac{\d}{\d x}\right).\psi(x)=0
\eeq
where ${\cal E}(x,y)$ is up to some shifts, the polynomial $E(x,y)$ introduced above in \eqref{spectr}.
This ODE is called the ``spectral curve" and plays a central role.

\subsection{Loop equations in $g_s$ expansion}

The leading order at large $\beta$ and fixed $N$ of loop equation \eqref{loopequation} (notice that we have $W_1^{(0)}(x)=V'_1(x)-Y(x)$ and $U_0^{(0)}(x,y)$ includes a shift by $x-V'_2(y)$) is:
%
%Now that we have the first loop equation \eqref{loopequation}, we can use the formal series expansion \eqref{Wn}, \eqref{Un} and \eqref{Pn} to project our loop equation into each power of $g_s$ and thus obtain an infinite set of equations that will help us to solve these loop equations. We also remind here that in the topological expansion we have performed a shift of the function $U_0^{(0)}(x,y)$ by $x-V'_2(y)$. Let's first look at the coefficients in front of the lowest power of $\beta$, we find: (we remind the reader that $Y(x):=(V'_1(x)-W_1^{(0)}(x))$):

\beq \label{Firstloopequation}
\left(y-Y(x)+\frac{T}{N}\partial_x\right)U_0^{(0)}(x,y)=(V_1'(x)-y)(V_2'(y)-x)-P_0^{(0)}(x,y)+T+\frac T N
\eeq
which is equivalent to (See \eqref{spectr}):
\beq \encadremath{
\left(y-Y(x)+\frac{T}{N}\partial_x\right)U_0^{(0)}(x,y)=E(x,y)}
\eeq
This is the master loop equation, also called the ``spectral curve". It generalizes the hermitian 2-matrix models's spectral curve \cite{staudacher, KazakovIsing, Boulatov, eynm2m, KazMar}.

\bigskip

If we expand in powers of $y$:
\beq
U_0^{(0)}(x,y) = \sum_{k=0}^{d_2} y^k\,\,U_{0,k}^{(0)}(x)
\virg
E(x,y) = \sum_{k=0}^{d_2+1} y^k\,\,E_k(x)
\eeq
equation \eqref{Firstloopequation} reads:

%

%\subsection{Polynomial expansion in $y^k$}

%Since we know that $U_n(x,y;x_1,\dots,x_n)$, $P_n(x,y;x_1,\dots,x_n)$ and $E(x,y)$ are polynomials of degree respectively $d_2-1$, $d_2-1$ and $d_2+1$ in $y$ (Note that $U_0^{(0)}(x,y)$ is of degree $d_2$ since we have specially added $V_2'(y)$), we can write them in the basis of monomials $y^k$ as:
%\bea \label{yprojection}
%U_n(x,y;x_1,\dots,x_n)&=& \sum_{k=0}^{d_2} U_{n,k}(x;x_1,\dots,x_n) y^k \cr
%P_n(x,y;x_1,\dots,x_n)&=& \sum_{k=0}^{d_2-1} P_{n,k}(x;x_1,\dots,x_n) y^k \cr
%E(x,y)&=& \sum_{k=0}^{d_2+1} E_k(x) y^k \cr
%U_n^{(g)}(x,y;x_1,\dots,x_n)&=& \sum_{k=0}^{d_2} U_{n,k}^{(g)}(x;x_1,\dots,x_n) y^k \cr
%P_n^{(g)}(x,y;x_1,\dots,x_n)&=& \sum_{k=0}^{d_2-1} P_{n,k}^{(g)}(x;x_1,\dots,x_n) y^k \cr
%\eea
%Since we have a lot of indices, we suggest the reader to see a subscript $_k$ and the lack of a variable $y$ as an indication that we have taken the coefficient of $y^k$. We now have all the needed definitions to get the loop equations. This will be done in the next section.

%

%%First, the projection of the master loop equation \eqref{loopequation} onto $y^k$ gives: (we use the notation \eqref{yprojection})

%%\beq \label{yprojected} U_{0,k-1}(x) +\left(\frac{T}{N}W(x)-V'_1(x)+\frac{T}{N}\partial_x\right)U_{0,k}(x)=\td{t}_kW(x)-xW(x)\delta_{k,0} -P_{0,k}(x)+N\delta_{k,0}-\frac{T}{N}U_{1,k}(x;x)\eeq

%It is useful to project the topologically expanded version of the master loop equation \eqref{Firstloopequation} and \eqref{loopequationg}, onto $y^k$. It gives:

\beq \encadremath{\label{yprojectedfirstloop} U_{0,k-1}^{(0)}(x) +\left(-Y(x)+\frac{T}{N}\partial_x\right)U_{0,k}^{(0)}(x) = E_k(x) 
%= V_1'(x)\td{t}_k -\td{t}_{k-1} +x\delta_{k,1} -xV_1'(x)\delta_{k,0}-P_{0,k}^{(0)} (x)
}\eeq
%and
%\bea \label{yprojectednextorders} &&U_{0,k-1}^{(g)}(x) +\left(-Y(x)+\frac{T}{N}\partial_x\right)U_{0,k}^{(g)}(x) =-\overset{g-1}{\underset{h=0}{\sum}}W_1^{(g-h)}(x)U_{0,k}^{(h)}(x)- P_{0,k}^{(g)}(x)\cr
%&&-U_{1,k}^{(g-1)}(x;x)+\frac{N}{T}\partial_x U_{0,k}^{(g-1)}(x)-\frac{N}{T}\delta_{g=1}\delta_{k=0}\cr
%\eea

\underline{Remark}: 
%In these equations, we have implicitly assumed that $U_{n,-1}=0$ (no term in $\frac{1}{y}$). Moreover remember that $U_{n}(x,y;x_1,\dots,x_n)$ is only a polynomial in $y$ of degree $d_2$. Therefore $\forall \,n \in \mathbb{N}:\, \, U_{n,d_2+1}(x;x_1,\dots,x_n)=0$ and so 
equation \eqref{yprojectedfirstloop} with $k=d_2+1$ gives us that:
\beq \label{d2}\encadremath{U_{0,d_2}^{(0)}(x)=-\td{t}_{d_2}}\eeq
In the same way, since $P_{0}^{(0)}(x,y)$ is only a polynomial in $y$ of degree $d_2-1$ we have also:
\beq \label{d2-1}\encadremath{U_{0,d_2-1}^{(0)}(x)=\td{t}_{d_2}(V'_1(x)-Y(x)) - \td{t}_{d_2-1}} \eeq
%The last two results are important since they will give us the starting point to obtain all the other correlation functions.

\subsection{The linear differential system}

%We are now interested in finding the function $U_0^{(0)}(x,y)$ or equivalently all the functions $U_{0,k}^{(0)}(x)$ for $0\leq k \leq d_2$. Note that the last section gives us already the coefficient in front of $y^{d_2}$ (i.e. $U_{0,d_2}^{(0)}(x)$) with \eqref{d2} and a relation between $U_{0,d_2-1}^{(0)}(x)$ and $Y(x)$ with \eqref{d2-1}. 
In order find the lower coefficients $U_{0,k}^{(0)}(x)$ with $k=0,\dots,d_2-2$ we introduce for convenience the functions $\psi_k(x)$ ($0\leq k \leq d_2$) defined by:

\beq \label{connection}
U_{0,k}(x) \overset{\text{def}}{=} \frac{\psi_k(x)}{\psi(x)} \,\,\, ,\,\,\, W_1^{(0)}(x)\overset{\text{def}}{=}\frac{T}{N} \frac{\psi'(x)}{ \psi(x)}=\frac{T}{N} \frac{\partial}{\partial x}\ln \psi(x)
\eeq
which is equivalent to say that:
\beq \psi(x)=e^{\frac{N}{T}\int^x W_1^{(0)}(z)dz }\,\,\, ,\,\,\, \psi_k(x)=\psi(x)U_{0,k}^{(0)}(x)\eeq
By definition, the function $\psi(x)$ is only determined up to a global multiplicative constant which can be fixed by specifying the lower bound of the integral in the last formula. 
%Observe also that we can define the topological expansion of $\psi_{k}(x)$ similarly to $U_{0,k}(x)$:
%\beq \psi_k(x)=\frac{N}{T}\sum_{g=0}^\infty g_s^{2g} \psi_k^{(g)}(x)\eeq
%We stress here that $\psi(x)$ is connected directly to $W_1^{(0)}(x)$ and not to $W_1(x)$, that is to say that it only involve the leading order (regarding the topological expansion $g_s\to 0$) of the first correlation function.

\medskip
\medskip

%The main reason to introduce the functions $\psi(x)$ and $\psi_k(x)$ is that 
The first loop equation \eqref{yprojectedfirstloop} can now be rewritten into a \textbf{linear} differential equation (ODE):
\bea \label{recursion}
\psi_{k-1}(x) +\left(\frac{T}{N}\partial_x- V'_1(x)\right)\psi_k^{(0)}(x) 
%&=& \left(V_1'(x)\td{t}_k -\td{t}_{k-1} +x\delta_{k,1} -xV_1'(x)\delta_{k,0}-P_{0,k}^{(0)} (x)\right) \psi(x)\cr
&=&E_k(x) \psi(x)
\eea
and using $\psi_{d_2}(x)=-\td{t}_{d_2} \psi(x)$,
%Also, remember that for $k=d_2$ and $k=d_2-1$ we had some extra knowledge regarding the values of $U^{(0)}_{0,k}(x)$ (given by \eqref{d2} and\eqref{d2-1}). With the new functions $\psi_k(x)$ it gives:
%\beq \label{psid2}
%\psi_{d_2}(x)=-\td{t}_{d_2} \psi(x)
%\eeq
%and
%\beq \label{psid2-1}
%\psi_{d_2-1}(x)=\td{t}_{d_2} \frac{T}{N} \psi'(x) - \td{t}_{d_2-1} \psi(x)
%\eeq
%
%Using \eqref{psid2}, we can replace $\psi(x)$ by its relationship to $\psi_{d_2}(x)$ in \eqref{recursion}. Eventually 
this leads to a set of linear equations that can be written into a matrix form: (equivalent to \eqref{recursion})
\bea \label{www}
\frac{T}{N} \frac{d}{dx} 
\begin{pmatrix}\psi_{d_2}(x) \cr \psi_{d_2-1}(x) \cr \vdots \cr \psi_0(x)\end{pmatrix}
&=&
\begin{pmatrix}
\frac{\td{t}_{d_2-1}}{\td{t}_{d_2}} & -1 &0 &\dots & 0 \cr 
-\frac{E_{d_2-1}(x)}{ \td{t}_{d_2}} & V'_1(x) & \ddots &\ddots & \vdots \cr 
 \vdots&0 & \ddots & \ddots & 0 \cr 
-\frac{E_{1}(x)}{ \td{t}_{d_2}} & \vdots&\ddots & \ddots & -1 \cr 
-\frac{E_{0}(x)}{ \td{t}_{d_2}} &0 & \dots& 0& V'_1(x) \cr 
\end{pmatrix}
\,\,
\begin{pmatrix}\psi_{d_2}(x) \cr \psi_{d_2-1}(x) \cr \vdots \cr \psi_0(x)\end{pmatrix}\cr
&=&\mathcal{C}(x)\begin{pmatrix}\psi_{d_2}(x) \cr \psi_{d_2-1}(x) \cr \vdots \cr \psi_0(x)\end{pmatrix}
\eea
Therefore, we have obtained a \textbf{system of first order linear ordinary differential equations given by a companion-like matrix}. The determinant of the matrix $\mathcal{C}(x)$ is given by the standard formula for the determinant of a companion matrix:
\beq \encadremath{
- \td{t}_{d_2}\,\det((y-V'_1(x))Id+\mathcal{C}(x)) = E(x,y) }
\eeq
That is to say that the characteristic polynomial of the matrix $\mathcal{C}(x)$ is exactly our function $E(x,y)$.

There is no known notion of integrable system associated to the arbitrary $\beta$ two matrix model, but the fact that the spectral curve turns out to be a characteristic polynomial (also called spectral determinant), is very suggestive from the point of view of integrability \cite{BertolaMarchal, Marco2, BBT}, this could be a promising route, and needs to be further investigated...

\subsection{ODE satisfied by $\psi(x)$}

The matrix equation \eqref{www} can be used to find a linear ODE satisfied by the function $\psi(x)$. Indeed we can rewrite it in following way:
\bea \left(V_1'(x)-\frac{T}{N} \partial_x\right)\psi_{d_2}(x)&=& \frac{E_{d_2}(x)}{ \td{t}_{d_2}}\psi_{d_2}(x)+\psi_{d_2-1}(x)\cr
\left(V_1'(x)-\frac{T}{N} \partial_x\right)\psi_{d_2-1}(x)&=&\frac{E_{d_2-1}(x)}{ \td{t}_{d_2}}\psi_{d-1}(x)+\psi_{d_2-2}(x)\cr
&\dots& \cr
\left(V_1'(x)-\frac{T}{N} \partial_x\right)\psi_{1}(x)&=& \frac{E_{1}(x)}{ \td{t}_{d_2}}\psi_{d_2}(x)+\psi_{0}(x)\cr
\left(V_1'(x)-\frac{T}{N} \partial_x\right)\psi_{0}(x)&=& \frac{E_{0}(x)}{\td{t}_{d_2}}\psi_{d_2}(x)\cr
\eea

Multiplying line $k$ on the left by $\left(V_1'(x)-\hbar \partial_x\right)^{d_2-k}$ and summing the whole set of equations give that:
\beq \left(V_1'(x)-\frac{T}{N} \partial_x\right)^{d_2+1}\psi(x)-\sum_{k=0}^{d_2}\left(V_1'(x)-\hbar \partial_x\right)^k \left(E_{k}(x) \psi(x)\right)=0\eeq
which can be rewritten as:
\beq \label{ODE}\encadremath{ \hat y=\frac{T}{N} \partial_x \,\,\,\, :\,\,\,\,  \left(E(x,V_1'(x)-\hat y)\right)\psi(x)\overset{\text{def}}{=}\left(\sum_{k=0}^{d_2+1} (V_1'(x)-\hat y)^k\,E_k(x)\,\, \right) \psi(x)=0} \eeq
Note in particular that since $x$ and $\hat y$ do not commute ($[\hat y,x]=\frac{T}{N}$), we must specify the position of the $\hat y$ compared to $x$. In the last formula, it is implicitly assumed that powers of $V'_1-\hat y$ are always to the left of all powers of $x$, and:
\beq \left(V_1'(x)-\hat y\right)^k= \underbrace{\left(V_1'(x)-\hat y\right)\left(V_1'(x)-\hat y\right)\dots \left(V_1'(x)-y\right)}_{k \text{ times}}\eeq

Eventually we find that $\psi(x)$ must satisfy a linear ODE of order $d_2+1$, with polynomial coefficients, given by the quantized spectral curve $E(x,\hat y)$. This generalizes the Schr\"{o}dinger equation in the case of the arbitrary-$\beta$ one-matrix model \cite{MoiBertrand}.

\subsection{Quantum spectral curve}

The spectral curve $E(x,Y(x))=0$ in the hermitian case, gets replaced by
\beq
[\hat y,x]=\frac{T}{N} \qquad , \quad E(x,V'_1(x)-\hat y)\,.\,\psi(x)=0.
\eeq
This can be interpreted as having now  a {\bf quantum} spectral curve.

\subsection{Non-unicity of the solutions of the loop equations}

In order to identify a solution of loop equations, it remains to compute the polynomial coefficients $E_k(x)$ of this ODE, and choose a solution $\psi$.
 
Then, as soon as $\psi(x)$ is chosen, we can find $\psi_{d_2}(x)$ by using $\psi_{d_2}(x)=-\td{t}_{d_2} \psi(x)$. The other functions $\psi_k(x)$ follow by a descending recursion with the help of \eqref{recursion}. Eventually, the connection between $\psi_k(x)$ and $U_{k}^{(0)}(x)$ is given by \eqref{connection} so that the polynomial $U^{(0)}(x,y)$ is known.

One may think that the knowledge of the loop equations would be enough to determine all the correlation functions by solving them properly. But we remind the reader here that in general the loop equations (even in the hermitian case) have infinitely many solutions. 

\medskip

Even if \eqref{www} gives us some linear differential equations with polynomial coefficients, the main problem, also present in the hermitian case, is that the function $E(x,y)$ is not known at the moment. Indeed, in the definition of $E(x,y)$, we face the polynomial $P_0^{(0)}(x,y)$ whose coefficients are unknown. 
The coefficients of the polynomial $P^{(0)}_0(x,y)$ are not given by the loop equations, they have to be determined by some other informations which we have not used so far.

Therefore, in order to pick the proper solution of the loop equations one need to understand the missing information in the loop equations. When dealing with convergent matrix models, the answer is quite simple: the missing information is about the choice of the contour of integration of the model. Indeed, we have assumed here that the eigenvalues of the matrices were real, but one could take a different one-dimensional contour and get the same loop equations as the ones we have derived here (as soon as there is no hard edges, i.e. no boundary terms in integration by parts). Nevertheless, different integration paths would lead to different correlation functions and thus to different solutions of the loop equations. In the case of formal matrix models, the path of integration and the associated convergence is pointless and some other considerations have to fix the choice of the solution of the loop equations (in the hermitian case it is the notion of ``filling fractions"). In this article we will restrict ourselves (specifically at equation \eqref{BetheAnsatz}) to a specific kind of solutions of the loop equations because it corresponds to the less technical possible case. 

Let us just mention that, in the hermitian case, the homology space of possible integration paths, has exactly the same dimension as the number of unknown coefficients of the polynomial $P_0^{(0)}(x,y)$, and basically, every choice of $P_0^{(0)}(x,y)$ is acceptable, and corresponds to a given integration path for the eigenvalues. However, finding the integration path from the knowledge of $P_0^{(0)}(x,y)$ is not so easy in general.

In the non-hermitian case $\beta\neq 1$, it is not even clear what the homology space of possible integration paths is, and a good understanding of the missing information, its connections to the initial model and of the generalization of the notion of ``filling fractions" for our case is still missing and is postponed for a future article.

\section{The Bethe ansatz \label{ansatz}}

\subsection{WKB approximation}

Here, we work at large $\beta$ and fixed $N$ (working at large $\beta$ or small $g_s$ when $N$ is fixed is completely equivalent since they are related by $g_s=\frac{T}{N\sqrt{\beta}}$), so that our integral:
\beq
Z = \int dx_1\dots dx_N\,\,dy_1\dots dy_N\,\, \ee{-\frac{1}{g_s^2}\,\, {\cal A}[X,Y]}
\eeq
where $X={\rm diag}(x_1,\dots,x_N)$, $Y={\rm diag}(y_1,\dots,y_N)$, with the action
\bea\label{eqdefAction}
{\cal A}[X,Y] &=& \frac{T}{N}\,\, (\Tr V_1(X)+\Tr V_2(Y)) - 2\frac{T^2}{N^2}\ln {\Delta(X)}- 2\frac{T^2}{N^2}\ln {\Delta(Y)} \cr
&&- \underset{\beta \to \infty}{\text{lim}}\,g_s^2\,\ln \left( I_{\beta}(X,Y)\right) + O(g_s)
\eea
can be computed by standard saddle point approximation, and in particular, we find that 
\beq
\frac N T W_1^{(0)}(x) = \underset{\beta \to \infty\,,\, N \text{ fixed}}{\text{lim}} \,\,\, W_1(x) = \sum_{i=1}^N \frac 1 {x-\bar x_i}
\eeq
i.e. it is a rational fraction with $N$ poles $\bar x_i$ which are the saddle points for $X$, i.e. the points such that
\beq
\left.\frac\partial{\partial x_i}\right|_{x_i=\bar x_i} {\cal A}_{\beta=\infty}[X,Y] = 0
\qquad , \qquad 
\left.\frac\partial{\partial y_i}\right|_{y_i=\bar y_i} {\cal A}_{\beta=\infty}[X,Y] = 0.\eeq

Therefore, $W_1^{(0)}(x) = \frac{T}{N}\,\,\frac{\psi'}{\psi}$ must be a rational function with $N$ poles, i.e. $\psi(x)$ is a polynomial of degree $N$. This is the case we will develop below and that corresponds to the generalization of the solution presented in \cite{MoiBertrand} in the case of a Schr\"{o}dinger equation.

However, we would like to stress again that this is not the only possible regime. Indeed, we could also try to study directly the ``topological regime" where $N\to \infty$, $T$ fixed, and $\beta\to 0$ in such a way that $\hbar=\frac{T}{N}\left(1 -\frac{1}{\beta}\right)$ remains finite. This regime is more interesting for applications to string theory and AGT conjecture \cite{AGT, HitchinSystems, Fuji, Dmodules, ACDKV}.
In this case, $\psi(x)$ would not be a polynomial, instead, the coefficients of $E(x,y)$, i.e. the choice of $W_1^{(0)}(x)$ would be dictated by asymptotic behaviors in accordance with the regime studied.
This is more challenging, we have done it for the 1-matrix model case \cite{MoiLeonidBertrand, MoiLeonidBertrand2}, and we plan to do it for multi-matrix models in the near future.

\textbf{Therefore in the rest of the article, we restrict ourselves to the ``Dirac" case (the terminology ``Dirac" is used here to remind that the density of eigenvalues is a discrete measure, i.e. a sum of $\delta$-functions, i.e. a ``Dirac comb"), where $N$ is fixed and $\beta \to \infty$. In particular it correspond to assume that $\psi(x)$ is a polynomial (whose degree is the number $N$ of eigenvalues).}

\subsection{Introduction of the ansatz in the equations}

So, let us look for a solution where $\psi(x)$ is polynomial in $x$. Then it is clear from the recursion relation \eqref{recursion} that all the $\psi_k(x)$'s are also polynomials. We will use the following notation:
\beq \label{BetheAnsatz}
\encadremath{\psi(x) = \prod_{i=1}^N (x-s_i)}
\eeq
So that:
\beq
Y(x) = V'_1(x) - \frac{T}{N}\sum_{i=1}^N \frac{1}{x-s_i}
\eeq
Here we have simply assumed that $\psi(x)$ is a monic (remember that $\psi(x)$ is only determined up to a multiplicative constant) polynomial of degree $N$ and labeled $(s_i)_{i=1,\dots,N}$ its complex zeros. Note that this ansatz is very restrictive, since usually a linear ODE of the kind \eqref{ODE} does not admit polynomial solution if the coefficients $E_k(x)$ are generic. Then, from the definition of $U_0(x,y)$, it is also trivial that $U_{0,k}^{(0)}(x)$ can only have simple poles at the $s_i$'s. Taking into account the term in $V_2'(y)-x$ in the definition we have:
\beq \label{Uk}
\encadremath{\forall \, 0\leq k\leq d_2 : \, \, U_{0,k}^{(0)}(x) = \sum_{i=1}^N \frac{u_{k,i}}{x-s_i} - \td{t}_{k} + \delta_{k,0} x}
\eeq
where the $u_{k,i}$ are at the moment unknown coefficients.

\subsection{Computation of the $s_i$'s and of the $u_{k,i}$'s}

Putting this ansatz back into the recursion relations \eqref{recursion} for the $U_{0,k}^{(0)}(x)$ and identifying the coefficient in $\frac{1}{x-s_i}$, we get for all $k$ and $i$: (Note that the double poles cancel)
\beq \label{ukirec}
u_{k-1,i} - (V'_1(s_i)-\frac{T}{N}\sum_{j\neq i}\frac{1}{s_i-s_j}) u_{k,i} + \frac{T}{N} \left(\sum_{j\neq i} \frac{u_{k,j}}{s_i-s_j} - \td{t}_{k} + \delta_{k,0} s_i\right)  = 0
\eeq

In order to solve this recursion we introduce the following matrices of size $N \times N$:
\beq
\encadremath{S={\rm diag}(s_1,\dots,s_N)
\virg
B = 
\left\{ 
\begin{array}{l}
B_{i,i} = V'_1(s_i)-\frac{T}{N}\underset{j\neq i}{\sum}\frac{1}{s_i-s_j} \cr
B_{i,j} = -\frac{T}{ N(s_i-s_j)}
\end{array}
\right.}
\eeq
and define the vectors :
\beq
\encadremath{\vec{e}=\begin{pmatrix}1\cr 1\cr \vdots \cr 1\end{pmatrix} \virg \vec{u}_{k}=\begin{pmatrix}u_{k,1}\cr u_{k,2}\cr \vdots \cr u_{k,N}\end{pmatrix}}
\eeq
Then the previous set of equations turns into:
\beq
\vec{u}_{k-1} - B \vec{u}_{k} = \frac{T}{N} (\td{t}_{k} - \delta_{k,0} S) \vec{e} 
\eeq
Remember again that we have some extra knowledge for $k=d_2$ and $k=d_2-1$ (Cf. \eqref{d2} and \eqref{d2-1}): 
\beq
\vec{u}_{d_2}=\vec{0}
\virg
\vec{u}_{d_2-1}=\frac{T}{N}\td{t}_{d_2}\vec{e}.
\eeq
Starting from $\vec{u}_{d_2-1}=\frac{T}{N} \td{t}_{d_2}\vec{e}$ and using the recursion relation, we have:
$$\vec{u}_{d_2-2}=B\vec{u}_{d_2-1}+ \frac{T}{N} \td{t}_{d_2-1}\vec{e}=\frac{T}{N}\left(\td{t}_{d_2}B\vec{e}+\td{t}_{d_2-1}\vec{e}\right)$$
$$\vec{u}_{d_2-3}=B\vec{u}_{d_2-2}+ \frac{T}{N} \td{t}_{d_2-2}\vec{e}=\frac{T}{N}\left(\td{t}_{d_2}B^2\vec{e}+\td{t}_{d_2-1}B\vec{e}+\td{t}_{d_2-2}\vec{e}\right)$$
down to
$$\vec{u}_{0}=B\vec{u}_{1}+ \frac{T}{N} \td{t}_{1}\vec{e}=\frac{T}{N}\left(\td{t}_{d_2}B^{d_2-1}\vec{e}+\td{t}_{d_2-1}B^{d_2-2}\vec{e}+\dots+\td{t}_{1}\vec{e}\right)$$
$$\label{u_0} -B\vec{u}_0=\frac{T}{N}(\td{t}_0-S)\vec{e}$$
In fact all the previous relations can be summarized into the matrix form:

\beq \label{u_{k,i}}  \vec{u}_k=\frac{T}{N} \sum_{p=0}^{d_2-k-1} \td{t}_{k+p+1} B^p \vec{e} \eeq
or else in components:
\beq \label{uki}
u_{k,i}=\frac{T}{N}\sum_{j=1}^N\sum_{p=0}^{d_2-k-1} \td{t}_{k+p+1} (B^p)_{i,j} \eeq

%Hence, since we know every $(u_{d_2,i})_{i=1,\dots,N}$ and every $(u_{d_2-1,i})_{i=1,\dots,N}$, we can compute explicitly all the coefficients $u_{k,i}$ for all values of $0\leq k \leq d_2$ and $1 \leq i \leq N$.

\medskip
\medskip
\medskip

Eventually using the fact that $-B\vec{u}_0=\frac{T}{N}(\td{t}_0-S)\vec{e}$ we eventually get:
\beq 
(V'_2(B)-S)\vec{e}=0
\eeq
On the other hand, from the definition of $B$ we have:
\beq
0 = \vec{e}^{\,t}(V'_1(S)-B)
\virg
[B,S] = \frac{T}{N} (\vec{e} \vec{e}^{\,t} - {\rm Id}) 
\eeq
Therefore we obtain a system of equations determining the roots $(s_i)_{1\leq i \leq N}$:
\beq \label{Bethe ansatz}
\encadremath{
(V'_2(B)-S)\vec{e}=0
\virg
\vec{e}^{\,t}(V'_1(S)-B)=0
\virg
[S,B] = \frac{T}{N} ({\rm Id}- \vec{e} \vec{e}^{\,t})
}
\eeq

The last system of equations \eqref{Bethe ansatz} gives us a complete set of equations to compute all the $s_i$'s. In fact, this corresponds to a  Bethe ansatz \cite{BabBetGaudin} found in \cite{MoiBertrand} for the arbitrary-$\beta$ one-matrix model (where the authors got in the same context: $V'(s_i)=\frac{T}{N}\underset{j\neq i}{\sum}\frac{1}{s_i-s_j}$).

\medskip
{\underline{Remark:}} \eqref{Bethe ansatz} is a set of algebraic equations which determine the $s_i$'s, and the solution is in general not unique.
The choice of a solution, is also a choice of a saddle point for computing our eigenvalue integral at large $\beta$, and thus it is related to a choice of integration path for the eigenvalue integral.
The number of solutions of \eqref{Bethe ansatz}, coincides with the dimension of the homology space of all possible integration paths.

\subsection{Rewriting $U_0^{(0)}(x,y)$ in terms of $S$ and $B$}

With the previous results, we have that:
\bea U_0^{(0)}(x,y)&=&x-V'_2(y)+\sum_{k=0}^{d_2}\sum_{i=1}^{N} \frac{u_{k,i}}{x-s_i}y^k \cr
 &=&x-V'_2(y)+\frac{T}{N}\sum_{k=0}^{d_2}\sum_{i,j=1}^{N}\sum_{p=0}^{d_2-k-1} \td{t}_{k+p+1}\frac{1}{x-s_i} (B^p)_{i,j}y^k \cr
\eea
Moreover we have:
\beq \label{comb}\sum_{k=0}^{d_2}\sum_{p=0}^{d_2-k-1}\td{t}_{k+p+1}B^p y^k=\sum_{k=0}^{d_2}\sum_{p=0}^{k-1}\td{t}_k y^p B^{k-1-p}=\frac{ V'_2(y)-V'_2(B)}{y-B}\eeq
so that:
\bea U_0^{(0)}(x,y)&=&x-V'_2(y)+\frac{T}{N}\sum_{i,j=1}^{N}\left(\frac{ V'_2(y)-V'_2(B)}{y-B}\right)_{i,j}\frac{1}{x-s_i}  
%\cr &=&\frac{T}{N}\vec{e}^{\,\, t} \frac{1}{x-S}\frac{ V'_2(y)-V'_2(B)}{y-B}\vec{e}+x-V'_2(y)
\eea
Hence in the context of the Bethe ansatz, we can express easily the function $U_0^{(0)}(x,y)$ by:
\beq \encadremath{U_0^{(0)}(x,y)=\frac{T}{N}\vec{e}^{\,\, t} \frac{1}{x-S}\frac{ V'_2(y)-V'_2(B)}{y-B}\vec{e}+x-V'_2(y)}
\eeq

\subsection{Rewriting $P_0^{(0)}(x,y)$ in terms of $S$ and $B$}

We start with equation \eqref{Firstloopequation} giving that:
\beq \left(y-V_1'(x)+W(x)+\frac{T}{N}\partial_x\right)U_0^{(0)}(x,y)=(V_1'(x)-y)(V_2'(y)-x)-P_0^{(0)}(x,y)+T+\frac{T}{N}\eeq
We can compute the l.h.s. with the help of the definition \eqref{Uk}:
\bea &&\left(y-V_1'(x)+W(x)+\frac{T}{N}\partial_x\right)U_0^{(0)}(x,y)= (V_1'(x)-y)(V_2'(y)-x)+\frac{T}{N}\cr
&&+ (y-V_1'(x))\sum_{k=0}^{d_2}\sum_{i=1}^N \frac{u_{k,i}}{x-s_i} y^k -\frac{T}{N}\sum_{k=0}^{d_2}\sum_{i=1}^N \frac{u_{k,i}}{(x-s_i)^2} y^k\cr
&&+\frac{T}{N} \sum_{k=0}^{d_2}\sum_{i=1}^N \frac{u_{k,i}}{(x-s_i)(x-s_j)} y^k +\frac{T}{N} (x-V_2'(y))\sum_{j=1}^N \frac{1}{x-s_j}\cr
&&=(V_1'(x)-y)(V_2'(y)-x)+\frac{T}{N}+ \sum_{k=0}^{d_2}\sum_{i=1}^N \frac{u_{k-1,i}}{x-s_i} y^k -\sum_{k=0}^{d_2}\sum_{i=1}^N \frac{u_{k,i}V_1'(x)}{x-s_i} y^k \cr
&&-\frac{T}{N}\sum_{k=0}^{d_2}\sum_{i=1}^N \frac{u_{k,i}}{(x-s_i)^2} y^k +\frac{T}{N} \sum_{k=0}^{d_2}\sum_{i=1}^N \frac{u_{k,i}}{(x-s_i)(x-s_j)} y^k\cr
&& +T +\frac{T}{N} \sum_{i=1}^N \frac{(s_i-V_2'(y))}{x-s_i}
\eea
Note that we have used the fact that $u_{d_2,i}=0$. With the help of the recursion relation for the $u_{k,i}$'s \eqref{ukirec} we find:
\bea
&&\left(y-V_1'(x)+W(x)+\frac{T}{N}\partial_x\right)U_0^{(0)}(x,y)=(V_1'(x)-y)(V_2'(y)-x)+\frac{T}{N}+T \cr
&&+\sum_{k=0}^{d_2}\sum_{i=1}^N \frac{(V_1'(s_i)-V_1'(x))u_{k,i}}{x-s_i} y^k -\sum_{k=0}^{d_2}\sum_{j\neq i}^N \frac{u_{k,i}+u_{k,j}}{(x-s_i)(s_i-s_j)} y^k \cr
&&+\frac{T}{N}\sum_{k=0}^{d_2}\sum_{j \neq i}^N \frac{u_{k,i}}{(x-s_i)(x-s_j)} y^k
\eea

Observe now that we have the identity:
\beq \forall \,  0\leq k\leq d_2: \, -\sum_{j\neq i}^N \frac{u_{k,i}+u_{k,j}}{(x-s_i)(s_i-s_j)} +\sum_{j \neq i}^N \frac{u_{k,i}}{(x-s_i)(x-s_j)}=0\eeq
Moreover we have also (using \eqref{comb}):
\beq \sum_{k=0}^{d_2}\sum_{i=1}^N \frac{(V_1'(s_i)-V_1'(x))u_{k,i}}{x-s_i} y^k=-\frac{T}{N}\vec{e}^{\,\, t} \frac{V_1'(x)-V_1'(S)}{x-S} \frac{ V_2'(y)-V_2'(B)}{y-B} \vec{e}\eeq
so that we are left with:
\beq \encadremath{ P_0^{(0)}(x,y)=\frac{T}{N}\vec{e}^{\,\, t} \frac{V_1'(x)-V_1'(S)}{x-S} \frac{ V_2'(y)-V_2'(B)}{y-B} \vec{e} }\eeq

%\subsection{Conclusion}
 
Eventually, we can summarize the results in the following way. With \eqref{Bethe ansatz} we can compute all the $s_i$'s and then using the relations \eqref{u_{k,i}} we can get every $u_{k,i}$ ($0\leq k \leq d_2$ and $1 \leq i \leq N$). Hence we have found an algorithm to compute explicitly all $U_k^{(0)}(x)$'s or equivalently every $\psi_k(x)$'s (and $\psi(x)$), that is to say that we have the solution of the loop equation at the dominant order in the studied limit. In particular, assuming that $\psi(x)$ is polynomial in $x$ is sufficient to determine it completely which indicates that the structure of the loop equations are very rigid and hides integrable structure. 

\subsection{Yang-Yang Variational approach of the Bethe ansatz}\label{secYangYang}

%The last system of equations \eqref{Bethe ansatz} might seem rather surprising. But in fact we will show in this section that 
The Bethe ansatz equation \eqref{Bethe ansatz} can also be obtained from a variational approach with a Yang--Yang action. Consider the following functional :
\bea \frac{N}{T}\mathcal{S}(S,\td{S},A,\vec{u})&=& \tr V_1(S) +\tr V_2(\td{S})- \tr (SA\td{S}A^{-1}) - \frac{T}{N}\ln(\Delta(S))\cr
&&-\frac{T}{N}\ln(\Delta(\td{S}))+\frac{T}{N}\ln \det(A) -\frac{T}{N}\vec{u}^t \left(A\vec{e}-\vec{e}\right)\eea
where $S$ and $\td S$ are diagonal matrices of size $N\times N$, $A$ is a $N\times N$ invertible matrix, $\vec u$ is an $N$-dimensional vector, and remember that $\vec{e}^{\,t}=(1,\dots,1)$. We now look for the extremal values of $\mathcal{S}$. First taking the derivative relatively to $\vec{u}$ gives:
\beq \frac{\partial \mathcal{S}}{\partial \vec{u}}=0 \Leftrightarrow A\vec{e}=\vec{e} \Leftrightarrow A^{-1}\vec{e}=\vec{e} \eeq 
Performing the derivative relatively to $S={\rm diag}(s_1,\dots,s_N)$ yields:
\beq \label{rr}\frac{\partial \mathcal{S}}{\partial s_i}=0 \Leftrightarrow V_1'(s_i)- \left(A\td{S}A^{-1}\right)_{i,i}-\frac{T}{N}\sum_{j \neq i} \frac{1}{s_i-s_j}=0 \eeq 
Performing the derivative relatively to $\td{S}={\rm diag}(\td{s}_1,\dots,\td{s}_n)$ is the same:
\beq \label{kkk}\frac{\partial \mathcal{S}}{\partial \td{s}_i}=0 \Leftrightarrow V_2'(\td{s}_i)- (A^{-1}SA)_{i,i}-\frac{T}{N}\sum_{j \neq i} \frac{1}{\td{s}_i-\td{s}_j}=0 \eeq
Performing the derivative relatively to $A=(A_{i,j})_{i,j}$ is more difficult, consider an infinitesimal variation $A\to A+\delta A$, that yields:
\bea \label{aa}\delta \mathcal{S}&=&-\tr(S\delta A \td{S} A^{-1})+\tr(SA\td{S}A^{-1}\delta A A^{-1})-\frac{T}{N}\tr(A \delta A)+ \frac{T}{N}\vec{u}^{\, t} \delta A \vec{e}\cr
&=&-\tr((\delta A) A^{-1}\left([A\td{S}A^{-1},S]+\frac{T}{N} (Id- A\vec{e}\vec{u}^{\, t})\right))\cr
\eea
thus since $A$ is invertible and $A\vec e=\vec e$:
\beq \label{ff}\frac{\d \mathcal{S}}{\d A}=0 \Leftrightarrow [A\td{S}A^{-1},S]+\frac{T}{N}(Id-  \vec{e}\vec{u}^{\, t})=0\eeq
But since $S$ and $\td S$ are diagonal we have $[A\td{S}A^{-1},S]_{i,i}=0$, and the diagonal term of  \eqref{ff} gives
%$[A\td{S}A^{-1},S]-\frac{T}{N} Id+ \frac{T}{N} A\vec{e}\vec{u}^{\, t}=0$ and $A\vec{e}=\vec{e}$ imply that
 $\forall i\,:\, 1=(\vec{e} \vec{u}^{\, t})_{i,i}$ i.e. we have $\forall i\, :\, u_i=1$ that is to say 
 $$\vec{u}=\vec{e}.$$ 
Then a non diagonal element of the relation \eqref{ff} gives:
\beq \label{uu}\forall i\neq j\,:\, \left(A\td{S}A^{-1}\right)_{i,j}=\frac{T}{N(s_i-s_j)}\eeq
Combining \eqref{rr} and \eqref{uu} gives that \textbf{at the extremum we must have $A\td{S}A^{-1}=B$.}
Then, we can rewrite \eqref{aa} as:
\bea \delta \mathcal{S}&=&-\tr(S\delta A \td{S} A^{-1})+\tr(SA\td{S}A^{-1}\delta A A^{-1})-\frac{T}{N}\tr(A \delta A)+ \frac{T}{N}\vec{u}^{\, t} \delta A \vec{e}\cr
&=&-\tr(A^{-1}(\delta A)\left([\td{S},A^{-1}SA]+\frac{T}{N} (Id- \vec{e}\vec{u}^{\, t}A)\right))\cr
\eea
Therefore the extremum equation $\frac{\delta \mathcal{S}}{\delta A}=0$ gives also that 
\beq \label{yy}[\td{S},A^{-1}SA]+\frac{T}{N} (Id- \vec{e}\vec{u}^{\, t}A)=0\eeq
Again, since $S$ and $\td{S}$ are diagonal, we have $[\td S,A^{-1} S A]_{i,i}=0$, which implies that 
$\forall i\, :\, (\vec{e}\vec{u}^{\, t}A)_{i,i}=1$. Since at the extremum $\vec{u}=\vec{e}$, the last equation gives us that at the extremum:  
\beq \sum_{k=1}^N A_{k,j}=1 \, \Leftrightarrow\,  \vec{e}^{\,\,t} A=\vec{e}^{\,\,t} \eeq
Let's now introduce the matrix $\td{B}=A^{-1}SA$. Then taking a non-diagonal element of \eqref{yy}, we find:
\beq \forall j\neq i: \, \td{B}_{i,j}=(A^{-1}SA)_{i,j}=\frac{T}{N(\td{s}_i-\td{s}_j)}\eeq 
We can now rewrite \eqref{rr} as:
\bea &&\td{B}\vec{e}=A^{-1}SA\vec{e}=V'_2(\td{S})\vec{e}\cr
&\Rightarrow& A\td{B}\vec{e}=A V'_2(\td{S})\vec{e}\cr
&\Rightarrow& A\td{B}A^{-1}\vec{e}=A V'_2(\td{S})A^{-1}\vec{e}\cr
&\Rightarrow &S\vec{e}=V'_2(B)\vec{e}\cr
\eea

\textbf{Therefore if we combine all previous results we find that an extremum of $\mathcal{S}$ must satisfies the following identities:} 
\begin{itemize}
 \item $\vec{u}=\vec{e}$ 
 \item $A \vec{e}=\vec{e}$ and $\vec{e}^{\,\,t} A=\vec{e}^{\,\,t}$ 
 \item $[A\td{S}A^{-1},S]=\frac{T}{N}(-Id+ \vec{e}\vec{e}^{\,\, t})$ and $[\td{S},A^{-1}SA]+\frac{T}{N} (Id- \vec{e}\vec{e}^{\, t})=0$
 \item $A\td{S}A^{-1}=B$ and $S\vec{e}=V'_2(B)\vec{e}$
\end{itemize}
We recover all the equations of \eqref{Bethe ansatz} giving us a variational way of deriving all quantities. Note also that by using \eqref{rr} and \eqref{uu} we have: $B=A^{-1}\td{S}A$ leading to $\td{S}\vec{e}=V_1'(\td{B})\vec{e}$ where remember that $\td{B}=A^{-1}SA$. The last relations show that the system is completely symmetric in tilde quantities and non-tilde quantities. Eventually the relation $\td{S}\vec{e}=V_1'(\td{B})\vec{e}$ allows to determine directly the $\td{s}_i$ similarly to the fact that $S\vec{e}=V'_2(B)\vec{e}$ determines the $s_i$ at the extremum. The determination of $A$ at the extremum is unfortunately more complicated. Indeed, $S\vec{e}=V'_2(B)\vec{e}$ and $\td{S}\vec{e}=V_1'(\td{B})\vec{e}$ determines the $s_i$ and the $\td{s}_j$ at the extremum and then the matrices $B$ and $\td{B}$ at the extremum. The matrix $A$ can then be computed by the relations $B=A^{-1}\td{S}A$ and $A\vec{e}=\vec{e}$ or equivalently by $\td{B}=A^{-1}SA$ and $A\vec{E}=\vec{e}$. This leads to the system of equations:
\bea \forall \, i,j \,\,:\,\, 0&=&\sum_{k\neq j}\frac{A_{i,k}-A_{i,j}}{s_k-s_j}+A_{i,j}\left(V_1'(s_j)-\td{s}_i\right)\cr
1&=&\sum_{k=1}^N A_{i,k} \eea

\medskip

\br
It looks like the Yang--Yang action should be the $\beta\to\infty$ limit of the actual action ${\cal A}[X,Y]$ defined in \eqref{eqdefAction}.
Unfortunately, since the angular integral $I_\beta(X,Y)$ is not very well known, this cannot be proved at the moment. But vice versa, that would give us some knowledge of the angular integral $I_\beta$.
\er

\subsection{Computation of $W_2^{(0)}(x,x')$ and $W_3^{(0)}(x,x',x'')$ via the variational approach }

Since $W_{n+1}^{(0)} = \frac{d W_n^{(0)}}{d V_1} $, and since we know $W_1^{(0)}(x) = \frac{T}{N}\sum_i \frac{1}{x-s_i}$ where the $s_i$'s are obtained by the Bethe ansatz i.e. from the extremization of the Yang-Yang action ${\cal S}$, we can easily find $W_2^{(0)}$ and $W_3^{(0)}$.
%The knowledge of a reformulation of our problem in terms of a variational approach has many advantages. In particular as we will see in the next sections, it is possible to use it to get the first correlation functions $W_2^{(0)}(x_1,x_2)$ and $W_3^{(0)}(x_1,x_2,x_3)$ and especially to prove that they are symmetric functions. 
We will see later \ref{section 8} another way of getting formulas for these functions which can be extended by recursion but having the main disadvantage that the symmetry of the functions is not obvious at all.

\subsubsection{Computation of $W_2^{(0)}(x,x')$ via the variational approach \label{secVar20}}

Starting from
$$W_1^{(0)}(x)=\underset{i=1}{\overset{N}{\sum}}\frac{1}{x-s_i}
\virg 
W_2^{(0)}(x,x')=\frac{\partial}{\partial V_1(x')}W_1^{(0)}(x)$$
we get:
$$ W_2^{(0)}(x,x') = \sum_i \frac{1}{(x-s_i)^2}\,\,\frac{\d s_i}{\d V_1(x')}.$$
It remains to compute the quantities $\frac{\partial s_i}{\partial V_1(x')}$. 
For that we use that the $s_i$'s are the extrema of the functional $\mathcal{S}(s,\td{s},A,u)$.

%Let 

%In the last section we got an explicit functional:
%\beq \mathcal{S}(s,\td{s},A,u)\eeq
%whose extremum $(s_\text{extr},\td{s}_\text{extr},T_\text{extr},u_\text{extr})$ corresponds precisely to the solution of our matrix model limit. For example, $s_\text{extr}$ are the roots of the function $\psi(x)$. Moreover, we know that we can generate $W_2^{(0)}(x,x')$ with $W_1^{(0)}(x)$ by derivation relatively to $V_1(x')$ in the sense presented by \eqref{insertionop}:
%$$W_1^{(0)}(x)=\underset{i=1}{\overset{N}{\sum}}\frac{1}{x-s_i}
%\virg 
%W_2^{(0)}(x,x')=\frac{\partial}{\partial V_1(x')}W_1^{(0)}(x)$$
%which gives:
%$$ W_2^{(0)}(x,x') = \sum_i \frac{1}{(x-s_i)^2}\,\,\frac{\d s_i}{\d V_1(x')}$$
%we can compute $W_2^{(0)}(x,x')$ if can we can compute the quantities $\frac{\partial s_i}{\partial V_1(x')}$.
%In order to have clearer notation, 

We introduce the variable $R=(s,\td{s},A,u)$ as a ``global" variable to avoid detailing each of the cases. It is a vector of dimension $N+N+N^2+N=3N+N^2$. Therefore, the functional $\mathcal{S}(R)$ is a functional of $R$ whose extremum $R_\text{extr}$ gives us the solution of our model.
Thus we have for every variation $\delta$ and every component $i$:
\bea 0&=&\delta\left(\left( \frac{\partial\mathcal{S}}{\partial R_i}\right)_{|R=R_\text{extr}}\right)\cr
&=&\left( \delta\frac{\partial \mathcal{S}}{\partial R_i}\right)_{|R=R_\text{extr}}+\sum_{j=1}^{3N+N^2} \left(\frac{\partial^2\mathcal{S}}{\partial R_i\partial R_j}\right)_{|R=R_\text{extr}} \delta \left((R_\text{extr})_j\right)\cr
\eea
The first equality comes from the fact that $\frac{\partial\mathcal{S}}{\partial R_i}(R=R_\text{extr})=0$ since we place ourselves at the extremum. Let's introduce the Hessian of the functional at the extremum point: 
\beq \mathcal{H}_{r,s}\overset{\text{def}}{=}\left(\frac{\partial^2\mathcal{S}}{\partial R_r\partial R_s}\right)_{|R=R_\text{extr}}\eeq
It is a $3N+N^2 \times 3N+N^2$ matrix whose shape is given in appendix \ref{appendixmatricehessienne}. The last identity gives us that:

\bea\label{gg}  \left( \delta\frac{\partial \mathcal{S}}{\partial R_i}\right)_{|R=R_\text{extr}}&=&-\sum_{j=1}^{3N+N^2} \left(\frac{\partial^2\mathcal{S}}{\partial R_i\partial R_j}\right)_{|R=R_\text{extr}} \delta \left((R_\text{extr})_j\right)\cr
&\Leftrightarrow&\cr
 \delta\left( (R_\text{extr})_i\right)&=&-\sum_{j=1}^{3N+N^2} \left(\mathcal{H}^{-1}\right)_{i,j} \left( \delta\frac{\partial \mathcal{S}}{\partial R_j}\right)_{|R=R_\text{extr}}\cr\eea

Let's now specialize the last formula for the case when the variation $\delta$ is $\frac{\partial}{\partial V_1(x)}$. Since the potential $V_1$ only appears in the term $\Tr V_1(S)$ in the functional $\mathcal{S}$ it gives:
\beq \frac{\partial \mathcal{S}}{\partial V_1(x)}=-\Tr \frac{1}{x-S}\eeq
Since it only depends on $S$ (and not $\td{S},T$ or $u$), we see that in the last sum of \eqref{gg}, $j$ only varies from $1$ to $N$, corresponding to the variables $s_j$'s (i.e. $\frac{\partial}{\partial s_j} \frac{\partial}{\partial V_1(x)} \mathcal{S}=-\frac{1}{(x-s_j)^2}$ and $ \forall \, j>N: \frac{\partial}{\partial R_j} \frac{\partial}{\partial V_1(x)} \mathcal{S}=0$). Therefore we find (taking $1\leq i\leq N$):
\beq \label{ooooo}
\encadremath{\frac{\partial s_i}{\partial V_1(x)}=\sum_{k=1}^{N} \left(\mathcal{H}^{-1}\right)_{i,k} \frac{1}{(x-s_k)^2} }\eeq
Note that this last result is only a small fraction of the results contained in the formula:
\beq \delta\left( (R_\text{extr})_i\right)=-\sum_{j=1}^{3N+N^2} \left(\mathcal{H}^{-1}\right)_{i,j} \left( \delta\frac{\partial \mathcal{S}}{\partial R_j}\right)_{|R=R_\text{extr}} \eeq
depending on the choice of variation $\delta$ and the component $i$ you take. With the help of formula \eqref{ooooo}, it is straightforward to compute $W_2^{(0)}(x,x')$:
\beq \encadremath{
W_2^{(0)}(x,x')=\frac{T}{N}\sum_{i,j=1}^N \frac{\left(\mathcal{H}^{-1}\right)_{i,j}}{(x-s_i)^2(x'-s_j)^2}}\eeq

In particular, in this formalism, it is clear that \textbf{$(x,x') \mapsto W_2^{(0)}(x,x')$ is a symmetric function} since $\mathcal{H}$ and its inverse are symmetric matrices.  

\subsubsection{Computation of $W_3^{(0)}(x_1,x_2,x_3)$ via the variational approach \label{secVar30}}

We use the same method:
% as in the last section, we can apply the the derivation relatively to the potential on $W_2^{(0)}$ to get $W_3^{(0)}$ by the formula: 
$$W_3^{(0)}(x,y,z)=\frac{\partial}{\partial V_1(z)} W_2^{(0)}(x,y)$$
%The only difficult step in the computation in to compute the quantities $\left(\frac{\partial \mathcal{H}_{i,j}}{\partial V_1(x'')}\right)_{i,j=1..N}$. 
For any variation $\delta$ and every component $k$, we have:
 
\bea \delta\left( \left(\mathcal{H}^{-1}_{|\text{extr}}\right)_{i,j} \right)&=&\sum_{k=1}^{3N+N^2} \frac{\partial \left(\mathcal{H}^{-1}\right)_{i,j}}{\partial R_k} \, \delta( (R_{\text{extr}})_k) +\left(\delta \left(\mathcal{H}^{-1}\right)_{i,j}\right)\cr
&=&\sum_{k=1}^{3N+N^2} \frac{\partial \left(\mathcal{H}^{-1}\right)_{i,j}}{\partial R_k} \, \delta( (R_{\text{extr}})_k) -\left( \left(\mathcal{H}^{-1}\left(\delta\mathcal{H}\right) \mathcal{H}^{-1}\right)_{i,j}\right)\cr
\eea
We choose $\delta = \frac{d}{d V_1(x'')}$ and thus
$$
\delta {\mathcal H}_{i,j} = \frac{\d}{\d R_i}\,\frac{\d}{\d R_j}\,\frac{\d}{\d V_1(x'')}\,{\mathcal S}
= -\, \frac{\d}{\d R_i}\,\frac{\d}{\d R_j}\,\Tr \frac{1}{x''-S}
$$
i.e.
\beq
\delta {\mathcal H}_{i,j} = \left\{\begin{array}{ll}
- \frac{2\delta_{i,j}}{(x''-s_i)^3}  & \qquad {\rm if}\,\, i\leq N,\,\, j\leq N \cr
0 & \qquad {\rm otherwise, \, i.e.\,}\, i>N\,\,{\rm or}\,\,j>N.
\end{array}\right.
\eeq
%
%If we take $1\leq i,j\leq N $ and $\delta=\frac{\partial}{\partial V_1(x'')}$, then only the term in $\tr V_1(S)$ contributes in the last term, and we get:
%\beq \forall \, 1\leq i,j\leq N \, :\,\, \frac{\partial \mathcal{H}_{i,j}}{\partial V_1(x'')} =- \frac{2\delta_{i,j}}{(x''-s_i)^3} \virg 1\leq i\leq N \, , \, j>N:\frac{\partial \mathcal{H}_{i,j}}{\partial V_1(x'')} =0 \eeq
Moreover, we have
$$
\frac{\partial \left(\mathcal{H}^{-1}\right)_{i,j}}{\partial R_k}
= - \sum_{p,q=1}^{3N+N^2} \left(\mathcal{H}^{-1}\right)_{i,p} \frac{\partial^3 {\mathcal{S}}}{\d R_p\,\d R_q\,\partial R_k} \,\,\left(\mathcal{H}^{-1}\right)_{q,j}
$$
and
$$
\frac{\d R_k}{\d V_1(x'')} = - \sum_{l=1}^{3N+N^2} \left(\mathcal{H}^{-1}\right)_{k,l} \,\frac{\d^2 {\mathcal S}}{\d R_l\,\d V_1(x'')}
= \sum_{l=1}^{N} \left(\mathcal{H}^{-1}\right)_{k,l} \,\frac{1}{(x''-s_l)^2}
$$

%
%Therefore we get:
%\beq - \left(\mathcal{H}^{-1}\left(\frac{\partial}{\partial V_1(x'')}\mathcal{H}\right) \mathcal{H}^{-1}\right)_{i,j}=\sum_{k=1}^N
%\left(\mathcal{H}^{-1}\right)_{i,k}\left(\mathcal{H}^{-1}\right)_{k,j}\frac{2}{(x''-s_k)^3}\eeq
%Since we already know $\frac{\partial}{\partial V_1(x'')}( (R_{\text{extr}})_k), \forall k\leq 3N+N^2$ with \eqref{oooo} 
Then we easily find:
\begin{equation}
\label{W30}
\fbox{$
   \begin{array}{rcl} 
W_3^{(0)}(x,x',x'')&=&\frac{2T}{N}\underset{i,j,k=1}{\overset{N}{\sum}}\frac{\left(\mathcal{H}^{-1}\right)_{i,j}\left(\mathcal{H}^{-1}\right)_{i,k}}{(x-s_i)^3(x'-s_j)^2(x''-s_k)^2} +\frac{2T}{N}\underset{i,j,k=1}{\overset{N}{\sum}}\frac{\left(\mathcal{H}^{-1}\right)_{i,j}\left(\mathcal{H}^{-1}\right)_{j,k}}{(x-s_i)^2(x'-s_j)^3(x''-s_k)^2}\cr
&&+\frac{2T}{N}\underset{i,j,k=1}{\overset{N}{\sum}}\frac{\left(\mathcal{H}^{-1}\right)_{i,k}\left(\mathcal{H}^{-1}\right)_{k,j}}{(x-s_i)^2(x'-s_j)^2(x''-s_k)^3}
-\frac{T}{N}\underset{i,j,k=1}{\overset{N}{\sum}} \frac{C_{i,j,k}}{(x-s_i)^2(x'-s_j)^2(x''-s_k)^2}\cr
\end{array}
   $}
\end{equation}
where the $C_{i,j,k}$'s are linked with the third derivative of the functional $\mathcal{S}$:
\beq
 C_{i,j,k} \overset{\text{def}}{=}\sum_{\alpha,\beta,\gamma=1}^{3N+N^2} \left(\mathcal{H}^{-1}\right)_{i,\alpha}\left(\mathcal{H}^{-1}\right)_{j,\beta}\left(\mathcal{H}^{-1}\right)_{k,\gamma} \frac{\partial^3 \mathcal{S}}{\partial R_\alpha \partial R_\beta \partial R_\gamma}_{|R=R_\text{extr}}
\eeq
It is again clear that \textbf{ $(x,x',x'')\mapsto W_3^{(0)}(x,x',x'')$ is a symmetric function of its variables}.

\section{The topological recursion \label{section 8}}

In the last sections we have seen how to compute $U_0^{(0)}(x,y)$, $P_0^{(0)}(x,y)$ and $W_1^{(0)}(x)$ in the context of the Bethe ansatz. Then by a recursive application of the loop insertion operators $\frac{\partial}{\partial V_1(z)}$ we can compute easily all the $W_n^{(0)}$'s $U_n^{(0)}$'s and $P_n^{(0)}$'s. Thus, for now we have all desired quantities at the dominant order (i.e. $g_s=0$) but we still miss the subleading corrections $g \geq 1$. It is the purpose of this section to present a recursive algorithm to get all the subleading corrections.

The algorithm is very similar to the ``topological recursion" considered in \cite{EO, eynloop1mat, EKK}, and is in fact more comparable to another version of the topological recursion for the hermitian 2-matrix model, described in \cite{EO2MMmatrix}.

\subsection{Higher loop equations}

Evaluating the coefficients of $g_s^{2g}$ in \eqref{loopequation} gives the higher loop equations for $g\geq 1$:
\begin{equation}
\label{loopequationg}
\fbox{$
   \begin{array}{rcl}
&&\left(y-Y(x)+\frac{T}{N}\partial_x\right)U_0^{(g)}(x,y)+W_1^{(g)}(x)U_0^{(0)}(x,y)=-\overset{g-1}{\underset{h=1}{\sum}}
 W_1^{(g-h)}(x)U_0^{(h)}(x,y)\cr
&&-U_1^{(g-1)}(x,y;x)+\frac N T\partial_x U_{0}^{(g-1)}(x,y) - P_0^{(g)}(x,y)-\frac N T\delta_{g=1} \end{array}
   $}
\end{equation}

We can also project the higher order loop equations \eqref{loopequationn} onto $g_s^{2g}$ (with $n\geq 1$) to find (we note $\vec{\xi}=\{\xi_1,\dots,\xi_n\}$ for short):
\begin{equation} \label{GeneralizedLoop}
\addtolength{\fboxsep}{10pt}
			\boxed{
				\begin{split}
				&\left(y-Y(x)+\frac{T}{N}\partial_x\right)U_n^{(g)}(x,y;\vec{\xi})+U_n^{(0)}(x,y;\vec{\xi})W_1^{(g)}(x) \cr
&=- \sum_{h=1}^{g-1} U_n^{(h)}(x,y;\vec{\xi})W_1^{(g-h)}(x)+ \frac{N}{T}\partial_xU_n^{(g-1)}(x,y;\vec{\xi})- U_{n+1}^{(g-1)}(x,y;x,\vec{\xi})\cr
&-\sum_{h=0}^{g}\underset{ I\subset \vec{\xi},\, I\neq \emptyset}{\sum} W_{1+|I|}^{(g-h)}(x,I)\,U_{n-|I|}^{(h)}(x,y;\vec\xi\setminus I)- P_n^{(g)}(x,y;\vec{\xi})-\underset{k=1}{\overset{n}{\sum}} \frac \partial {\partial \xi_k}\cr
&\left( \frac{U_{n-1}^{(g)}(x,y;\xi_1,\dots,\xi_{k-1},\xi_{k+1},\dots,\xi_n)-U_{n-1}^{(g)}(\xi_k,y;\xi_1,\dots,\xi_{k-1},\xi_{k+1},\dots,\xi_n)}{x-\xi_k}\right) \cr
				\end{split}
			}
	\end{equation}

We shall also write those equations for each power of $y^k$, using
\beq
U_n^{(g)}(x,y;\vec \xi) = \sum_{k=0}^{d_2-1} y^k\,\,U_{n,k}^{(g)}(x;\vec \xi)
\eeq	
\beq
P_n^{(g)}(x,y;\vec \xi) = \sum_{k=0}^{d_2-1} y^k\,\,P_{n,k}^{(g)}(x;\vec \xi)
\eeq

\subsection{Matrix form of the loop equations for $g \geq 1$} 

The goal of this subsection is to rewrite the loop equations \eqref{loopequationg} ($0\leq k \leq d_2+1$)
\bea &&\label{LL} U_{0,k-1}^{(g)}(x) +\left(-Y(x)+\frac{T}{N}\partial_x\right)U_{0,k}^{(g)}(x) =-\overset{g-1}{\underset{h=0}{\sum}}W_1^{(g-h)}(x)U_{0,k}^{(h)}(x)- P_{0,k}^{(g)}(x)\cr
&&-U_{1,k}^{(g-1)}(x;x)+\frac{N}{T}\partial_x U_{0,k}^{(g-1)}(x)-\frac{N}{T}\delta_{g=1}\delta_{k=0}\cr
\eea 
into a matrix form suitable for our algorithm.

First by taking the case $k=d_2+1$ and $k=d_2$ we obtain that: 
\beq \label{W1g}\forall \,g>0 \, :\, U_{0,d_2}^{(g)}(x)=0 \virg U_{0,d_2-1}^{(g)}(x)=\td{t}_{d_2}W_1^{(g)}(x) \eeq
%In particular we observe that the corrections of the first correlation function $W_1^{(g)}(x)$ simply correspond to $U_{0,d_2-1}^{(g)}(x)$. In other words, if we can determine all $U_{0,k}^{(g)}(x)$, then we will obtain all $W_1^{(g)}(x)$. 

Now let's introduce the following $d_2 \times d_2$ matrix:
\beq 
\mathcal{D}(x)= \begin{pmatrix}
-Y(x) & 0           &\ldots & 0 &\frac{U_{0,0}^{(0)}(x)}{\td{t}_{d_2}} \cr
1           & -Y(x) & \dots & 0 &\frac{U_{0,1}^{(0)}(x)}{\td{t}_{d_2}} \cr
\ddots      & \ddots      &       &   & \dots                             \cr
0           & \ldots      &       & 1 & -Y(x)+\frac{U_{0,d_2-1}^{(0)}(x)}{\td{t}_{d_2}} \cr
\end{pmatrix}
\eeq 
Then we can rewrite the loop equations \eqref{LL} into the following matrix form:
\begin{equation} \label{Matrixform}
\addtolength{\fboxsep}{10pt}
			\boxed{
				\begin{split}
			&\left( \frac{T}{N} \partial_x + \mathcal{D}(x) \right) 
\begin{pmatrix}U_{0,0}^{(g)}(x) \cr U_{0,1}^{(g)}(x) \cr \vdots \cr U_{0,d_2-1}^{(g)}(x)\end{pmatrix}
=
-\begin{pmatrix}P_{0,0}^{(g)}(x) \cr P_{0,1}^{(g)}(x) \cr \vdots \cr P_{0,d_2-1}^{(g)}(x)\end{pmatrix}-\begin{pmatrix}U_{1,0}^{(g-1)}(x;x) \cr U_{1,1}^{(g-1)}(x;x) \cr \vdots \cr U_{1,d_2-1}^{(g-1)}(x;x)\end{pmatrix}\cr
&-\sum_{l=1}^{g-1} \frac{U_{0,d_2-1}^{(l)}(x)}{\td{t}_{d_2}} \begin{pmatrix}U_{0,0}^{(g-l)}(x) \cr U_{0,1}^{(g-l)}(x) \cr \vdots \cr U_{0,d_2-1}^{(g-l)}(x)\end{pmatrix}
+\frac{N}{T}\partial_x \begin{pmatrix} U_{0,0}^{(g-1)}(x)\cr U_{0,1}^{(g-1)}(x)\cr \vdots \cr U_{0,d_2-1}^{(g-1)}(x)\end{pmatrix}
-\frac{N}{T}\delta_{g=1}\begin{pmatrix}1\cr0\cr \vdots\cr 0\end{pmatrix}\cr
				\end{split}
			}
	\end{equation}

In the same way, we can also rewrite the projection of the higher loop equations \ref{GeneralizedLoop} onto $y^k$ as:

\bea
&&U_{n,k-1}^{(g)}(x;\vec{\xi})+\left(\frac{T}{N}\partial_x -Y(x)\right)U_{n,k}^{(g)}(x;\vec{\xi})=- \sum_{h=0}^{g-1} U_{n,k}^{(h)}(x;\vec{\xi})W_1^{(g-h)}(x)\cr
&&-\sum_{h=0}^{g}\underset{ I\sqcup J= \vec{\xi},\, I\neq \emptyset}{\sum} W_{1+|I|}^{(g-h)}(x,I)\,U_{|J|,k}^{(h)}(x;J)- P_{n,k}^{(g)}(x;\vec{\xi})\cr
&&+ \frac{N}{T}\partial_xU_{n,k}^{(g-1)}(x;\vec{\xi}) - U_{n+1,k}^{(g-1)}(x;x,\vec{\xi})-\underset{j=1}{\overset{n}{\sum}} \frac \partial {\partial \xi_j}\cr
&&\left( \frac{U_{n-1,k}^{(g)}(x,\xi_1,\dots,\xi_{j-1},\xi_{j+1},\dots,\xi_n)-U_{n-1,k}^{(g)}(\xi_j,\xi_1,\dots,\xi_{j-1},\xi_{j+1},\dots,\xi_n)}{x-\xi_j}\right) \cr
\eea

Evaluating the coefficients in $k=d_2+1$ and $k=d_2$ leads to:
\beq \label{Wng}\encadremath{\forall \,n\geq 1\,,\, \forall \, g\geq 0: \, U_{n,d_2}^{(g)}(x,\vec{\xi})=0 \,\,\,\,\, ,\,\,\,\,\, U_{n,d_2-1}^{(g)}(x,\vec{\xi})=\td{t}_{d_2}W_{n+1}^{(g)}(x,\vec{\xi}) }\eeq
and thus leads to the linear system:

\begin{equation} \label{GeneralizedMatrixform}
\addtolength{\fboxsep}{10pt}
			\boxed{
				\begin{split}
			&\left( \frac{T}{N} \partial_x + \mathcal{D}(x) \right) 
\begin{pmatrix}U_{n,0}^{(g)}(x;\vec{\xi}) \cr U_{n,1}^{(g)}(x;\vec{\xi}) \cr \vdots \cr U_{n,d_2-1}^{(g)}(x;\vec{\xi})\end{pmatrix}
=-\begin{pmatrix} U_{n+1,0}^{(g-1)}(x;x,\vec{\xi})\cr U_{n+1,1}^{(g-1)}(x;x,\vec{\xi})\cr \vdots\cr
U_{n+1,d_2-1}^{(g-1)}(x;x,\vec{\xi})\end{pmatrix}
\cr
&+\frac{N}{T}\partial_x\begin{pmatrix} U_{n,0}^{(g-1)}(x;\vec{\xi})\cr U_{n,1}^{(g-1)}(x;\vec{\xi})\cr \vdots\cr
U_{n,d_2-1}^{(g-1)}(x;\vec{\xi})\end{pmatrix}-\begin{pmatrix}P_{n,0}^{(g)}(x;\vec{\xi}) \cr P_{n,1}^{(g)}(x;\vec{\xi}) \cr \vdots \cr P_{n,d_2-1}^{(g)}(x;\vec{\xi})\end{pmatrix}\cr
&-\sum_{h=0}^g\sum_{I\sqcup J= \vec{\xi},\, /\, (I,h)\neq\{(\emptyset,g)\cup (\vec{\xi},0)\}} W_{1+|I|}^{(g-h)}(x,I)\begin{pmatrix} U_{|J|,0}^{(h)}(x;J)\cr U_{|J|,1}^{(h)}(x;J)\cr \vdots \cr U_{|J|,d_2-1}^{(h)}(x;J)\end{pmatrix}\cr
&-\underset{j=1}{\overset{n}{\sum}} \frac \partial {\partial \xi_j}\begin{pmatrix} \frac{U_{n-1,0}^{(g)}(x,\xi_1,\dots,\xi_{j-1},\xi_{j+1},\dots,\xi_n)-U_{n-1,0}^{(g)}(\xi_j,\xi_1,\dots,\xi_{j-1},\xi_{j+1},\dots,\xi_n)}{x-\xi_j} \cr
\frac{U_{n-1,1}^{(g)}(x,\xi_1,\dots,\xi_{j-1},\xi_{j+1},\dots,\xi_n)-U_{n-1,1}^{(g)}(\xi_j,\xi_1,\dots,\xi_{j-1},\xi_{j+1},\dots,\xi_n)}{x-\xi_j}\cr
\vdots\cr
\frac{U_{n-1,d_2-1}^{(g)}(x,\xi_1,\dots,\xi_{j-1},\xi_{j+1},\dots,\xi_n)-U_{n-1,d_2-1}^{(g)}(\xi_j,\xi_1,\dots,\xi_{j-1},\xi_{j+1},\dots,\xi_n)}{x-\xi_j}\end{pmatrix}
				\end{split}
}
\end{equation}

Observe now that the r.h.s. of the last two systems \eqref{Matrixform} \eqref{GeneralizedMatrixform} is composed of lower orders in $g$ and an unknown polynomial vector $P_{n,k}^{(g)}$. The l.h.s. is always the same at every order in $g$ and for every $n$. Therefore, if we can find a way to get rid of the unknown polynomials $P_{n,k}^{(g)}$'s in the inversion of the system, we see that we can perform a recursion and get all the corrections recursively. We will do so in the following sections by introducing a suitable kernel $K(x_0,x)$ and using some residue methods (which will automatically get rid of the polynomials).

\subsection{Inversion of the linear system: kernels $K(x_0,x)$ and $G(x_0,x)$}

%In this section, we explain how we can invert the previous system and get rid of the unknown polynomials $P_k^{(g)}(x)$. 
Suppose that we can find matrices $K(x_0,x)$ and $G(x_0,x)$ of size $d_2 \times d_2$ such that:
\begin{itemize} \label{conditions}
\item They satisfy the relation:
\beq\label{K} \left(\mathcal{D}^t(x)-\frac{T}{N} \partial_x \right) K(x_0,x)=G(x_0,x)\eeq
\item The function $G(x_0,x)$ has the form:
\beq G(x_0,x)=\frac{{\rm Id}_N}{x_0-x} +\sum_{j=1}^N \frac{A_j(x_0)}{x-s_j} \label{GG}\eeq
where the $A_j(x_0)$'s are $d_2 \times d_2$ matrices.
\item The function $x \mapsto K(x_0,x)$ is analytic at every $x=s_i$
\end{itemize}

If we can find such matrices, then we claim that we can invert any system of the form \eqref{Matrixform}: $ \left(\frac{T}{N} \partial_x +\mathcal{D}(x) \right) \vec{u}(x)=\vec{v}(x)$, where $\vec{v}(x)$ is assumed to be known up to a polynomial and $\vec{u}(x)$ is assumed to have only poles at the $s_i$'s and behaves like $O\left(\frac{1}{x}\right)$ when $x \to \infty$. Indeed under these assumptions we have:

\bea
\vec{u}^{\,t}(x_0)
&=& -\Res_{x \to x_0} \vec{u}^{\,t}(x)G(x_0,x)\cr
&=&\sum_{i=1}^N \Res_{x \to s_i} \vec{u}^{\,t}(x)G(x_0,x)\cr
&=&\sum_{i=1}^N \Res_{x \to s_i} \vec{u}^{\,t}(x)\left(\mathcal{D}^t(x)-\frac{T}{N} \vec{\partial_x} \right)K(x_0,x)\cr
&=&\sum_{i=1}^N \Res_{x \to s_i} \vec{u}^{\,t}(x)\left(\mathcal{D}^t(x)+\frac{T}{N} \overleftarrow{\partial_x} \right)K(x_0,x)\cr
&=&\sum_{i=1}^N \Res_{x \to s_i} \vec{v}^{\,t}(x)K(x_0,x)\cr
\eea
Note that in the last equality, the polynomial part of $\textbf{v}(x)$ does not contribute only the terms in $\vec v(x)$ which have poles at the $s_i$'s contribute. Hence we have successfully inverted our system:
\beq \label{residue}\encadremath{ \vec{u}(x_0)=\sum_{i=1}^N \Res_{x \to s_i} K^t(x_0,x)\vec{v}(x)}\eeq

\subsection{The topological recursion \label{sectoprec}}

Therefore with the help of the results in the last section we can compute all correlation functions by the recursion:

\begin{equation} \label{OurRecursion}
\addtolength{\fboxsep}{10pt}
			\boxed{
				\begin{split}
			&\begin{pmatrix}U_{0,0}^{(g)}(x_0) \cr U_{0,1}^{(g)}(x_0) \cr \vdots \cr U_{0,d_2-1}^{(g)}(x_0)\end{pmatrix}
=\sum_{i=1}^N \Res_{x \to s_i} K^t(x_0,x) \Big[
\sum_{l=1}^{g-1} \frac{U_{0,d_2-1}^{(l)}(x)}{\td{t}_{d_2}} \begin{pmatrix}U_{0,0}^{(g-l)}(x) \cr U_{0,1}^{(g-l)}(x) \cr \vdots \cr U_{0,d_2-1}^{(g-l)}(x)\end{pmatrix}\cr
&
+\begin{pmatrix}U_{1,0}^{(g-1)}(x;x) \cr U_{1,1}^{(g-1)}(x;x) \cr \vdots \cr U_{1,d_2-1}^{(g-1)}(x;x)\end{pmatrix}-\frac{N}{T}\partial_x \begin{pmatrix} U_{0,0}^{(g-1)}(x)\cr U_{0,1}^{(g-1)}(x)\cr \vdots \cr U_{0,d_2-1}^{(g-1)}(x)\end{pmatrix}
 \Big]\cr 
				\end{split}
			}
	\end{equation}

or more generally for higher order correlation functions:

\begin{equation} \label{GeneralizedRecursion}
\addtolength{\fboxsep}{10pt}
			\boxed{
				\begin{split}
			&\begin{pmatrix}U_{n,0}^{(g)}(x_0,\vec\xi) \cr U_{n,1}^{(g)}(x_0,\vec\xi) \cr \vdots \cr U_{n,d_2-1}^{(g)}(x_0,\vec\xi)\end{pmatrix}
=\sum_{i=1}^N \Res_{x \to s_i} K^t(x_0,x) \Big[\cr
&\begin{pmatrix} U_{n+1,0}^{(g-1)}(x;x,\vec{\xi})\cr U_{n+1,1}^{(g-1)}(x;x,\vec{\xi})\cr \vdots\cr
U_{n+1,d_2-1}^{(g-1)}(x;x,\vec{\xi})\end{pmatrix}-\frac{N}{T}\partial_x \begin{pmatrix} U_{n,0}^{(g-1)}(x;\vec{\xi})\cr U_{n,1}^{(g-1)}(x;\vec{\xi})\cr \vdots\cr U_{n,d_2-1}^{(g-1)}(x;\vec{\xi})\end{pmatrix}\cr
&+\sum_{h=0}^g\sum_{I\sqcup J= \vec{\xi},\, /\, (I,h)\neq\{(\emptyset,g)\cup (\vec{\xi},0)\}} W_{1+|I|}^{(g-h)}(x,I)\begin{pmatrix} U_{|J|,0}^{(h)}(x;J)\cr U_{|J|,1}^{(h)}(x;J)\cr \vdots \cr U_{|J|,d_2-1}^{(h)}(x;J)\end{pmatrix}\cr
&+\underset{j=1}{\overset{n}{\sum}} \frac{1}{(x-\xi_j)^2}\,\,\begin{pmatrix} {U_{n-1,0}^{(g)}(x,\xi_1,\dots,\xi_{j-1},\xi_{j+1},\dots,\xi_n)} \cr
{U_{n-1,1}^{(g)}(x,\xi_1,\dots,\xi_{j-1},\xi_{j+1},\dots,\xi_n)}\cr\vdots\cr {U_{n-1,d_2-1}^{(g)}(x,\xi_1,\dots,\xi_{j-1},\xi_{j+1},\dots,\xi_n)}\end{pmatrix}\Big]\cr 
\end{split}
			}
\end{equation}

\subsection{Determination of the kernel $K(x_0,x)$ \label{secK}}

The determination of the kernels $K(x_0,x)$ and $G(x_0,x)$ is presented in appendix \ref{AppendixComp}. We present in details some recursive formulas to obtain the various $K(x_0,s_i)$ and the derivatives $K'(x_0,s_i),\dots, K^{(n)}(x_0,s_i)$ in the appendix. The results are the following. If we define the following vectors (of size $d_2$):
\beq \vec{r}=\begin{pmatrix}0\cr \vdots \cr 0\cr 1\end{pmatrix} \,\,\,\, , \,\,\,\, \vec{w}_i=\frac{1}{\td{t}_{d_2}} \begin{pmatrix}u_{0,i}\cr\vdots\cr u_{d_2-1,i}\end{pmatrix}\eeq
Then the matrices $A_i(x_0)$ defining $G(x_0,x)$ are connected to the $K(x_0,s_i)$ by:
\beq\encadremath{A_i(x_0)=\left(\frac{T}{N}+\vec{r}\vec{w}_i^t\right)K(x_0,s_i)}\eeq
The matrices $K(x_0,s_i)$ are determined as a solution of the following system:
\beq \encadremath{\mathcal{M}\vec{K}_q=\vec{f}_q}\eeq
where
\beq
\vec{K_q}=\begin{pmatrix} K(x_0,s_1)_{0,q} \cr \vdots \cr K(x_0,s_N)_{0,q} \cr ----\cr K(x_0,s_1)_{1,q} \cr \vdots \cr K(x_0,s_N)_{1,q}\cr---- \cr \vdots\cr ----\cr  K(x_0,s_1)_{d_2-1,q} \cr \vdots \cr K(x_0,s_N)_{d_2-1,q} \cr\end{pmatrix} \,\,,\,\, 
\vec{f_q}=\begin{pmatrix} \vec{0} \cr----\cr \vdots\cr ----\cr  \vec{0} \cr----\cr \frac{1}{x_0-s_1} \cr \vdots\cr \frac{1}{x_0-s_N}\cr ----\cr \vec{0} \cr----\cr \vec{0}\cr----\cr \frac{u_{q,1}}{(x_0-s_1)^2} \cr \vdots\cr \frac{u_{q,N}}{(x_0-s_N)^2} \end{pmatrix}\,\,,\,\, \vec{f_{d_2-1}}=\frac{T}{N} \td{t}_{d_2}\begin{pmatrix} \vec{0} \cr----\cr \vdots\cr----\cr \vec{0}\cr----\cr \frac{1}{(x_0-s_1)^2} \cr \vdots\cr \frac{1}{(x_0-s_N)^2} \end{pmatrix}  \eeq
where the $\frac{1}{x_0-s_i}$ in $\vec f_q$ are in position $Nq+i$. (Note that $\vec{f}_{d_2-1}$ does not have these terms).
\beq
\mathcal{M}=\begin{pmatrix}
-B^t &Id& 0 & \ldots &0 \cr
0 & -B^t & Id & &0 \cr
\vdots & \ddots & \ddots &\vdots &\vdots\cr
0& \ldots &  &-B^t & Id \cr
B_0 & B_1& \ldots & B_{d_2-2} & B_{d_2-1}\cr
\end{pmatrix}\eeq
where all the matrices $B_k$, $\text{Id}$ and $0$ are symmetric matrices of size $N\times N$ given by:

\beq B_0= \frac{T}{N} \begin{pmatrix}(B_0)_{1,1}&  \frac{u_{0,1}+u_{0,2}}{(s_1-s_2)^2} &\ldots & \frac{u_{0,1}+u_{0,N}}{(s_1-s_N)^2}\cr
 \frac{u_{0,1}+u_{0,2}}{(s_1-s_2)^2} & \ddots & \ddots  & \vdots \cr
\vdots & \ddots &  &   \frac{u_{0,N-1}+u_{0,N}}{(s_{N-1}-s_N)^2}\cr
 \frac{u_{0,N}+u_{0,1}}{(s_N-s_1)^2} &  \ldots &  \frac{u_{0,N}+u_{0,N-1}}{(s_N-s_{N-1})^2} &(B_0)_{N,N}\end{pmatrix}
\eeq
with $(B_0)_{i,i}=-\frac{N}{T}V''_1(s_i)u_{0,i}-\underset{j \neq 1}{\sum} \frac{u_{0,i}+u_{0,j}}{(s_i-s_j)^2} +1$.
 
And $\forall k>0$: 
\beq B_k= \begin{pmatrix}(B_k)_{1,1}& \frac{T}{N} \frac{u_{k,1}+u_{k,2}}{(s_1-s_2)^2} &\ldots & \frac{T}{N} \frac{u_{k,1}+u_{k,N}}{(s_1-s_N)^2}\cr
\frac{T}{N} \frac{u_{k,1}+u_{k,2}}{(s_1-s_2)^2} &  & \ddots & \vdots \cr
\vdots & \ddots & \ddots &  \frac{T}{N} \frac{u_{k,N-1}+u_{k,N}}{(s_{N-1}-s_N)^2}\cr
\frac{T}{N} \frac{u_{k,N}+u_{k,1}}{(s_N-s_1)^2} &  \ldots & \frac{T}{N} \frac{u_{k,N}+u_{k,N-1}}{(s_N-s_{N-1})^2} &(B_k)_{N,N} \end{pmatrix}
\eeq 
with $(B_k)_{i,i}=-V''_1(s_i)u_{k,i}-\frac{T}{N}\underset{j \neq i}{\sum} \frac{u_{k,i}+u_{k,j}}{(s_i-s_j)^2}$.

If we define
$$ C_k = \sum_{j=k}^{d_2-1} B_j {B^t}^{j-k}
%\virg
%C = C_0^{-1}\otimes {\rm Id}_{d_2\times d_2}
%=\begin{pmatrix}
%C_0^{-1} & & & \cr
%& C_0^{-1} & & \cr
%& & \ddots & \cr
%& & & C_0^{-1}\cr
%\end{pmatrix}
$$
we have
\beq
{\cal M}^{-1} = \begin{pmatrix}
0 & 0 & \dots & & & 0 \cr 
{\rm Id} & \ddots && & & \vdots \cr  
B^t & {\rm Id} & \ddots && & \vdots \cr
{B^t}^2 & B^t & {\rm Id} & \ddots & & \vdots \cr
\vdots & & \ddots& \ddots &\ddots & \vdots \cr
{B^t}^{d_2-2} & \dots &{B^t}^2 & B^t &{\rm Id} & 0 \cr
\end{pmatrix}
- \begin{pmatrix} {\rm Id} \cr B^t \cr {B^t}^2 \cr \vdots \cr {B^t}^{d_2-1} \end{pmatrix}\,
C_0^{-1}
\begin{pmatrix}
C_1 &
C_2 &
\dots &
C_{d_2-1} &
{\rm Id}
\end{pmatrix}
\eeq
That gives if $q<d_2-1$:
\bea
K_{p,q}(x_0,s_i) 
&=& \sum_j \theta(p> q)\,((B^t)^{p-q-1})_{i,j}\frac{1}{x_0-s_j}  \cr
&& - \sum_j  ({B^t}^p C_0^{-1} C_q)_{i,j}  \frac{1}{x_0-s_j} 
 - \sum_j  ({B^t}^p C_0^{-1})_{i,j}  \frac{u_{q,j}}{(x_0-s_j)^2}
\eea
where $\theta(p> q)$ is the heavyside function, equal to $1$ if $p> q$ and $0$ otherwise.
And when $q=d_2-1$:
\beq
K_{p,d_2-1}(x_0,s_i) 
= - \frac{T}{N}\,\td t_{d_2}\,\,\sum_j  ({B^t}^p C_0^{-1})_{i,j}  \frac{1}{(x_0-s_j)^2}
\eeq

Eventually the higher derivatives in $x$ of $K(x_0,x)$ evaluated at $x=s_i$ are obtained recursively with formulas \eqref{assumption}, \eqref{K'}, \eqref{K''}, \eqref{Kn} and \eqref{Knbis} derived in details in appendix. Note also that to complete these formulas, we need to prove that the system of equations is consistent. This involves a special identity which is proved in appendix \ref{Cancellation}. \textbf{As a conclusion, we have obtained here a recursion to compute all correlations functions $U_n^{(g)}(x,y;\vec{x})$}. This corresponds to a generalization of the topological recursion \cite{eynloop1mat, EO} for the arbitrary-$\beta$ two-matrix model studied here. Moreover, when $d_2=2$, one can recover the case of the arbitrary-$\beta$ one-matrix model developped in \cite{MoiBertrand}.

\subsection{Leading free energy}

We clearly have from the saddle-point approximation method, that
\beq
f_0=-\frac{T}{N}\,\mathcal{S}.
\eeq
where the Yang--Yang action ${\cal S}$ 
\bea {\cal S} &=& \tr V_1(S) +\tr V_2(\td{S})- \tr (SA\td{S}A^{-1}) - \frac{T}{N}\ln(\Delta(S))\cr
&&-\frac{T}{N}\ln(\Delta(\td{S}))+\frac{T}{N}\ln \det(A) -\frac{T}{N}\vec{u}^t \left(A\vec{e}-\vec{e}\right) 
\eea
is evaluated at its extremum.

The link \eqref{linktopoWkb} between the WKB expansion and the standard topological expansion gives
\beq
F_0(\frac{T}{N}) = f_0(N) = -\mathcal{S}
\eeq

\subsection{Subleading $g=1$}

To the first subleading order, our topological recursion gives
\beq
\vec U^{(1)}_0(x_0) = \sum_i \Res_{x\to s_i}\,\,K^t(x_0,x)\,\left(\vec U^{(0)}_1(x;x) - \frac N T \partial_x\,\vec U^{(0)}_0(x)\right)
\eeq
and we remind that the $d_2-1^{\rm th}$ component of the vector $\vec U^{(1)}_0(x_0)$ is the function $W_1^{(1)}(x_0)$.

In order to perform the computation, we first need to compute $\vec U^{(0)}_1(x;x)$, which is also computed by the topological recursion and is worth
\beq
\vec U^{(0)}_1(x_0,\xi) = \frac N T\,\, \sum_i \Res_{x\to s_i}\,\frac{1}{(x-\xi)^2}\,\,K^t(x_0,x)\,\vec U^{(0)}_0(x)\,\,
\eeq
i.e.
\beq
U^{(0)}_{1,k}(x_0,\xi) = \frac N T\,\sum_{l=0}^{d_2-1}\, \sum_{i=1}^N \,\frac{1}{(s_i-\xi)^2}\,\,[K(x_0,s_i)]_{l,k}\,u_{l,i}\,\,
\eeq
and we need it at $x_0=\xi$ i.e.
\beq
U^{(0)}_{1,k}(x,x) = \frac N T\,\sum_{l=0}^{d_2-1}\, \sum_{i=1}^N \,\frac{1}{(s_i-x)^2}\,\,[K(x,s_i)]_{l,k}\,u_{l,i}\,\,
\eeq
Notice that $K(x,s_i)$ may have double poles at  $x=s_j$, and thus $U^{(0)}_{1,k}(x,x)$ may have poles of order up to 4.
This implies that $\vec U^{(1)}_0(x_0)$ can be expressed in terms of $K(x_0,s_i)$, $K'(x_0,s_i)$, $K'''(x_0,s_i)$.

\medskip
From the WKB approximation, we know that we must have
\beq
f_1 = -\frac{1}{2}\ln\det{{\cal A}''},
\eeq
where ${\cal A}$ is the action (see \eqref{eqdefAction}), which is approximated by the Yang-Yang action ${\cal S}$ in the limit $\beta\to\infty$, and ${\cal H}$ is the Hessian of the Yang--Yang action ${\cal S}$, so this leads us to conjecture that
\beq 
f_1=-\frac{1}{2}\ln\det{{\cal H}}.
%\left(V_1''(s)V_2''(V_1'(s))-1\right)-\frac{N^2F_{0}}{T^2}=-\frac{1}{2}\ln\left(V_1''(s)V_2''(\td{s})-1\right)-\frac{N^2F_{0}}{T^2}
\eeq 

%We had not been able to compute completely $W_1^{(1)}(x)$ and hence $F^{(1)}$ with our formalism in the general case. Nevertheless, 
We find that:
\beq\encadremath{ W_1^{(1)}(x)=\frac{U_{d_2-1}}{\td{t}_{d_2}}=-\sum_j \left[\frac{\mathcal{H}^{-1}_{j,j}}{(x_0-s_j)^3}+\frac{B_j}{(x_0-s_j)^2}\right] }\eeq
where we could explicitly verify the lack of a simple pole term. The coefficients $B_j$ are given by:
\bea B_j&=&\sum_k Z^{-1}_{j,k}\Big[ -\frac{1}{2}\sum_{u,v} \frac{d^2[V_2'(B)]}{d s_u ds_v}\mathcal{H}_{u,v}+\sum_i V_1'(s_i)V_{k,i}-\sum_{i,n}B_{k,i}V_{i,n}\cr
&&+\sum_{p=1}^{d_2}\sum_{q,n,m} [B^{p-1}]_{k,q}Q_{q,n}^m \frac{d \Phi_{m,p-1}}{d V_1'(s_n)}\Big]\cr
&&+2\sum_{i,j\neq i,r}\frac{\td{t}_{d_2}}Z^{-1}_{j,r}[B^{d_2-1}]_{r,i}\eea
The technical details (and the definitions of quantities involved in the last formula) of this computation can be found in appendix \ref{F_1} with other possible expressions for $B_j$ given in \eqref{B_j} or \eqref{B_j}.

\subsection{Example}

As an example, we present explicit computations of the method in the case when $N=1$ in appendix \eqref{Oneroot}. In particular in this case we are able to determine $W_1^{(1)}(x)$ explicitly and to compute the corresponding $F_1$. In particular we show in this case that it satisfies the equation:
\beq 
f_1=-\frac{1}{2}\ln\det{{\cal H}}
%\left(V_1''(s)V_2''(V_1'(s))-1\right)-\frac{N^2F_{0}}{T^2}=-\frac{1}{2}\ln\left(V_1''(s)V_2''(\td{s})-1\right)-\frac{N^2F_{0}}{T^2}
\eeq

\subsection{WKB computation}

The WKB approximation, consists in expanding the integrand in the vicinity of a saddle point.
To leading order, one gets a Gaussian integral, and corrections, are moments of the Gaussian integral, and they are computed by Wick's theorem.
In principle, one can write:
\beq
Z = \int dX\,dY\,\ee{-\frac{1}{g_s^2}\,{\cal A}[X,Y]}
\eeq
and expand:
\beq
X=\bar X+ g_s\,{\cal X}
\virg
Y=\bar Y + g_s\,{\cal Y}
\eeq
and expand
\beq
{\cal A}[X,Y] = {\cal A}[\bar X,\bar Y] + \frac{1}{2}g_s^2\,{\cal A}''[{\cal X},{\cal Y}] + g_s^3\,\delta{\cal A}[{\cal X},{\cal Y}]
\eeq
where the Hessian ${\cal A}''$ is the quadratic form of second derivatives of $\underset{g_s\to 0}{\text{lim}} {\cal A}$, and 
$\delta{\cal A}$ can be computed by its Taylor expansion in $g_s$.

We thus get
\beq
Z = \int dX\,dY\,\ee{-\frac{1}{g_s^2}\,{\cal A}[X,Y]}
\sim \frac{\ee{-\frac{1}{g_s^2}{\cal A}[\bar X,\bar Y] }}{\sqrt{\det {\cal A}''}}\,\, \frac{\int d{\cal X}d{\cal Y}\,\,\ee{-\frac{1}{2}{\cal A}''[{\cal X},{\cal Y}] } \,\,\ee{-g_s\,\delta{\cal A}[{\cal X},{\cal Y}]} }{\int d{\cal X}d{\cal Y}\,\,\ee{-\frac{1}{2}{\cal A}''[{\cal X},{\cal Y}] } }
\eeq
i.e. we recover
\beq
f_0 = -\lim_{\beta\to\infty}\,{\cal A}[\bar X,\bar Y]
\eeq
and
\beq
f_1  = -\lim_{\beta\to\infty}\,\frac{1}{2}\,\det\,{\cal A}'',
\eeq
and in principle, a standard WKB approximation using Wick's theorem, should allow one to find $f_2, f_3, \dots$ explicitly. Then, $W_n^{(g)}$ can be recovered by applying $\frac{d}{dV_1}$ repeatedly. However, this is long and tedious, and difficult to do in a systematic way.
One of the difficulties, is that one has to compute the large $\beta$ expansion of $I_\beta(X,Y)$, which is not very well known at the present time.

So, our loop equation method is an alternative to WKB, it can be performed in a systematic way, and gives all orders in the $g_s$ expansion.
Also, notice that Wick's theorem gives an expansion of $Z$ in powers of $g_s$, whereas our method gives directly the expansion of $\ln Z$.

\section{Conclusion}

We have presented here the loop equations of the two-matrix arbitrary-$\beta$ matrix models and a recursive way to solve them for a specific polynomial Bethe ansatz. This work is a generalization of the arbitrary-$\beta$ one matrix model with the same kind of ansatz as developed in \cite{MoiBertrand}. This work is also part of the ``quantum algebraic geometry" project started for the one-matrix model in \cite{MoiBertrand,MoiLeonidBertrand,MoiLeonidBertrand2}. In particular, we have proved that in the two-matrix arbitrary-$\beta$ matrix model, we find a ``quantum curve" \eqref{ODE} of arbitrary degree. For now we have only solved this model under the assumption of a polynomial ansatz for the solution corresponding to a specific limit of the model ($N$ and $T$ fixed while $g_s\to0$ (i.e. $\beta\to \infty$)). However, we expect the notions developed in \cite{MoiLeonidBertrand,MoiLeonidBertrand2} to generalize well for the two-matrix arbitrary-$\beta$ matrix model as it has been the case here. This future work considering generic solutions of the ``quantum curve" \eqref{ODE} is under progress at the moment. Then one may wonder if properties of integrability and of spectral invariants \cite{BertolaMarchal,EO,CEO} known for the hermitian case can be generalized in the arbitrary-$\beta$ case.

\section*{Acknowledgements}
We would like to thank Ga\"{e}tan Borot, Leonid Chekhov and Nicolas Orantin for useful discussions and correspondences. 
The work of B. E. is partly supported by the ANR project GranMa ``Grandes Matrices Al\'{e}atoires" ANR-08-BLAN-0311-01, by the European Science Foundation through the Misgam program, by the Quebec government with the FQRNT, and the CERN. The work of O. M. is supported by MS081 start-up grant offered by the University of Alberta and the government of Canada.

\renewcommand{\thesection}{\Alph{section}}
\setcounter{section}{0} 

\appendix
\section{Hessian matrix of the Yang--Yang function \label{appendixmatricehessienne}}

Using the notation of the paragraph regarding the variational approach and using \eqref{rr}, we have:
\beq \forall j \neq i : \left(\frac{\partial^2 \mathcal{S}}{\partial s_i \partial s_j}\right)_{|\text{extr}}=-\frac{T}{N}\frac{1}{(s_i-s_j)^2}\eeq
\beq \forall 1\leq i\leq N : \left(\frac{\partial^2 \mathcal{S}}{\partial s_i \partial s_i}\right)_{|\text{extr}}=V_1''(s_i)+\frac{T}{N}\sum_{j\neq i}\frac{1}{(s_i-s_j)^2}\eeq
\beq \forall 1\leq i,j\leq N : \left(\frac{\partial^2 \mathcal{S}}{\partial s_i \partial \td{s}_j}\right)_{|\text{extr}}=A_{i,j}(A^{-1})_{j,i}\eeq
\beq \forall 1\leq i,j\leq N :\left(\frac{\partial^2 \mathcal{S}}{\partial s_i \partial u_j}\right)_{|\text{extr}}=0\eeq
\bea \forall 1\leq i,r,s\leq N : \left(\frac{\partial^2 \mathcal{S}}{\partial s_i \partial A_{r,s}}\right)_{|\text{extr}}&=&\delta_{r,i}\td{s}_s(A^{-1})_{s,i}-\sum_{p=1}^N A_{i,p}\td{s}_p (A^{-1})_{p,r}(A^{-1})_{s,i}\cr
&=&\delta_{r,i}\td{s}_s(A^{-1})_{s,i}-\sum_{p=1}^N B_{i,r}(A^{-1})_{s,i}\eea
Where we have used here:
\beq \frac{\partial (A^{-1})_{p,q}}{\partial A_{r,s}}=-(A^{-1})_{p,r} (A^{-1})_{s,q}\eeq
We can do the same kind of computation starting from \eqref{kkk}:
\beq \forall j \neq i : \left(\frac{\partial^2 \mathcal{S}}{\partial \td{s}_i \partial \td{s}_j}\right)_{|\text{extr}}=-\frac{T}{N}\frac{1}{(\td{s}_i-\td{s}_j)^2}\eeq
\beq \forall 1\leq i\leq N : \left(\frac{\partial^2 \mathcal{S}}{\partial \td{s}_i \partial \td{s}_i}\right)_{|\text{extr}}=V_2''(\td{s}_i)+\frac{T}{N}\sum_{j\neq i}\frac{1}{(\td{s}_i-\td{s}_j)^2}\eeq
\beq \forall 1\leq i,j\leq N :\left(\frac{\partial^2 \mathcal{S}}{\partial \td{s}_i \partial u_j}\right)_{|\text{extr}}=0\eeq
\bea \forall 1\leq i,r,s\leq N : \left(\frac{\partial^2 \mathcal{S}}{\partial \td{s}_i \partial A_{r,s}}\right)_{|\text{extr}}&=&\delta_{s,i}\td{s}_r(A^{-1})_{i,r}-\sum_{p=1}^N A_{p,i}\td{s}_p (A^{-1})_{s,p}(A^{-1})_{i,r}\cr
&=&\delta_{s,i}\td{s}_r(A^{-1})_{i,r}-\sum_{p=1}^N \td{B}_{s,i}(A^{-1})_{i,r}\eea
The computation relatively to $u$ is easy:
\beq \forall 1\leq i,r,s\leq N : \left(\frac{\partial^2 \mathcal{S}}{\partial u_i \partial A_{r,s}}\right)_{|\text{extr}}=-\frac{T}{N}\delta_{r,i}\eeq
\beq \forall 1\leq i,j\leq N:\left(\frac{\partial^2 \mathcal{S}}{\partial u_i \partial s_j}\right)_{|\text{extr}}=\left(\frac{\partial^2 \mathcal{S}}{\partial u_i \partial \td{s}_j}\right)_{|\text{extr}}=\left(\frac{\partial^2 \mathcal{S}}{\partial u_i \partial u_j}\right)_{|\text{extr}}=0\eeq
Eventually the most technical one is the double derivative relatively to $A$:
\bea &&\forall\,1\leq a,b,r,s\leq N : \left(\frac{\partial^2 \mathcal{S}}{\partial A_{a,b} \partial A_{r,s}}\right)_{|\text{extr}}
=-(A^{-1})_{s,a}-(A^{-1})_{s,a} \left(\td{S}A^{-1}S\right)_{b,r} \cr
&&-(A^{-1})_{b,r} \left(A^{-1}SB\right)_{s,a}\eea

Eventually the shape of the matrix looks like:
\beq
\mathcal{H}=\begin{pmatrix}
             G_{N\times N} & X_{N \times N} & Y_{N \times N^2}& 0_{N \times N}\\ 
             X^t& \td{G}_{N \times N} & Z_{N \times N^2}& 0_{N \times N}\\
             Y^t & Z^t & \mathcal{T}_{N^2 \times N^2} & I_{N^2 \times N}\\
             0_{N \times N} & 0_{N \times N} & I^t &0_{N \times N}
            \end{pmatrix}
\eeq
Where the matrix $G$ is given by:
\bea G_{i,i}&=&V_1'(s_i)+\frac{T}{N}\sum_{j \neq i} \frac{1}{(s_i-s_j)^2}\cr
G_{i,j}&=&- \frac{T}{N(s_i-s_j)^2}\eea
and the matrix $\td{G}$ is its dual:
\bea \td{G}_{i,i}&=&V_2'(\td{s}_i)+\frac{T}{N}\sum_{j \neq i} \frac{1}{(\td{s}_i-\td{s}_j)^2}\cr
\td{G}_{i,j}&=&-\frac{T}{N(\td{s}_i-\td{s}_j)^2}\eea
The matrix $I_{N^2 \times N}$ is is a $N^2 \times N$ matrix with $N$ times $-\frac{T}{N}$ in each column and zeros elsewhere. If we sort the $A_{r,s}$'s in the following way: $(A_{1,1},\dots,A_{1,N},\dots,A_{2,1},\dots A_{2,N},\dots, A_{N,N})$ then the matrix looks like:
\beq I_{N^2 \times N}=-\frac{T}{N} \begin{pmatrix}
                              \vec{e} &\vec{0} &\dots &\dots &\vec{0}\\
\vec{0} & \vec{e} &\vec{0}& \ddots &\vec{0}\\
\vdots & & \ddots & & \vdots\\
\vdots & \ddots & & \vec{e} & \vec{0}\\
\vec{0} &\dots & & \vec{0} & \vec{e}
                             \end{pmatrix}
\eeq
The other matrices $X$, $Y$ and $Z$ are expressed by the previous computations, but since they do not have a compact or interesting form we do not mention them here.

\section{Finding kernels $K(x_0,x)$ and $G(x_0,x)$ \label{AppendixComp}}

\subsection{Decomposition of the matrix $\mathcal{D}^t(x)$}

Before presenting the way to find the kernels $K(x_0,x)$ and $G(x_0,x)$, we need to introduce some notations. First, since $\mathcal{D}^t(x)$ is going to play an important role, we decompose it in the following way: 

\beq \encadremath{\mathcal{D}^t(x)=-Y(x)\text{Id}+\L +\frac{1}{\td{t}_{d_2}}\vec{r} \vec{U}^t(x)}\eeq
where we have defined the vectors and matrices of dimension $d_2$ by:
\beq 
\vec{r}=\begin{pmatrix}0\cr0\cr \vdots\cr 0\cr 1\end{pmatrix} \virg \vec{U}=\begin{pmatrix}U_{0,0}^{(0)}(x)\cr U_{0,1}^{(0)}(x)\cr\vdots\cr U_{0,d_2-2}^{(0)}(x)\cr U_{0,d_2-1}^{(0)}(x)\end{pmatrix} \virg 
\L=  \begin{pmatrix}
0 & 1 & \dots & 0 \cr 
\vdots &\ddots & \ddots &  \cr 
\vdots & &\ddots & 1 \cr 
0 &\ldots & & 0 \cr 
\end{pmatrix}
\eeq

Note here that all the matrices defined here are already known since they only involve leading order quantities determined earlier. In particular we have:
\beq
Y(x) = V'_1(x)-\frac{T}{N} \sum_{i=1}^N \frac{1}{x-s_i}
\eeq
and:
\beq U_{0,d_2}^{(0)}(x)=-\td{t}_{d_2} \, ,\, U_{0, d_2-1}^{(0)}(x)=\td{t}_{d_2}(V_1'(x)-Y(x))-\td{t}_{d_2-1} \,,\, U_{0,k}^{(0)}(x)= \sum_{i=1}^N \frac{u_{k,i}}{ x-s_i} - \td{t}_{k} + \delta_{k,0} x\eeq
where the coefficients $u_{k,i}$ are known and satisfy \eqref{u_{k,i}}.
In the following equations, we will also need to write down a series expansion of $\vec{U}$ around $x=s_i$. Therefore, it is natural to introduce the following vectors:
\beq \label{V_i}\vec{w}_i=\frac{1}{\td{t}_{d_2}}\begin{pmatrix}u_{0,i}\cr \vdots \cr u_{d_2-1,i}\end{pmatrix} \virg    \vec{V}_i=\frac{1}{\td{t}_{d_2}} \begin{pmatrix}\underset{j\neq i}{\sum} \frac{u_{0,j}}{s_i-s_j}-\td{t}_0+s_i \cr\underset{j\neq i}{\sum}  \frac{u_{1,j}}{s_i-s_j}-\td{t}_1\cr \vdots \cr  \underset{j\neq i}{\sum}  \frac{u_{d_2-1,j}}{s_i-s_j}-\td{t}_{d_2-1}\end{pmatrix} \eeq

\beq
\vec{V}_i^{1}=\frac{1}{\td{t}_{d_2}}\begin{pmatrix}-\underset{j\neq i}{\sum} \frac{u_{0,j}}{(s_i-s_j)^2}+1 \cr -\underset{j\neq i}{\sum} \frac{u_{1,j}}{(s_i-s_j)^2}\cr \vdots\cr -\underset{j\neq i}{\sum} \frac{u_{d_2-1,j}}{(s_i-s_j)^2} \end{pmatrix} \virg \vec{V}_i^{2}=\frac{1}{\td{t}_{d_2}}\begin{pmatrix}\underset{j\neq i}{\sum} \frac{u_{0,j}}{(s_i-s_j)^3} \cr \underset{j\neq i}{\sum} \frac{u_{1,j}}{(s_i-s_j)^3}\cr \vdots\cr \underset{j\neq i}{\sum} \frac{u_{d_2-1,j}}{(s_i-s_j)^3} \end{pmatrix} \eeq
which correspond to the first terms of the expansion:
\beq \td{t}_{d_2}\vec{U}= \frac{\vec{w}_i}{x-s_i}+ \vec{V}_i+ \vec{V}^1_i (x-s_i)+ \vec{V}^2_i (x-s_i)^2 + O\left(x-s_i\right)^3\eeq 

Remember that $u_{d_2-1,i}=\frac{T}{N} \td{t}_{d_2}$ so that we have the following important relations:
\beq \label{w_i r} \encadremath{ \vec{r}^{\,\,t} \vec{w_i}=\frac{T}{N} \text{Id} \virg \vec{w_i}^t \vec{r}=\frac{T}{N} \text{Id}}\eeq

\subsection{Solving the system: Finding $G(x_0,x)$ and $K(x_0,x)$}

In order to apply our scheme of recursion we do not need to compute explicitly $K(x_0,x)$ but only to be able to determine $K(x_0,s_i)$, $K'(x_0,s_i)$ and so on (where a prime means a derivative relatively to the second variable: $K'(x_0,x)=\partial_x K(x_0,x)$). Indeed, if we know this quantities then we can compute all the residues we need in our scheme and invert the differential system. This is reassuring since it was clear that $K(x_0,x)$ was not uniquely defined since it is a solution of a differential equation. We will first determine the kernel $G(x,x_0)$ by computing the matrices $A_i(x_0)$ presented in \eqref{GG}. \textbf{Note that in order to have simpler expressions, we will label the components of the matrices $K(x_0,x)$ and $G(x_0,x)$ from $0$ to $d_2-1$ instead of the standard $1$ to $d_2$.}

The strategy is the following: we will identify the coefficients of order $(x-s_i)^k$ in equation \eqref{K}.

\medskip
\medskip
\underline{Terms in $(x-s_i)^{-1}$:}
\medskip
\medskip

A straightforward computation gives:
\beq \label{A i function of K} \encadremath{A_i(x_0)=\left(\frac{T}{N}+\vec{r}\vec{w_i}^t\right)K(x_0,s_i)}\eeq
Multiplying on the left by $\vec{w_i}^t$ and using \eqref{w_i r} we find that:
\beq 2\frac{T}{N} \vec{w_i}^tK(x_0,s_i)=\vec{w_i}^tA_i(x_0)\eeq
so that:
\beq \encadremath{ \frac{T}{N} K(x_0,s_i)=\left(1-\frac{N}{2T}\vec{r}\vec{w_i}^t\right) A_i(x_0)}\eeq
Therefore, we see that the knowledge of $K(x_0,s_i)$ is completely equivalent to the knowledge of $A_i(x_0)$. In terms of components we see that only the last line is special:
\beq \forall q\leq d_2\, , \, \forall p < d_2-1 :\,\, \left(A_i(x_0)\right)_{p,q}=\frac{T}{N}\left(K(x_0,s_i)\right)_{p,q}\eeq
\beq \forall q\leq d_2: \,\,\left(A_i(x_0)\right)_{d_2-1,q}=\frac{2T}{N} \left(K(x_0,s_i)\right)_{d_2-1,q}+\frac{1}{\td{t}_{d_2}}\sum_{k=0}^{d_2-2} u_{k,i}\left(K(x_0,s_i)\right)_{k,q}\eeq

We need to go to the next order in order to determine $K(x_0,s_i)$ or equivalently $A_i(x_0)$:

\medskip
\medskip
\underline{Terms in $(x-s_i)^{0}$:}
\medskip
\medskip

Looking at the constant coefficient of the expansion gives:
\beq \label{ordre 0}\left(\L-V'_1(s_i)+\frac{T}{N} \sum_{j\neq i} \frac{1}{s_i-s_j}+\vec{r}\vec{V_i}^t\right)K(x_0,s_i)+\vec{r}\vec{w_i}^t K'(x_0,s_i)=\frac{Id}{x_0-s_i} +\sum_{j\neq i} \frac{A_j(x_0)}{s_i-s_j}\eeq

\medskip
\medskip
\underline{Terms in $(x-s_i)^{1}$:}
\medskip
\medskip

\bea \label{ordre 1}
&&\left(-V'_1(s_i)+ \sum_{j\neq i} \frac{T}{N(s_i-s_j)}+\L+\vec{r}\vec{V_i}^t\right)K'(x_0,s_i)\cr
&&+\left(-V''_1(s_i)- \sum_{j\neq i} \frac{T}{N(s_i-s_j)^2}+\vec{r}\vec{V}_i^{1\, t}\right)K(x_0,s_i)\cr &&-\frac{T}{2N}K''(x_0,s_i)+\vec{r}\vec{w_i}^t\frac{K''(x_0,s_i)}{2}=\frac{Id}{(x_0-s_i)^2}-\sum_{j \neq i} \frac{A_j(x_0)}{(s_i-s_j)^2}\eea

We see that the first equation gives can determine the first lines of $K(x_0,s_i)$ as soon as we know the last line of $K(x_0,s_i)$. In the last equation, we see that we can get rid of the second derivatives by multiplying on the left by $\vec{r}\vec{w}_i^t$ and using $\vec{w}_i^t\vec{r}=\frac{T}{N}$. It gives:
\bea \label{ordre 1 projected}
&&\vec{r}\vec{w}_i^t\left(-V'_1(s_i)+ \sum_{j\neq i} \frac{T}{N(s_i-s_j)}+\L\right)K'(x_0,s_i)+\frac{T}{N}\vec{r}\vec{V_i}^tK'(x_0,s_i)\cr
&&+\vec{r}\vec{w}_i^t\left(-V''_1(s_i)-\sum_{j\neq i} \frac{T}{N(s_i-s_j)^2}\right)K(x_0,s_i)+\frac{T}{N}\vec{r}\vec{V}_i^{1\, t}K(x_0,s_i)\cr &&=\frac{\vec{r}\vec{w}_i^t}{(x_0-s_i)^2}-\sum_{j \neq i} \frac{\vec{r}\vec{w}_i^tA_j(x_0)}{(s_i-s_j)^2}\eea

The last equation is only interesting on the last line, because all other lines trivially give zero (because we multiply on the left by $\vec{r}$). Now we note the following identity for the component $(d_2-1,q)$:
\bea \label{simplification}&&\left(\vec{r}\vec{w}_i^t\left(-V'_1(s_i)+ \sum_{j\neq i} \frac{T}{N(s_i-s_j)}+\L\right)K'(x_0,s_i)+\frac{T}{N}\vec{r}\vec{V_i}^tK'(x_0,s_i) \right)_{d_2-1,q}\cr
&&=\left(-V'_1(s_i)+ \sum_{j\neq i} \frac{T}{N(s_i-s_j)}\right)\sum_{k=0}^{d_2-1}\frac{u_{k,i}}{\td{t}_{d_2}}K'(x_0,s_i)_{k,q}+
\sum_{k=1}^{d_2-1}\frac{u_{k-1,i}}{\td{t}_{d_2}}K'(x_0,s_i)_{k,q}\cr
&&+\frac{T}{N\td{t}_{d_2}}\sum_{k=0}^{d_2-1}\left(\underset{j\neq i}{\sum} \frac{u_{k,j}}{s_i-s_j}-\td{t}_k+s_i\delta_{k,0}\right)K'(x_0,s_i)_{k,q}\cr
&&=\frac{1}{\td{t}_{d_2}}\sum_{k=0}^{d_2-1}\Big(u_{k-1,i}-V'_1(s_i)u_{k,i}+ \frac{T}{N}\sum_{j\neq i} \frac{ u_{k,i}+u_{k,j}}{s_i-s_j}+\frac{T}{N}(-\td{t}_k+s_i\delta_{k,0}) \Big)K'(x_0,s_i)_{k,q}\cr
&&=0
\eea

Where we have used the relation for the $u_{k,i}$'s \eqref{ukirec}. Therefore, we can simplify \eqref{ordre 1 projected} into a much simpler way:
\bea \label{ordre 1 simplifie}
&&\vec{r}\vec{w}_i^t\left(-V''_1(s_i)-\sum_{j\neq i} \frac{T}{N(s_i-s_j)^2}\right)K(x_0,s_i)+\frac{T}{N}\vec{r}\vec{V}_i^{1\, t}K(x_0,s_i)\cr
&&=\frac{\vec{r}\vec{w}_i^t}{(x_0-s_i)^2}-\sum_{j \neq i} \frac{\vec{r}\vec{w}_i^tA_j(x_0)}{(s_i-s_j)^2}\eea

This time, we see that all the derivatives $K'(x_0,s_i)$ have vanished. Observe that we can replace $A_j(x_0)$ by some $K(x_0,s_i)$'s using \eqref{A i function of K}.
\bea
&&\vec{r}\vec{w}_i^t\left(-V''_1(s_i)-\sum_{j\neq i} \frac{T}{N(s_i-s_j)^2}\right)K(x_0,s_i)+\frac{T}{N}\vec{r}\vec{V}_i^{1\, t}K(x_0,s_i)\cr
&&=\frac{\vec{r}\vec{w}_i^t}{(x_0-s_i)^2}-\frac{T}{N}\sum_{j \neq i} \frac{(\vec{r}\vec{w}_i^t+\vec{r}\vec{w}_j^t)K(x_0,s_j)}{(s_i-s_j)^2}\eea

Taking the component $(d_2-1,q)$ of the last equation we have:
\bea \label{Determination 1}
&&\sum_{k=0}^{d_2-1}\left(-V''_1(s_i)u_{k,i}-\frac{T}{N}\sum_{j\neq i} \frac{ u_{k,i}+u_{k,j}}{(s_i-s_j)^2}+\frac{T}{N}\delta_{k,0} \right)K(x_0,s_i)_{k,q}\cr
&&=\frac{u_{q,i}}{(x_0-s_i)^2}-\frac{T}{N}\sum_{k=0}^{d_2-1}\sum_{j \neq i} \frac{(u_{k,i}+u_{k,j}) K(x_0,s_j)_{k,q}}{(s_i-s_j)^2}\eea

Remember that from \eqref{ordre 0} we have some recursion for the other lines. $\forall p<d_2-1$ :
\bea \label{Determination 2}
&&\left(-V'_1(s_i)+\frac{T}{N} \sum_{j\neq i} \frac{1}{s_i-s_j}\right)K(x_0,s_i)_{p,q}+K(x_0,s_i)_{p+1,q}-\frac{T}{N}\sum_{j\neq i} \frac{K(x_0,s_j)_{p,q}}{s_i-s_j}\cr
&&=\frac{\delta_{p,q}}{x_0-s_i} \eea
Therefore, we can group \eqref{Determination 1} and \eqref{Determination 2} into a matrix form. Let's introduce the following vectors:
\beq
\vec{K_q}=\begin{pmatrix} K(x_0,s_1)_{0,q} \cr \vdots \cr K(x_0,s_N)_{0,q} \cr ----\cr K(x_0,s_1)_{1,q} \cr \,\,,\,\, \cr K(x_0,s_N)_{1,q}\cr---- \cr \vdots\cr ----\cr  K(x_0,s_1)_{d_2-1,q} \cr \vdots \cr K(x_0,s_N)_{d_2-1,q} \cr\end{pmatrix} \,\,,\,\, 
\vec{f_q}=\begin{pmatrix} \vec{0} \cr----\cr \vdots\cr ----\cr  \vec{0} \cr----\cr \frac{1}{x_0-s_1} \cr \vdots\cr \frac{1}{x_0-s_N}\cr ----\cr \vec{0} \cr----\cr \vec{0}\cr----\cr \frac{u_{q,1}}{(x_0-s_1)^2} \cr \vdots\cr \frac{u_{q,N}}{(x_0-s_N)^2} \end{pmatrix}\,\,,\,\, \vec{f_{d_2-1}}=\frac{T}{N} \td{t}_{d_2}\begin{pmatrix} \vec{0} \cr----\cr \vdots\cr----\cr \vec{0}\cr----\cr \frac{1}{(x_0-s_1)^2} \cr \vdots\cr \frac{1}{(x_0-s_N)^2} \end{pmatrix}  \eeq
where the $\frac{1}{x_0-s_i}$ are in position $Nq+i$. (Note that $\vec{f}_{d_2-1}$ does not have these terms).
\beq
\mathcal{M}=\begin{pmatrix}
-B^t &Id& 0 & \ldots &0 \cr
0 & -B^t & Id & &0 \cr
\vdots & \ddots & \ddots &\vdots &\vdots\cr
0& \ldots &  &-B^t & Id \cr
B_0 & B_1& \ldots & B_{d_2-2} & B_{d_2-1}\cr
\end{pmatrix}\eeq
where all the matrices $B_k$, $\text{Id}$ and $0$ are symmetric matrices of size $N\times N$ given by:

\beq B_0= \frac{T}{N} \begin{pmatrix}(B_0)_{1,1}&  \frac{u_{0,1}+u_{0,2}}{(s_1-s_2)^2} &\ldots & \frac{u_{0,1}+u_{0,N}}{(s_1-s_N)^2}\cr
 \frac{u_{0,1}+u_{0,2}}{(s_1-s_2)^2} & \ddots & \ddots  & \vdots \cr
\vdots & \ddots &  &   \frac{u_{0,N-1}+u_{0,N}}{(s_{N-1}-s_N)^2}\cr
 \frac{u_{0,N}+u_{0,1}}{(s_N-s_1)^2} &  \ldots &  \frac{u_{0,N}+u_{0,N-1}}{(s_N-s_{N-1})^2} &(B_0)_{N,N}\end{pmatrix}
\eeq
with $(B_0)_{i,i}=-\frac{N}{T}V''_1(s_i)u_{0,i}-\underset{j \neq 1}{\sum} \frac{u_{0,i}+u_{0,j}}{(s_i-s_j)^2} +1$.
 
And $\forall k>0$: 
\beq B_k= \begin{pmatrix}(B_k)_{1,1}& \frac{T}{N} \frac{u_{k,1}+u_{k,2}}{(s_1-s_2)^2} &\ldots & \frac{T}{N} \frac{u_{k,1}+u_{k,N}}{(s_1-s_N)^2}\cr
\frac{T}{N} \frac{u_{k,1}+u_{k,2}}{(s_1-s_2)^2} &  & \ddots & \vdots \cr
\vdots & \ddots & \ddots &  \frac{T}{N} \frac{u_{k,N-1}+u_{k,N}}{(s_{N-1}-s_N)^2}\cr
\frac{T}{N} \frac{u_{k,N}+u_{k,1}}{(s_N-s_1)^2} &  \ldots & \frac{T}{N} \frac{u_{k,N}+u_{k,N-1}}{(s_N-s_{N-1})^2} &(B_k)_{N,N} \end{pmatrix}
\eeq 
with $(B_k)_{i,i}=-V''_1(s_i)u_{k,i}-\frac{T}{N}\underset{j \neq i}{\sum} \frac{u_{k,i}+u_{k,j}}{(s_i-s_j)^2}$

Then, we can rewrite the equations \eqref{Determination 1} and \eqref{Determination 2} into the nice matrix form:
\beq \label{matrix form} \encadremath{ \forall \, 0\leq q\leq d_2-1 : \,  \mathcal{M} \vec{K_q}=\vec{f_q}}\eeq

Hence, by a simple matrix inversion, we can find all the $K(x_0,s_i)$ and then using \eqref{A i function of K} find the kernel $G(x_0,x)$. Note in particular that the last column of $K(x_0,s_i)$ and hence $G(x_0,x)$ only have terms in $\frac{1}{(x_0-s_i)^2}$. In particular, if we define the ``Bergmann kernel'' $B(x_0,x)=-\frac{1}{2}\partial_x G(x_0,x)$, we observe that it is a symmetric function. Eventually, note that in the case $d_2=2$ (i.e. the one-matrix model) we recover the results of the one-matrix model as described in \cite{MoiBertrand} since we have a scalar problem corresponding to the component $(d_2-1,d_2-1)$ of our problem.

\subsection{Computation/choice of $K'(x_0,s_i)$ and $K''(x_0,s_i)$}

Now that the $K(x_0,s_i)$'s and the kernel $G(x_0,x)$ have been determined, the next step is to find the higher derivatives $K'(x_0,s_i)$, $K''(x_0,s_i)$, $\dots$ in order to be able to apply the residue formula \eqref{residue}. Unfortunately, since $K(x_0,x)$ is defined as as solution of a differential equation, it is not determined uniquely. For instance we can write the loop equation projected at orders $(x-s_i)^0$, $(x-s_i)^1$ and $(x-s_i)^2$. Using notations \eqref{V_i} we find:

\underline{order 0}:
\beq \label{0}\left(\L-V'_1(s_i)+\frac{T}{N} \sum_{j\neq i} \frac{1}{s_i-s_j}+\vec{r}\vec{V_i}^t\right)K(x_0,s_i)+\vec{r}\vec{w_i}^t K'(x_0,s_i)=\frac{Id}{x_0-s_i} +\sum_{j\neq i} \frac{A_j(x_0)}{s_i-s_j}\eeq
\\
\underline{order 1}:
\bea \label{1}
&&\left(-V'_1(s_i)+ \sum_{j\neq i} \frac{T}{N(s_i-s_j)}+\L+\vec{r}\vec{V_i}^t\right)K'(x_0,s_i)\cr
&&+\left(-V''_1(s_i)- \sum_{j\neq i} \frac{T}{N(s_i-s_j)^2}+\vec{r}\vec{V}_i^{1\, t}\right)K(x_0,s_i)\cr &&-\frac{T}{2N}K''(x_0,s_i)+\vec{r}\vec{w_i}^t\frac{K''(x_0,s_i)}{2}=\frac{Id}{(x_0-s_i)^2}-\sum_{j \neq i} \frac{A_j(x_0)}{(s_i-s_j)^2}\eea
\\
\underline{order 2}:
\bea \label{2}
&&\left(-V'_1(s_i)+ \sum_{j\neq i} \frac{T}{N(s_i-s_j)}+\L+\vec{r}\vec{V_i}^t\right)\frac{K''(x_0,s_i)}{2}\cr
&&+\left(-V''_1(s_i)- \sum_{j\neq i} \frac{T}{N(s_i-s_j)^2}+\vec{r}\vec{V}_i^{1\, t}\right)K'(x_0,s_i)\cr
&&+\left(-\frac{V'''_1(s_i)}{2}+ \sum_{j\neq i} \frac{T}{N(s_i-s_j)^3}+\vec{r}\vec{V}_i^{2\, t}\right)K(x_0,s_i)\cr &&-\frac{T}{3N}K'''(x_0,s_i)+\vec{r}\vec{w_i}^t\frac{K'''(x_0,s_i)}{6}=\frac{Id}{(x_0-s_i)^3}+\sum_{j \neq i} \frac{A_j(x_0)}{(s_i-s_j)^3}\eea

Remember that if we multiply on the left by $\vec{r}\vec{w}_i^t$ we kill the terms $\left(-V'_1(s_i)+ \underset{j\neq i}{\sum} \frac{T}{N(s_i-s_j)}+\L+\vec{r}\vec{V_i}^t\right)$. We clearly see that now that the $K(x_0,s_i)$'s are known, we have some arbitrariness in solving this set of equations. Moreover, it is not obvious that this set of equations is compatible with our determination of $K'(x_0,s_i)$'s in the last section. Fortunately, we prove in appendix \ref{Cancellation} that this is the case. For our residue scheme, we can choose whatever kernel $K(x_0,x)$ we want providing that it satisfies the assumption described earlier. Here we will choose it so that:
\beq \label{assumption}\encadremath{\forall \, p < d_2-1 \, ,\, \forall \, q,i : K'(x_0,s_i)_{p,q}=0 \virg \forall \, q,i : K''(x_0,s_i)_{d_2-1,q}=0}\eeq

We claim that this assumption is compatible with our previous set of equations. Indeed, since the matrices $K'(x_0,s_i)$ have vanishing $d_2-2$ first lines, we can extract from \eqref{0}, the missing components:
\bea
&&\frac{T}{N} K'(x_0,s_i)=\underset{j\neq i}{\sum} \frac{A_j(x_0)_{d_2-1,q}}{s_i-s_j}+\left(V_1'(s_i)-\underset{j\neq i}{\sum} \frac{T}{N(s_i-s_j)}\right)K(x_0,s_i)_{d_2-1,q}\cr
&&+\frac{\delta_{q,d_2-1}}{x_0-s_i}+\underset{k=0}{\overset{d_2-1}{\sum}} \left(\underset{j\neq i}{\sum} \frac{u_{k,j}}{s_i-s_j} -\td{t}_k+\delta_{k,0}s_i\right)K(x_0,s_i)_{k,q}\cr
&=&\frac{\delta_{q,d_2-1}}{x_0-s_i}+V_1'(s_i)K(x_0,s_i)_{d_2-1,q} -\frac{T}{N}\underset{j\neq i}{\sum}\frac{K(x_0,s_i)_{d_2-1,q}-K(x_0,s_i)_{d_2-1,q}}{s_i-s_j}\cr
&+&\underset{k=0}{\overset{d_2-1}{\sum}} \left(\underset{j\neq i}{\sum} \frac{u_{k,j}}{s_i-s_j}\left(K(x_0,s_i)_{k,q}-K(x_0,s_j)_{k,q}\right)+(-\td{t}_k+\delta_{k,0}s_i)K(x_0,s_i)_{k,q}\right)\cr  \eea

This expression is also equivalent to the following (multiply \eqref{0} on the left by $\vec{r}\vec{w}_i^t$ and use \eqref{simplification} to get rid of $K(x_0,s_i)$):
\beq \label{K'}\fbox{$
   \begin{array}{rcl}
 \left(\frac{T}{N}\right)^2 K'(x_0,s_i)_{d_2-1,q}&=&\frac{u_{q,i}}{\td{t}_{d_2}(x_0-s_i)} +\underset{k=0}{\overset{d_2-1}{\sum}}\underset{j\neq i}{\sum}\frac{u_{k,i}A_j(x_0)_{k,q}}{\td{t}_{d_2}(s_i-s_j)}\cr
&=& \frac{u_{q,i}}{\td{t}_{d_2}(x_0-s_i)} +\frac{T}{N}\underset{k=0}{\overset{d_2-1}{\sum}}\underset{j\neq i}{\sum}\frac{(u_{k,i}+u_{k,j})K(x_0,s_j)_{k,q}}{\td{t}_{d_2}(s_i-s_j)}\cr\end{array}
   $} 
\eeq
Then we can compute the first lines of $K''(x_0,s_i)$ using \eqref{1}: $\forall p<d_2-1$ :
\beq \label{K''}\fbox{$
   \begin{array}{rcl}
\frac{T}{2N}K''(x_0,s_i)_{p,q}&=& \delta_{p,d_2-2}K'(x_0,s_i)_{d_2-1,q}-\left(V''_1(s_i)+ \underset{j\neq i}{\sum} \frac{T}{N(s_i-s_j)^2}\right)K(x_0,s_i)_{p,q}\cr
&&-\frac{\delta_{p,q}}{(x_0-s_i)^2}+\underset{j\neq i}{\sum} \frac{A_j(x_0)_{p,q}}{(s_i-s_j)^2}\cr
 &=& \delta_{p,d_2-2}K'(x_0,s_i)_{d_2-1,q}-V''_1(s_i)K(x_0,s_i)_{p,q}-\frac{\delta_{p,q}}{(x_0-s_i)^2}\cr
&&-\frac{T}{N}\underset{j\neq i}{\sum}\frac{K(x_0,s_i)_{p,q}-K(x_0,s_j)_{p,q}}{(s_i-s_j)^2}\end{array}
   $} 
\eeq
where the r.h.s. is known. We see also that in \eqref{1}, the last line of $K''(x_0,s_i)$ vanishes identically so it cannot be determined and we can choose it to be zero to simplify the computations. These degree of freedom were also present in the one-matrix model \cite{MoiBertrand} where $K''(x_0,s_i)$ (which was scalar) remained unknown. Moreover, if we count the degrees of freedom, we see that we have exactly $d_2\,^2$ unknowns in the determination of $K(x_0,x)$ which corresponds to a unknown matrix of size $d_2\times d_2$ as expected for a linear differential equation of the form \eqref{K}.

\subsection{Recursion to determine $K^{(n)}(x_0,s_i)$}

Now that we have the matrices $K(x_0,s_i)$, $K'(x_0,s_i)$ and $K''(x_0,s_i)$, we can perform a recursion to get the higher derivatives. To obtain them, one needs to look at the terms of the expansion of order $(x-s_i)^n$. The main problem here is not conceptual, but rather notational. We introduce the following expansions:
\bea Y(x)&=&\frac{T}{N(x-s_i)}+ \sum_{p=0}^\infty \left(\frac{V_1^{(p+1)}(s_i)}{p!}- \sum_{j\neq i} \frac{T(-1)^p }{N(s_i-s_j)^{p+1}}\right)(x-s_i)^p\cr
&:=&\sum_{p=-1}^\infty Y_{(x-s_i)^p}(x-s_i)^p\eea 

From the definition of $U_{0,k}^{(0)}(x)=\underset{i=1}{\overset{N}{\sum}} \frac{u_{k i}}{x-s_i} -\td{t}_k +\delta_{k,0}x$ we have also:
\bea  
U_{0,k}^{(0)}(x)&=&\frac{u_{k i}}{x-s_i} +\left(\sum_{j\neq i} \frac{u_{k j}}{s_i-s_j} -\td{t}_k +\delta_{k,0}s_i\right)+\left(\sum_{j\neq i} \frac{-u_{k j}}{(s_i-s_j)^2}+\delta_{k,0}\right)(x-s_i) \cr
&& +\sum_{p=2}^\infty\left(\sum_{j\neq i} \frac{(-1)^p u_{k j}}{(s_i-s_j)^{p+1}}\right)(x-s_i)^p:=\sum_{p=-1}^\infty U_{0,k,(x-s_i)^p}^{(0)}(x-s_i)^p  \eea 
We can regroup them into the following vectors:
\beq \vec{U}_{(x-s_i)^p}=\begin{pmatrix}U_{0,0,(x-s_i)^p}^{(0)}\cr U_{1,0,(x-s_i)^p}^{(0)}\cr \vdots \cr U_{d_2-1,0,(x-s_i)^p}^{(0)}\end{pmatrix}\eeq
Note for example that we have in the former notations $\vec{w}_i=\vec{U}_{(x-s_i)^{-1}}$, $\vec{V}_i=\vec{U}_{(x-s_i)^{0}}$, $\vec{V}_i^{1}=\vec{U}_{(x-s_i)^{1}}$,...

We will also need an expansion in $(x-s_i)^p$ of the $x \mapsto G(x_0,x)$ function:
\bea G(x_0,x)&=&\frac{A_i(x_0)}{x-s_i}+\sum_{p=0}^\infty \left(\frac{ Id}{(x_0-s_i)^{p+1}}+\sum_{j \neq i} \frac{(-1)^p A_j(x_0)}{(s_i-s_j)^{p+1}}\right)(x-s_i)^p\cr
&:=&\sum_{p=-1}^\infty G_{(x-s_i)^p}(x_0)(x-s_i)^p \eea

Now it is easy to write the equality of the power $n\geq 1$ of the differential equation $\left(\mathcal{D}^t(x)-\frac{T}{N} \partial_x\right)K(x_0,x)=G(x_0,x)$:

\medskip
\medskip
\underline{Terms in $(x-s_i)^{n}$:}
\medskip
\medskip

\bea \label{order n}
G_{(x-s_i)^n}(x_0)&=&-n\frac{T}{N}\frac{K^{(n+1)}(x_0,s_i)}{(n+1)!}-\sum_{l=0}^{n}Y_{(x-s_i)^{n-l}}\frac{K^{(l)}(x_0,s_i)}{l!}+\L \frac{K^{(n)}(x_0,s_i)}{n!}\cr
&&+\vec{r}\vec{w_i}^t \frac{K^{(n+1)}(x_0,s_i)}{(n+1)!} +\frac{1}{\td{t}_{d_2}} \vec{r} \sum_{l=0}^n \vec{U}^t_{(x-s_i)^{n-l}}\frac{K^{(l)}(x_0,s_i)}{l!} 
\eea

We can project this equation on the first $d_2-1$ lines. Any term with a $\vec{r}$ does not contribute. We find for $p<d_2-1$:
\beq \label{Kn}\fbox{$
   \begin{array}{rcl}
&&n\frac{T}{N}\frac{K^{(n+1)}(x_0,s_i)_{p,q}}{(n+1)!}=-G_{(x-s_i)^n}(x_0)_{p,q}\cr
&&-\underset{l=0}{\overset{n}{\sum}}Y_{(x-s_i)^{n-l}}\frac{K^{(l)}(x_0,s_i)_{p,q}}{l!}+\frac{K^{(n)}(x_0,s_i)_{p+1,q}}{n!}\cr
\end{array}
   $} 
\eeq

Note that the r.h.s is completely known by recursion since it involves derivatives only up to order $n$. Then since we know the first lines, we can compute the missing last one by projecting the equation on the last line:
\beq \label{Knbis}\fbox{$
   \begin{array}{rcl}
&&(n-1)\frac{T}{N}\frac{K^{(n+1)}(x_0,s_i)_{d_2-1,q}}{(n+1)!}=-G_{(x-s_i)^n}(x_0)_{d_2-1,q}\cr
&&-\underset{l=0}{\overset{n}{\sum}}Y_{(x-s_i)^{n-l}}\frac{K^{(l)}(x_0,s_i)_{d_2-1,q}}{l!}+\underset{k=0}{\overset{d_2-2}{\sum}}\frac{u_{k,i}}{\td{t}_{d_2}}\frac{K^{(n+1)}(x_0,s_i)_{k,q}}{(n+1)!}\cr
&& +\frac{1}{\td{t}_{d_2}} \underset{k=0}{\overset{d_2-1}{\sum}}\underset{l=0}{\overset{n}{\sum}} \vec{U}_{k,(x-s_i)^{n-l}}\frac{K^{(l)}(x_0,s_i)_{k,q}}{l!}  \end{array}
   $} 
\eeq

Again, we note that the r.h.s. is known by recursion since it involves only derivatives up to order $n$. Moreover the coefficient $(n-1)\frac{T}{N}$ is never zero when $n>1$ so that we can compute $K^{(n+1)}(x_0,s_i)_{d_2-1,q}$. Therefore we can determine with both \eqref{Kn} and \eqref{Knbis} all the components of $K^{(n+1)}(x_0,s_i)$ as soon as we know the lower orders. Since we know $K(x_0,s_i)$, $K'(x_0,s_i)$ and $K''(x_0,s_i)$ \textbf{we can compute by recursion every derivative $K^{(n)}(x_0,s_i)$ and thus implement the scheme given by \eqref{recursion} to determine all functions $U_{0,k}^{(g)}(x)$ and then by taking the last component of this vector \eqref{W1g} the correlation functions $W_1^{(g)}(x)$}.

\section{Compatibility of the system at order $0$ \label{Cancellation}}

In this appendix, we will show that equations given by the identification of the coefficient $(x-s_i)^0$ given before do not add any new constraints. Indeed, from the equation \eqref{ordre 0} we get:
\bea&&\left(-V'_1(s_i)+\frac{T}{N} \sum_{j\neq i} \frac{1}{s_i-s_j}+\L+\vec{r}\vec{V}_i\right)K(x_0,s_i)\cr
&&+\vec{r}\vec{w}_i^t K'(x_0,s_i)=\frac{Id}{x_0-s_i} +\sum_{j\neq i} \frac{(\frac{T}{N}+\vec{r}\vec{w}_j^t)K(x_0,s_j)}{s_i-s_j}\eea
A possible worry is that one can eliminate the term $K'(x_0,s_i)$ in the previous equation by multiplying on the left by $\vec{r}\vec{w}_i^t$. Then one can obtain a closed system of equations for the $K(x_0,s_i)$'s which is different from the one we presented before. We will prove here that this does not happen.

Let's first project the equation by multiplying on the left by $\vec{r}\vec{w_i}^t$. The first term disappear because of the relation \eqref{simplification}. Therefore we find:
\beq 
\frac{T}{N}\vec{r}\vec{w_i}^t K'(x_0,s_i)=\frac{\vec{r}\vec{w_i}^t}{x_0-s_i} +\frac{T}{N}\sum_{j\neq i} \frac{(\vec{r}\vec{w_i}^t+\vec{r}\vec{w_j}^t)K(x_0,s_j)}{s_i-s_j}\eeq
From there we can extract $\vec{r}\vec{w_i}^t K'(x_0,s_i)$ and put it back into the first equation. We get a closed relation for the $K(x_0,s_i)$:
\bea&&\frac{T}{N}\left(-V'_1(s_i)+\frac{T}{N} \sum_{j\neq i} \frac{1}{s_i-s_j}+\L+\vec{r}\vec{V_i}\right)K(x_0,s_i)\cr
&&+\frac{\vec{r}\vec{w_i}^t}{x_0-s_i} +\frac{T}{N}\sum_{j\neq i} \frac{(\vec{r}\vec{w_i}^t+\vec{r}\vec{w_j}^t)K(x_0,s_j)}{s_i-s_j}\cr 
&&=\frac{T}{N(x_0-s_i)} +\frac{T}{N}\sum_{j\neq i} \frac{(\frac{T}{N}+\vec{r}\vec{w_j}^t)K(x_0,s_j)}{s_i-s_j}\eea
Some simplifications occur and we find:
\bea&&\frac{T}{N}\left(-V'_1(s_i)+\frac{T}{N} \sum_{j\neq i} \frac{1}{s_i-s_j}+\L+\vec{r}\vec{V_i}\right)K(x_0,s_i)\cr
&&=\frac{\frac{T}{N}-\vec{r}\vec{w_i}^t}{x_0-s_i} +\frac{T}{N}\sum_{j\neq i} \frac{(\frac{T}{N}-\vec{r}\vec{w_i}^t)K(x_0,s_j)}{s_i-s_j}\eea

We can project this relation into components. Note that the first $d_2-2$ lines do not give any new information since they are the same as \eqref{ordre 0}:
\beq 
\left(-V'_1(s_i)+\frac{T}{N} \sum_{j\neq i} \frac{1}{s_i-s_j}\right)K(x_0,s_i)_{p,q} -\frac{T}{N}\sum_{j \neq i} \frac{K(x_0,s_j)_{p,q}}{s_i-s_j} + K(x_0,s_i)_{p+1,q}=\frac{\delta_{p,q}}{x_0-s_i}\eeq

The projection on the last line is interesting:
\bea
&&\frac{T}{N}\left(-V'_1(s_i)+\frac{T}{N} \sum_{j\neq i} \frac{1}{s_i-s_j}\right)K(x_0,s_i)_{d_2-1,q}\cr
&& +\frac{T}{N\td{t}_{d_2}}\sum_{k=0}^{d_2-1}\left(\sum_{j\neq i}\frac{u_{k,j}}{s_i-s_j} -\td{t}_k+\delta_{k,0}s_i\right)K(x_0,s_i)_{k,q}\cr
&&=\frac{\frac{T}{N}\delta_{d_2-1,q}-\frac{u_{q,i}}{\td{t}_{d_2}}}{x_0-s_i} +\frac{T^2}{N^2}\sum_{j\neq i}\frac{K(x_0,s_j)_{d_2-1,q}}{s_i-s_j}\cr
&&- \frac{T}{N\td{t}_{d_2}}\sum_{k=0}^{d_2-1}\left(\sum_{j\neq i}\frac{u_{k,i}}{s_i-s_j}\right)K(x_0,s_j)_{k,q}\cr\eea

In the last term, the case $k=d_2-1$ cancels the previous one:
\bea
&&\frac{T}{N}\left(-V'_1(s_i)+\frac{T}{N} \sum_{j\neq i} \frac{1}{s_i-s_j}\right)K(x_0,s_i)_{d_2-1,q}\cr &&+\frac{T}{N\td{t}_{d_2}}\sum_{k=0}^{d_2-1}\left(\sum_{j\neq i}\frac{u_{k,j}}{s_i-s_j} -\td{t}_k+\delta_{k,0}s_i\right)K(x_0,s_i)_{k,q}\cr
&&=\frac{\frac{T}{N}\delta_{d_2-1,q}-\frac{u_{q,i}}{\td{t}_{d_2}}}{x_0-s_i} - \frac{T}{N\td{t}_{d_2}}\sum_{k=0}^{d_2-2}\left(\sum_{j\neq i}\frac{u_{k,i}}{s_i-s_j}\right)K(x_0,s_j)_{k,q}\cr\eea

We can put this system into a matrix $d_2N\times d_2N$ form. We introduce:
\beq
\vec{K_q}=\begin{pmatrix} K(x_0,s_1)_{1,q} \cr \vdots \cr K(x_0,s_N)_{1,q} \cr ----\cr K(x_0,s_1)_{2,q} \cr \vdots \cr K(x_0,s_N)_{2,q}\cr---- \cr \vdots\cr ----\cr  K(x_0,s_1)_{d_2-1,q} \cr \vdots \cr K(x_0,s_N)_{d_2-1,q} \cr\end{pmatrix} \,\,,\,\, 
\vec{g_q}=\begin{pmatrix} \vec{0} \cr----\cr \vdots\cr ----\cr  \vec{0} \cr----\cr \frac{1}{x_0-s_1} \cr \vdots\cr \frac{1}{x_0-s_N}\cr ----\cr \vec{0} \cr----\cr  \frac{-u_{q,1}}{\td{t}_{d_2}(x_0-s_1)} \cr \vdots\cr \frac{-u_{q,N}}{\td{t}_{d_2}(x_0-s_N)}\end{pmatrix}  
\,\,,\,\, \vec{g_{d_2-1}}=\vec{0}\eeq 
and 
\beq
\mathcal{P}=\begin{pmatrix}-B &Id& 0 & \ldots &0 \cr
0 & -B & Id & &0 \cr
\vdots & \ddots & \ddots &Id &\vdots\cr
0& \ldots &  &-B & Id \cr
C_0 & C_1& \ldots & C_{d_2-2} & C_{d_2-1}\cr
\end{pmatrix}\eeq
where all the matrices $C_k$, $Id$ and $0$ are of size $N\times N$ given by :
\beq C_k=\left\{ 
\begin{array}{l}
(C_k)_{i,i} =\frac{T}{N\td{t}_{d_2}}\left(\underset{j\neq i}{\sum}\frac{u_{k,j}}{s_i-s_j}-\td{t}_k+\delta_{k,0}s_i\right) \cr
(C_k)_{i,j} = \frac{T}{N} \frac{u_{k,i}}{\td{t}_{d_2}(s_i-s_j)}
\end{array}
\right.
\eeq
and a special case for $C_{d_2-1}$:
\beq C_{d_2-1}=\frac{T}{N} \, diag(-V_1'(s_i)+\frac{2T}{N} \sum_{j\neq i}\frac{1}{s_i-s_j} -\frac{\td{t}_{d_2-1}}{\td{t}_{d_2}})
\eeq

Then the two sets of equations are equivalent to the matrix equation:
\beq \encadremath{\forall q:\, \mathcal{P} \vec{K}_q=\vec{g}_q}\eeq

The main problem now is that this system is different from the one we got before and thus the system seems overdetermined.
Fortunately, the matrices, $\mathcal{M}$ and $\mathcal{P}$ and the vectors $\vec{f}_q$ and $\vec{g}_q$ have the same $d_2-2$ first lines. But the main difference is that $\vec{f}_q$ has double poles in $x_0=s_i$ on the last line whereas $\vec{g}_q$ has simple poles. Writing:
\beq  
\mathcal{P}\mathcal{M}^{-1} \vec{f}_q=\vec{g}_q\eeq
And looking at the last component gives that the two matrix systems are compatible if and only if the last column of $\mathcal{P}\mathcal{M}^{-1}$ is null. From the form (companion-like) of the matrices we have:
It is easy computation to see that the last column can be split in blocks of size $m$ :
\beq \mathcal{M}^{-1}=\begin{pmatrix}X&\dots&X&M_0\cr \vdots & & \vdots &\vdots\cr X&\dots &X &M_{d_2-1}\end{pmatrix}\eeq
where:
\beq M_k=B^k\left(B_0+B_1B+\dots +B_{d_2-1}B^{d_2-1}\right)^{-1}\eeq
so that:
Multiplying on the left by $\mathcal{P}$ gives that the two systems are compatible if and only if:
\beq \encadremath { C_0+C_1 B +\dots +C_{d_2-1}B^{d_2-1}=0}\eeq
Note that we have also the determinantal identity:
\beq \label{determinantal identity}
det(\mathcal{M})=det\begin{pmatrix}-X &Id& 0 & \ldots &0 \cr
0 & -X & Id & &0 \cr
\vdots & \ddots & \ddots &Id &\vdots\cr
0& \ldots &  &-X & Id \cr
A_0 & A_1& \ldots & A_{d-1} & A_{d}\cr
\end{pmatrix}=(-1)^{Nd}det\left(\sum_{i=0}^{d} A_iX^i\right)\eeq
valid for every matrices $A_i$ and every matrix $X$ of size $N \times N$.

This condition seems rather completely non trivial from the definition of $B$ and of $C_k$. But we will show in the following that from the definition of the $u_{k,i}$'s, the last equation is automatically satisfied. Note then that because of \eqref{determinantal identity} we obtain at the same time that $\mathcal{P}$ is not invertible and thus only \eqref{matrix form} gives us the $K(x_0,s_i)$'s. Let's now prove that we have always:
\beq C_0+C_1 B +\dots +C_{d_2-1}B^{d_2-1}=0\eeq
From the definition of the $u_{k,i}$'s we have:
\bea &&\frac{N}{T}\sum_{k=0}^{d_2-1} u_{k-1,i}(B^k)_{i,j}= \frac{N}{T}\sum_{k=0}^{d_2-2} u_{k,i}\left(V_1'(s_i)-\frac{T}{N}\sum_{r\neq i} \frac{1}{s_i-s_r} \right) (B^k)_{i,j}\cr
&&+\sum_{k=0}^{d_2-2}\sum_{r\neq i}  \frac{u_{k,i}}{s_i-s_r} (B^k)_{r,j}\eea

We obtain then that:
\bea \left(\sum_{k=0}^{d_2-2} C_kB^k\right)_{i,j}&=& \frac{T}{N\td{t}_{d_2}}\sum_{k=0}^{d_2-2}\left(\sum_{r\neq i} \frac{u_{k,r} (B^k)_{i,j} +u_{k,i} (B^k)_{r,j}}{s_i-s_r}-\td{t}_k (B^k)_{i,j} +s_i \delta_{i,j}\right)\cr
\eea
Note then that:
$$\sum_{r \neq i} \frac{u_{k,r}}{s_i-s_r}=\td{t}_k -\delta_{k,0}s_i -\frac{N}{T}u_{k-1,i} +\left(\frac{N V_1'(s_i)}{T}-\sum_{r\neq i}\frac{1}{s_i-s_r}\right)u_{k,i}$$
so that:
\bea \left(\sum_{k=0}^{d_2-2} C_kB^k\right)_{i,j}&=&\frac{T}{N\td{t}_{d_2}}\sum_{k=0}^{d_2-2}\Big[ \left(\td{t}_k -\delta_{k,0}s_i
+\left(\frac{NV_1'(s_i)}{T}-\sum_{r\neq i}\frac{1}{s_i-s_r}\right)u_{k,i}\right) (B^k)_{i,j}\cr
&&-\frac{N}{T}u_{k-1,i}(B^k)_{i,j}+ \frac{ u_{k,i}}{s_i-s_r}(B^k)_{r,j}-\td{t}_k (B^k)_{i,j}+s_i \delta_{i,j}\Big]\cr&=& \frac{1}{\td{t}_{d_2}}u_{d_2-2,i} (B^{d_2-1})_{i,j}\eea

Then we need to add the last term:
$$\left(C_{d_2-1}B^{d_2-1}\right)_{i,j}=\left(-\frac{T}{N} V_1'(s_i)+\frac{2T^2}{N^2}\sum_{j \neq i} \frac{1}{s_i-s_j}-\frac{T}{N}\frac{\td{t}_{d_2-1}}{\td{t}_{d_2}}\right)(B^{d_2-1})_{i,j}$$.

From the definition of $u_{d_2-1,2}$ we have also:
$$u_{d_2-2,i}=\sum_{k=1}^N (B^t)_{i,k} u_{d_2-1,k}+\frac{T}{N} \td{t}_{d_2-1}$$
so that:
$$\frac{1}{\td{t}_{d_2}}u_{d_2-2,i}-\frac{T}{N} V_1'(s_i)+\frac{2T^2}{N^2}\sum_{j \neq i} \frac{1}{s_i-s_j}-\frac{T}{N}\frac{\td{t}_{d_2-1}}{\td{t}_{d_2}}=0$$
and thus:
\beq 
\left(\sum_{k=0}^{d_2-1} C_kB^k\right)_{i,j}=\left(\frac{1}{\td{t}_{d_2}}u_{d_2-2,i}-\frac{T}{N} V_1'(s_i)+\frac{T^2}{N^2}\sum_{j \neq i} \frac{2}{s_i-s_j}-\frac{T}{N}\frac{\td{t}_{d_2-1}}{\td{t}_{d_2}}\right)(B^{d_2-1})_{i,j}=0\eeq

proving that only \eqref{matrix form} gives the correct formula to determine all the $K(x_0,s_i)$'s.

\section{Example: Only one root $s$ \label{N=1} \label{Oneroot}}

In order to illustrate our algorithm, we present here the case when there is only one root $s$ (we omit the subscript $s_1$). In this case, almost all the computation can be carried out analytically and in particular we show how to obtain $W_1^{(1)}(x)$ and $F_1$ in this context. The results presented here are direct applications of the formula presented in the article.
First we have from \eqref{uki}:
\beq B=V_1'(s) \virg u_{k,1}=\frac{T}{N}\sum_{p=0}^{d_2-k-1} \td{t}_{k+p+1} \left(V'_1(s)\right)^p\eeq
Then we have:
\beq U_0^{(0)}(x,y)= \frac{T}{N(x-s)}\sum_{p=0}^{d_2}\sum_{q=0}^{p-1}\td{t}_p y^{p-1-q}\left(V_1'(s)\right)^q \,\,+x-V_2'(y)\eeq
The action $\mathcal{S}$ is given by:
\beq \mathcal{S}(s,\td{s},A,u)=V_1(s)+V_2(\td{s})-s\td{s}+\frac{T}{N} \ln A -\frac{T}{N} u(A-1)\eeq
The extremum is obtained for $u=1$, $A=1$, $V_1'(s)=\td{s}$ and $V'_2(\td{s})=s=V_2'(V_1'(s))$.
The Hessian matrix is given by:
\beq \mathcal{H}(s,\td{s},A,u)=\begin{pmatrix}
                  V_1''(s) & -1 &0 &0 \\
-1 & V_2''(\td{s}) & 0 &0 \\
0&0&-\frac{T}{NA^2}& -\frac{T}{N} \\
0&0&-\frac{T}{N} &0
                 \end{pmatrix}
\eeq
At the extremum it gives:
\beq \mathcal{H}_{|\text{extr}}=\begin{pmatrix}
                  V_1''(s) & -1 &0 &0 \\
-1 & V_2''(V_1'(s)) & 0 &0 \\
0&0&-\frac{T}{N}& -\frac{T}{N} \\
0&0&-\frac{T}{N} &0
                 \end{pmatrix}
\eeq
which gives:
\beq \mathcal{H}^{-1}(s,\td{s},A,u)=\begin{pmatrix}
                  \frac{V_2''(\td{s})}{V_2''(\td{s})V_1''(s)-1} & \frac{1}{V_2''(\td{s})V_1''(s)-1} &0 &0 \\
\frac{1}{V_2''(\td{s})V_1''(s)-1} & \frac{V_1''(s)}{V_2''(\td{s})V_1''(s)-1} & 0 &0 \\
0&0&0& -\frac{N}{T} \\
0&0&-\frac{N}{T} &\frac{N}{TA^2}
                 \end{pmatrix}
\eeq

so that at the extremum we get:
\beq \mathcal{H}^{-1}\,_{|\text{extr}}=\begin{pmatrix}
                  \frac{V_2''(V_1'(s))}{V_2''(V_1'(s))V_1''(s)-1} & \frac{1}{V_2''(V_1'(s))V_1''(s)-1} &0 &0 \\
\frac{1}{V_2''(V_1'(s))V_1''(s)-1} & \frac{V_1''(s)}{V_2''(V_1'(s))V_1''(s)-1} & 0 &0 \\
0&0&0& -\frac{N}{T} \\
0&0&-\frac{N}{T} &\frac{N}{T}
                 \end{pmatrix}
\eeq
Hence we have with \eqref{ooooo}:
\beq \frac{\partial s}{\partial V_1(x)}=\frac{V_2''(V_1'(s))}{\left(V_2''(V_1'(s))V_1''(s)-1\right)(x-s)^2}\eeq
giving the formula for $W_2^{(0)}(x_1,x_2)$:
\beq W_2^{(0)}(x_1,x_2)=\frac{T}{N}\frac{V_2''(V_1'(s))}{\left(V_2''(V_1'(s))V_1''(s)-1\right)(x_1-s)^2(x_2-s)^2}\eeq
We can now compute the kernels $G(x_0,x)$ and $K(x_0,s_i)$. The matrix $\mathcal{M}$ (size $d_2\times d_2$) is given by:
\beq \mathcal{M}=\begin{pmatrix}
                   -V_1'(s)& 1 &0 &\dots &0 \\
0 & -V_1'(s) & 1 & \vdots \\
\vdots &\ddots &\ddots &\ddots & \\
0 & & &  -V_1'(s) &1 \\
-V_1''(s)u_{0,1}+\frac{T}{N} & -V_1''(s)u_{1,1} & \dots &-V_1''(s)u_{d_2-2,1} &-V_1''(s)u_{d_2-1,1}
                 \end{pmatrix}
\eeq
and the linear equation giving the $K(x_0,s)$ reads for $q=d_2-1 $:
\beq 
\mathcal{M}
\begin{pmatrix}
 K(x_0,s)_{0,d_2-1} \\ 
K(x_0,s)_{1,d_2-1}\\
\vdots \\ 
K(x_0,s)_{d_2-2,d_2-1}\\
K(x_0,s)_{d_2-1,d_2-1}
\end{pmatrix}
=\begin{pmatrix}
 0 \\ 
\vdots \\
 \vdots\\ 
0\\
\frac{1}{(x_0-s)^2}
\end{pmatrix}
\eeq

The expression of $\mathcal{M}^{-1}$ is quite complicated. But fortunately, the last column (which is the only one we need for the computation of $W_1^{(1)}$) is the most simple: 
\beq \left(\mathcal{M}^{-1}\right)_{p,d_2-1}=\frac{V_1'(s)^p}{ -V_1''(s)\left(\underset{k=0}{\overset{d_2-1}{\sum}}u_{k,1}(V_1'(s))^k\right)+\frac{T}{N}}\eeq

But since $u_{k,1}=\frac{T}{N} \underset{p=0}{\overset{d_2-k-1}{\sum}} \td{t}_{d_2+p+1}(V_1'(s))^p$, we find easily that:
\beq \encadremath{\left(\mathcal{M}^{-1}\right)_{p,d_2-1}=-\frac{N V_1'(s)^p}{T \left(V_1''(s)V''_2(V_1'(s))-1\right)} }
\eeq

We can compute $U_{1}^{(0)}(x,y;x)$ by derivation of $U_0^{(0)}(x,y)$ relatively to $V_1(x)$ with the formula $\frac{\partial U_0^{(0)}(x,y)}{\partial V_1(z)}=U_1^{(0)}(x,y;z)$. We find: 
\bea U_{1}^{(0)}(x,y;x)&=&\frac{T}{N}\frac{V_2''(V_1'(s))}{\left(V_2''(V_1'(s))V_1''(s)-1\right)(x-s)^4}\sum_{p=0}^{d_2}\sum_{q=0}^{p-1}\td{t}_p y^{p-1-q}\left(V_1'(s)\right)^q\cr
&&+\frac{T}{N(x-s)}\sum_{p=0}^{d_2}\sum_{q=0}^{p-1}\td{t}_p y^{p-1-q}q\left(V_1'(s)\right)^{q-1} \frac{\partial V_1'(s)}{\partial V_1(x)}\cr
&=&\frac{T}{N}\frac{V_2''(V_1'(s))}{\left(V_2''(V_1'(s))V_1''(s)-1\right)(x-s)^4}\sum_{p=0}^{d_2}\sum_{q=0}^{p-1}\td{t}_p y^{p-1-q}\left(V_1'(s)\right)^q\cr
&&+\frac{T}{N(x-s)}\sum_{p=0}^{d_2}\sum_{q=0}^{p-1}\td{t}_p y^{p-1-q}q\left(V_1'(s)\right)^{q-1} \Big(-\frac{1}{(x-s)^2}\cr
&&+\frac{V_1''(s)V_2''(V_1'(s))}{\left(V_2''(V_1'(s))V_1''(s)-1\right)(x-s)^2}\Big)\cr\eea 

If we project along the $k^{\text{th}}$ power of $y$, it gives:
\beq \label{oooo}\fbox{$
   \begin{array}{rcl}
U_{1,k}^{(1)}(x;x)&=&\frac{T}{N}\frac{V_2''(V_1'(s))}{\left(V_2''(V_1'(s))V_1''(s)-1\right)(x-s)^4}\underset{r=0}{\overset{d_2-k-1}{\sum}}\td{t}_{r+k+1}\left(V_1'(s)\right)^r\cr
&+&\frac{T}{N \left(V_2''(V_1'(s))V_1''(s)-1\right)}\underset{r=0}{\overset{d_2-k-1}{\sum}}r\td{t}_{r+k+1}\left(V_1'(s)\right)^{r-1} \frac{1}{(x-s)^3}\cr \end{array}
   $} 
\eeq
Therefore we have with \eqref{OurRecursion}:
\beq \begin{pmatrix}
      U_{0,0}^{(1)}(x)\\
\vdots\\
U_{0,d_2-1}^{(1)}(x)\\
     \end{pmatrix}=\Res_{z \to s} K^t(x,z)\Big[\begin{pmatrix}
      U_{1,0}^{(1)}(z;z)\\
\vdots\\
U_{1,d_2-1}^{(1)}(z;z)\\
     \end{pmatrix}-\frac{N}{T}\partial_z \begin{pmatrix}
      U_{0,0}^{(0)}(z)\\
\vdots\\
U_{0,d_2-1}^{(0)}(z)\\
     \end{pmatrix}\Big]
\eeq

We want to compute $W_1^{(1)}(x)=\frac{1}{\td{t}_{d_2}}U_{0,d_2-1}^{(1)}(x)$ that is to say that we only need the last column of the previous vector. If we write it in components we get:
\beq \label{s0}\encadremath{\td{t}_{d_2}W_1^{(1)}(x)=\Res_{z \to s} \sum_{k=0}^{d_2-1} K(x,z)_{k,d_2-1}\left(U_{1,k}^{(1)}(z;z)-\frac{N}{T}\partial_z U^{(0)}_{0,k}(z)\right)}\eeq

The contribution from $U^{(0)}_{0,k}(z)$ is easy to compute since it only involves a double pole at $x=s$. Regarding the fact that $K'(x_0,s)_{p,q}$ are null for $p<d_2-1$, there is only one term involved. An easy computation leads to:
\beq 
-\frac{N}{T}\Res_{z \to s} \sum_{k=0}^{d_2-1} K(x,z)_{k,d_2-1}\partial_z U^{(0)}_{0,k}(z)= \frac{N}{T}\td{t}_{d_2}\frac{1}{x-s}\eeq

The contribution from $U_{1,k}^{(1)}(z;z)$ is technically more involved. Indeed, since $U_{1,k}^{(1)}(z;z)$ has poles up to order $4$, we need to compute the derivatives up to $K'''(x_0,s)$. In particular, we only need the last column of the matrices. Since $m=1$, the formulas of the derivatives simplify a lot:

\beq \vec{w}=\frac{1}{\td{t}_{d_2}}\begin{pmatrix}u_{0,i}\cr \vdots \cr u_{d_2-1,i}\end{pmatrix} \,\,,\,\,    \vec{V}=\frac{1}{\td{t}_{d_2}} \begin{pmatrix}-\td{t}_0+s \cr-\td{t}_1\cr \vdots \cr  -\td{t}_{d_2-1}\end{pmatrix} \,\,,\,\,
\vec{V}^{1}=\begin{pmatrix}\frac{1}{\td{t}_{d_2}} \cr 0\cr \vdots\cr 0 \end{pmatrix} \,\,,\,\,
\vec{V}^{2}=\vec{0} \eeq

which gives:
\beq\fbox{$
   \begin{array}{rcl}
K(x_0,s)_{p,d_2-1}&=&-\td{t}_{d_2}\frac{V_1'(s)^p}{  \left(V_1''(s)V''_2(V_1'(s))-1\right)}\frac{1}{(x_0-s)^2}\cr
K'(x_0,s)_{d_2-1,q}&=&\frac{N^2u_{q,1}}{T^2\td{t}_{d_2}(x_0-s)}\cr
K''(x_0,s)_{p,d_2-1}&=&\frac{2N}{T}\left(\delta_{p,d_2-2}K'(x_0,s)_{d_2-1,d_2-1}-V_1''(s)K(x_0,s)_{p,d_2-1}\right) \cr
&=&\frac{2N^2\delta_{p,d_2-2}}{T^2(x_0-s)}+\frac{2N\td{t}_{d_2}V_1''(s)V_1'(s)^p}{T \left(V_1''(s)V''_2(V_1'(s))-1\right)(x_0-s)^2} \,\, (p<d_2-1)\cr \end{array}
   $} 
\eeq

Then \eqref{2} reduces for $p<d_2-1$ to:
\beq\fbox{$
   \begin{array}{rcl}
\frac{T}{3N}K'''(x_0,s)_{p,d_2-1}&=&-\frac{1}{2}V_1'(s)K''(x_0,s)_{p,d_2-1}+\frac{1}{2}K''(x_0,s)_{p+1,d_2-1}\cr
&&-\frac{V_1'''(s)}{2}K(x_0,s)_{p,d_2-1}\cr
 \end{array}
   $} 
\eeq

The last component is a bit more sophisticated. We can obtain it by multiplying \eqref{2} by $\vec{r}\vec{w}^t$ in order to get rid of $K''(x_0,s)$:
\bea \frac{T}{6N}\vec{r}\vec{w}^t K'''(x_0,s)&=&\left(-V_1''(s)\vec{r}\vec{w}^t+\frac{T}{N} \vec{r}\vec{V^1}\right)K'(x_0,s)\cr
&&-\frac{V_1'''(s)}{2}\vec{r}\vec{w}^tK(x_0,s)_{p,d_2-1}-\frac{\vec{r}\vec{w}^t}{(x_0-s)^3}\eea
which gives in components:
\beq\fbox{$
   \begin{array}{rcl}
\frac{T^2}{6N^2}K'''(x_0,s)_{d_2-1,d_2-1}&=&-\frac{T}{6N}\underset{k=0}{\overset{d_2-2}{\sum}}\frac{u_{k,1}}{\td{t}_{d_2}} K'''(x_0,s)_{k,d_2-1}-\frac{T}{N} V_1''(s)K'(x_0,s)_{d_2-1,d_2-1}\cr
&&-\frac{V_1'''(s)}{2}\underset{k=0}{\overset{d_2-1}{\sum}}\frac{u_{k,1}}{\td{t}_{d_2}}K(x_0,s)_{k,d_2-1}-\frac{T}{N(x_0-s)^3} \end{array}
   $} 
\eeq
Now we can compute $W_1^{(1)}(x_0)$. First we can compute the term coming from $\frac{1}{(x-s)^3}$ of \eqref{oooo} in \eqref{s0}:
\bea (1)&=&\frac{T}{N\left(V_2''(V_1'(s))V_1''(s)-1\right)}\sum_{k=0}^{d_2-1} \underset{r=0}{\overset{d_2-k-1}{\sum}}r\td{t}_{r+k+1}\left(V_1'(s)\right)^{r-1}\frac{K''(x_0,s)_{k,d_2-1}}{2}\cr
&&=\frac{1}{\left(V_2''(V_1'(s))V_1''(s)-1\right)}\sum_{k=0}^{d_2-1} \underset{r=0}{\overset{d_2-k-1}{\sum}}r\td{t}_{r+k+1}\left(V_1'(s)\right)^{r-1}\Big(\frac{N\delta_{k,d_2-2}}{T(x_0-s)}\cr
&&+\frac{\td{t}_{d_2}V_1''(s)V_1'(s)^k}{ \left(V_1''(s)V''_2(V_1'(s))-1\right)(x_0-s)^2}\Big)\cr
&&=\frac{N\td{t}_{d_2}}{T\left(V_2''(V_1'(s))V_1''(s)-1\right)(x_0-s)}+\frac{\td{t}_{d_2}V_1''(s)V_2'''(V_1'(s))}{2\left(V_1''(s)V''_2(V_1'(s))-1\right)^2(x_0-s)^2}\cr
\eea
Then we observe that we have to compute:
\bea (2)&=&\frac{T V_2''(V_1'(s))}{6N(V_1''(s)V_2''(V_1'(s))-1)}\sum_{k=0}^{d_2-1}\sum_{r=0}^{d_2-k-1} \td{t}_{r+k+1}\left(V_1'(s)\right)^r K'''(x_0,s)_{k,d_2-1}\cr
&=&  \frac{T V_2''(V_1'(s))}{6N(V_1''(s)V_2''(V_1'(s))-1)}\Big(\sum_{k=0}^{d_2-2}\sum_{r=0}^{d_2-k-1} \td{t}_{r+k+1}\left(V_1'(s)\right)^r K'''(x_0,s)_{k,d_2-1} \cr
&&+\td{t}_{d_2} K'''(x_0,s)_{d_2-1,d_2-1}\Big)\cr
&=& \frac{T V_2''(V_1'(s))}{6N(V_1''(s)V_2''(V_1'(s))-1)}\Big(\sum_{k=0}^{d_2-2}\sum_{r=0}^{d_2-k-1} \td{t}_{r+k+1}\left(V_1'(s)\right)^r K'''(x_0,s)_{k,d_2-1}\cr
&&-\sum_{k=0}^{d_2-2}\frac{N u_{k,1}}{T} K'''(x_0,s)_{k,d_2-1} -\frac{6N\td{t}_{d_2}}{T}V_1''(s)K'(x_0,s)_{d_2-1,d_2-1}\cr
&&-\frac{3N^2}{T^2}V_1'''(s)\sum_{k=0}^{d_2-1}u_{k,1}K(x_0,s)_{k,d_2-1} -\frac{6N\td{t}_{d_2}}{T(x_0-s)^3}\Big)\cr
&=&\frac{T V_2''(V_1'(s))}{6N(V_1''(s)V_2''(V_1'(s))-1)}\Big(-\frac{6N\td{t}_{d_2}}{T}V_1''(s)K'(x_0,s)_{d_2-1,d_2-1}\cr
&&-\frac{3N^2}{T^2}V_1'''(s)\sum_{k=0}^{d_2-1}u_{k,1}K(x_0,s)_{k,d_2-1} -\frac{6N\td{t}_{d_2}}{T(x_0-s)^3}\Big)\cr
&=&\frac{ V_2''(V_1'(s))}{(V_1''(s)V_2''(V_1'(s))-1)}\Big(-\td{t}_{d_2}V_1''(s)K'(x_0,s)_{d_2-1,d_2-1}\cr
&&-\frac{N}{2T}V_1'''(s)\sum_{k=0}^{d_2-1}u_{k,1}K(x_0,s)_{k,d_2-1} -\frac{\td{t}_{d_2}}{(x_0-s)^3}\Big)\cr
\eea
where we have used the definition of $u_{k,1}$ to cancel the terms in $K'''(x_0,s)$. Eventually we observe that:
\bea &&\sum_{k=0}^{d_2-1} u_{k,1}K(x_0,s)_{k,d_2-1}\cr
&&=-\frac{T\td{t}_{d_2}}{N(V_1''(s)V_2''(V_1'(s))-1)(x_0-s)^2}\sum_{k=0}^{d_2-1}u_{k,1}\left(V_1'(s)\right)^k\cr
&&=-\frac{T\td{t}_{d_2}}{N(V_1''(s)V_2''(V_1'(s))-1)(x_0-s)^2}\sum_{k=0}^{d_2-1}\sum_{r=0}^{d_2-1}\td{t}_{r+k+1}\left(V_1'(s)\right)^{r+k}\cr
&&=-\frac{T\td{t}_{d_2}}{N(V_1''(s)V_2''(V_1'(s))-1)(x_0-s)^2}\sum_{k=0}^{d_2-1}\sum_{p=k}^{d_2-1}\td{t}_{p+1}\left(V_1'(s)\right)^{p}\cr \\
&&=-\frac{T\td{t}_{d_2}}{N(V_1''(s)V_2''(V_1'(s))-1)(x_0-s)^2}\sum_{p=0}^{d_2-1}\sum_{k=0}^{p}\td{t}_{p+1}\left(V_1'(s)\right)^{p}\cr
&&=-\frac{T\td{t}_{d_2}V_2''(V_1'(s))}{N(V_1''(s)V_2''(V_1'(s))-1)(x_0-s)^2}\eea
Therefore we find:
\bea 
(2)&=&-\frac{N\td{t}_{d_2}V_1''(s) V_2''(V_1'(s))}{T(V_1''(s)V_2''(V_1'(s))-1)(x_0-s)}+\frac{\td{t}_{d_2} V_1'''(s)\left(V_2''(V_1'(s))\right)^2}{2\left(V_1''(s)V_2''(V_1'(s))-1\right)^2(x_0-s)^2}\cr
&&-\frac{ \td{t}_{d_2}V_2''(V_1'(s))}{(V_1''(s)V_2''(V_1'(s))-1)(x_0-s)^3}\eea
And eventually putting $(1)$ and $(2)$ together and the contribution from $\partial_z U_{0,k}^{(0)}(z)$, we get to:
\bea
&&W_1^{(1)}(x_0)=\frac{N}{T\left(V_2''(V_1'(s))V_1''(s)-1\right)(x_0-s)}\cr
&&+\frac{V_1''(s)V_2'''(V_1'(s))}{2\left(V_1''(s)V''_2(V_1'(s))-1\right)^2(x_0-s)^2}\cr
&&+\frac{ V_1'''(s)\left(V_2''(V_1'(s))\right)^2}{2\left(V_1''(s)V_2''(V_1'(s))-1\right)^2(x_0-s)^2}-\frac{N V_1''(s) V_2''(V_1'(s))}{T(V_1''(s)V_2''(V_1'(s))-1)(x_0-s)}\cr
&&-\frac{V_2''(V_1'(s))}{(V_1''(s)V_2''(V_1'(s))-1)(x_0-s)^3}+\frac{N}{T}\td{t}_{d_2}\frac{1}{x_0-s}\cr
&&\cr
&&=-\frac{V_2''(V_1'(s))}{(V_1''(s)V_2''(V_1'(s))-1)(x_0-s)^3}\cr
&&+\frac{V_1''(s)V_2'''(V_1'(s))+V_1'''(s)\left(V_2''(V_1'(s))\right)^2}{2\left(V_1''(s)V''_2(V_1'(s))-1\right)^2(x_0-s)^2} \cr
\eea
i.e.:
\beq\fbox{$
   \begin{array}{rcl}
W_1^{(1)}(x_0)&=&\frac{V_1''(s)V_2'''(V_1'(s))+V_1'''(s)\left(V_2''(V_1'(s))\right)^2}{2\left(V_1''(s)V''_2(V_1'(s))-1\right)^2(x_0-s)^2}\cr
&&-\frac{V_2''(V_1'(s))}{(V_1''(s)V_2''(V_1'(s))-1)(x_0-s)^3} \end{array}
   $} 
\eeq

Then a straightforward computation gives $F_{1}$ defined by $\frac{\partial}{\partial V_1(x)}F_{1}=W_1^{(1)}(x)$ with:
\beq \encadremath{F_{1}=-\frac{1}{2}\ln\left(V_1''(s)V_2''(V_1'(s))-1\right)=-\frac{1}{2}\ln\left(V_1''(s)V_2''(\td{s})-1\right)}\eeq
where $F_{0}$ is given by $-\mathcal{S}$ which reduces when $N=1$ to:
\beq \encadremath{ F_{0}= \frac{T}{N}\left(-V_1(s)-V_2(V_1'(s))+sV_1'(s)\right)=\frac{T}{N}\left(-V_1(s)-V_2(\td{s})+s\td{s}\right) } \eeq
and which satisfies $\frac{\partial}{\partial V_1(x)}F_{0}=W_1^{(0)}(x)=\frac{T}{N(x-s)}$.

\section{Computation of $W_1^{(1)}(x)$ in the general case \label{F_1}}

We have:
\beq [V'_2(B)]_{i}=\sum_{k=0}^{d_2} \td{t}_k(B^{(k)})_{i}=s_i \eeq
where \beq B_i^{(k)}=\sum_{j=1}^N [B^k]_{i,j}\eeq
The relation
\beq \frac{\delta\left( [V_2'(B)]_i-s_i\right)}{\delta V_1(x)} =0\eeq
implies \beq \sum_{j} Z_{i,j}\mathcal{H}_{j,k}=V_{i,k}\eeq
where:
\beq V_{i,j}=\frac{d [V'_2(B)]_i}{dV_1'(s_j)} \,\,\, ,\,\,\, Z_{i,j}=\frac{d \left([V_2'(B)]_i-s_i\right)}{d s_j}\eeq
The equation to solve is:
\bea&& U^{(1)}(x,\hat{y})\left[ \frac{T}{N}\sum_i \frac{1}{x-s_i}-V_1'(x)+\hat{y}\right]\cr
&&+\left[ \sum_{i,j} A_{i,j}\frac{V'_2(\hat{y})-V'_2(\td{s}_j)}{\hat{y}-\td{s}_j} \frac{1}{x-s_i}+x-V_2'(y)\right]W_1^{(1)}(x)\cr
&&+U_0^{(1)}(x,x,\hat{y})=-P^{(1)}(x,\hat{y})\eea
where:
\beq U_1^{(0)}(x,z,y)=\frac{\delta U_0^{(0)}(x,y)}{\delta V_1(z)}=\frac{\delta}{\delta V_1(z)}\left[\frac{T}{N} \sum_{i,j}A_{i,j}\frac{V'_2(y)-V'_2(\td{s}_j)}{y-\td{s}_j} \frac{1}{x-s_i}+x-V_2'(y)\right]\eeq
Since we must perform an expansion in $y$, we write:
\beq \frac{V'_2(y)-V'_2(\td{s}_j)}{y-s_j}=\sum_{k=0}^{d_2-1} E_{j,k}(\td{s}_j,\td{t}_n)y^k\eeq
and \beq \frac{T}{N}\sum_{j}A_{i,j} \frac{V'_2(y)-V'_2(\td{s}_j)}{y-s_j}=\sum_{k=0}^{d_2-1}\Phi_{i,k}y^k\eeq
with
\beq \Phi_{i,k}=\frac{T}{N}\sum_{j} A_{i,j}E_{j,k}(\td{s}_j,\td{t}_n)\eeq
Since:
\bea E_{j,d_2-1}(\td{s}_j,\td{t}_n)&=&\td{t}_{d_2}\cr
E_{j,p-1}(\td{s}_j,\td{t}_n)-\td{s}_jE_{j,p}(\td{s}_j,\td{t}_n)&=&\td{t}_p\,\, \forall 1\leq p= d_2-1\cr
-\td{s}_jE_{j,0}(\td{s}_j,\td{t}_n)&=&\td{t}_0-V_2'(\td{s}_j)\eea
we obtain:
\bea \Phi_{i,d_2-1}&=&\td{t}_{d_2}\cr
\Phi_{i,p-1}-\sum_k B_{i,k}\Phi_{k,p}&=&\td{t}_p \,\, \forall 1\leq p= d_2-1\cr
-\sum_{k}B_{i,k}\Phi_{k,0}&=&\td{t}_0- [V_2'(B)]_i=\td{t}_0-s_i\eea

We have the following properties:
\beq \Phi_{i,p}=\sum_{k=0}^{d_2-p-1}\td{t}_{k+p+1}B_i^{(k)}\eeq
We calculate:
\bea \sum_{p=0}^{d_2-1}\sum_{q,i}[B^p]_{r,q} \frac{dB_{q,i}}{d s_j}\Phi_{l,p}&=& \sum_{p=0}^{d_2-1}\sum_{q,i}[B^p]_{r,q} \frac{dB_{q,i}}{d s_j}\sum_{k=0}^{d_2-p-1}\td{t}_{k+p+1}B_i^{(k)}\cr
&=&\sum_{p=0}^{d_2-1}\sum_{q,i}[B^p]_{r,q} \frac{dB_{q,i}}{d s_j}\sum_{u=p+1}^{d_2}\td{t}_{u}B_i^{(u-p-1)}\cr
&=&\sum_{u=0}^{d_2}\sum_{p=0}^{u-1}\td{t}_u\left[\sum_{q,i}[B^p]_{r,q} \frac{dB_{q,i}}{d s_j}B_i^{(u-p-1)}\right]\cr
&=&\sum_{u=0}^{d_2} \td{t}_u \frac{d B_r^{(u)}}{d s_j}=\frac{d[V_2'(B)]_{r}}{d s_j}=Z_{r,j}+\delta_{r,j}\eea
Consequently,
\bea \frac{d^2 [V_2'(B)]_r}{d s_j ds_k}&=&\sum_{p=0}^{d_2-1}\sum_{q,i} \frac{d[B^p]_{r,q}}{d s_k} \frac{dB_{q,i}}{d s_j} \Phi_{i,p}\cr
&&+\sum_{p=0}^{d_2-1}\sum_{q,i}[B^p]_{r,q} \frac{d^2B_{q,i}}{ds_j ds_k} \Phi_{i,p}\cr
&&+\sum_{p=0}^{d_2-1}\sum_{q,i} [B^p]_{r,q} \frac{dB_{q,i}}{d s_j} \frac{d\Phi_{i,p}}{d s_k}\eea

We note that:
\bea &&\sum_{p=0}^{d_2-1}\sum_{q,i} \frac{d[B^p]_{r,q}}{d s_k} \frac{dB_{q,i}}{d s_j} \Phi_{i,p}=\sum_{p=1}^{d_2-1}\sum_{q,i} \frac{d[B^p]_{r,q}}{d s_k}\frac{d}{d s_j} [B_{q,i}\Phi_{i,p}]\cr
&&-\sum_{p=0}^{d_2-1}\sum_{q,i} \frac{d[B^p]_{r,q}}{d s_k} B_{q,i}\frac{d \Phi_{i,p}}{d s_j}\cr
&&= \sum_{p=0}^{d_2-1}\sum_{q,i} \frac{d[B^p]_{r,q}}{d s_k} \frac{d \Phi_{q,p-1}}{d s_j}-\sum_{p=0}^{d_2-2}\sum_{q,i} \frac{ d[B^{p+1}]_{r,i}}{d s_k} \frac{d \Phi_{i,p}}{d s_j}\cr
&&+\sum_{p=0}^{d_2-1}\sum_{q,i} [B^p]_{r,q} \frac{d B_{q,i}}{d s_k} \frac{d \Phi_{i,p}}{d s_j}\cr
&&=\sum_{p=0}^{d_2-1}\sum_{q,i} [B^p]_{r,q} \frac{d B_{q,i}}{d s_k} \frac{d \Phi_{i,p}}{d s_j}\eea
so that
\bea \frac{d^2 [V_2'(B)]_r}{d s_j ds_k}&=& \sum_{p=0}^{d_2-1}\sum_{q,i} [B^p]_{r,q}\frac{ d^2 B_{q,i}}{d s_j s_k} \Phi_{i,p}\cr
&&+\sum_{p=0}^{d_2-1}\sum_{q,i} [B^p]_{r,q}\left[ \frac{d B_{q,i}}{d s_j} \frac{d \Phi_{i,p}}{d s_k} +(j \leftrightarrow k)\right]\eea
Similarly,
\beq \sum_{p=0}^{d_2-1}\sum_{q,i} [B^p]_{r,q} \frac{ d B_{q,i}}{d V'_1(s_j)}\Phi_{i,p}=\sum_{p=0}^{d_2-1} [B^p]_{r,j}\Phi_{j,p}=\frac{d [V_2'(B)]_r}{d V_1'(s_j)}=V_{r,j}\eeq
Now we have
\beq U_1^{(0)}(x,z,y)=\sum_{p=0}^{d_2-1}y^p\sum_{i,k}\left[ \Delta_k \Phi_{i,p}\frac{1}{x-s_i}\frac{1}{(z-s_k)^2}+\Phi_{i,p}\mathcal{H}_{i,k}\frac{1}{(x-s_i)^2}\frac{1}{(z-s_k)^2}\right]\eeq
where
\beq \frac{\delta}{\delta V_1(z)} \Phi_{i,p}=\sum_k \Delta_k \Phi_{i,p}\frac{1}{(z-s_k)^2}\eeq
Then, $U_1^{(0)}(x,x,\hat{y})$ is obtained by changing $z \rightarrow x$ and $y\rightarrow \hat{y}=-\frac{T}{N}\frac{d}{dx}$ with the convention that $\hat{y}$ is written at the left of all $x$ dependent terms.
\beq  U_1^{(0)}(x,x,\hat{y})=\sum_{p=0}^{d_2-1}\hat{y}^p\sum_{i,k}\left[ \Delta_k \Phi_{i,p}\frac{1}{x-s_i}\frac{1}{(z-s_k)^2}+\Phi_{i,p}\mathcal{H}_{i,k}\frac{1}{(x-s_i)^2}\frac{1}{(z-s_k)^2}\right]\eeq

Also we write
\beq U^{(1)}(x,\hat{y})=\sum_{p=0}^{d_2-1}\hat{y}^pU_{p}(x)\eeq
so that
\beq U^{(1)}(x,\hat{y})\hat{y}=\hat{y}^{d_2}U_{d_2-1}(x)+ \sum_{p=0}^{d_2-2}\hat{y}^p\left[ U_p(x)+\frac{d U_{p+1}(x)}{dx}\right] +\frac{d U_1(x)}{dx}\eeq
The system to be solved is:
\bea
&&\hat{y}^{d_2-1}U_{d_2-1}(x)+ \sum_{p=0}^{d_2-2}\hat{y}^p\left[ U_p(x)+\frac{d U_{p+1}(x)}{dx}\right] +\sum_{p=0}^{d_2-1} \hat{y}^p U_{p}(x)\left[\sum_{i}\frac{1}{x-s_i}-V_1'(x)\right]\cr
&&+\left[\sum_{p=0}^{d_2-1}\hat{y}^p\sum_l \Phi_{i,p}\frac{1}{x-s_l}+x-\sum_{p=0}^{d_2}\hat{y}^p\td{t}_p\right]W_1^{(1)}(x)+\frac{d U_1(x)}{dx}\cr
&&=-\left(\sum_{p=0}^{d_2-1}\hat{y}^p\sum_{l,k}\left[\Delta_k\Phi_{l,p}\frac{1}{x-s_i}\frac{1}{(x-s_k)^2}+\Phi_{i,p}\mathcal{H}_{i,k}\frac{1}{(x-s_i)^2}\frac{1}{(x-s_k)^2}\right]\right)\eea

Clearly
\beq U_{d_2-1}(x)=\td{t}_{d_2}W_1^{(1)}(x)\,\,\,\, ,\,\,\,\, \frac{T}{N}\frac{d}{dx}\vec{u}+\mathcal{D}\vec{u}=-\vec{v}\eeq
where \beq \vec{u}=\begin{pmatrix} U_1\cr U_2\cr\vdots\cr U_{d_2-1}\end{pmatrix}\eeq
and
\beq \mathcal{D}=\begin{pmatrix} \frac{T}{N}\underset{i}{\sum} \frac{1}{x-s_i}-V_1'(x) &0&\dots&0&\frac{1}{\td{t}_{d_2}}\left[\underset{i}{\sum} \frac{\Phi_{i,0}}{x-s_i}+x-\td{t}_0\right]\cr
1&\ddots& \ddots& \vdots& \frac{1}{\td{t}_{d_2}}\left[\underset{i}{\sum} \frac{\Phi_{i,1}}{x-s_i}-\td{t}_1\right]\cr
0&\ddots&\ddots&0&\vdots\cr
\vdots& \ddots& 1&\frac{T}{N}\sum_i \frac{1}{x-s_i}-V_1'(x)&\frac{1}{\td{t}_{d_2}}\left[\underset{i}{\sum} \frac{\Phi_{i,p-1}}{x-s_i}-\td{t}_{p-1}\right]\cr
0&\dots&0&1& \frac{T}{N}\underset{i}{\sum}\frac{2}{x-s_i}-V_1'(x)-\frac{\td{t}_{d_2-1}}{\td{t}_{d_2}}\end{pmatrix}
\eeq

and 
\beq 
\vec{v}=\begin{pmatrix} \underset{i,k}{\sum} \left[ \Delta_k\Phi_{i,0}\frac{1}{x-s_i}\frac{1}{(x-s_k)^2}+\Phi_{i,0}\mathcal{H}_{i,k}\frac{1}{(x-s_i)^2}\frac{1}{(x-s_k)^2}\right]\cr
\underset{i,k}{\sum} \left[ \Delta_k\Phi_{i,1}\frac{1}{x-s_i}\frac{1}{(x-s_k)^2}+\Phi_{i,1}\mathcal{H}_{i,k}\frac{1}{(x-s_i)^2}\frac{1}{(x-s_k)^2}\right]\cr
\vdots\cr
\underset{i,k}{\sum} \left[ \Delta_k\Phi_{i,p-1}\frac{1}{x-s_i}\frac{1}{(x-s_k)^2}+\Phi_{i,p-1}\mathcal{H}_{i,k}\frac{1}{(x-s_i)^2}\frac{1}{(x-s_k)^2}\right]\cr
\vdots\cr
\td{t}_{d_2}\underset{i,k}{\sum}\mathcal{H}_{i,k}\frac{1}{(x-s_i)^2}\frac{1}{(x-s_k)^2}\end{pmatrix}
\eeq
We have
\bea W_1^{(1)}(x)&=&\frac{U_{d_2-1}}{\td{t}_{d_2}}\cr
&=&=\frac{1}{\td{t}_{d_2}}\sum_j\Res_{x\to s_j}\sum_{p=0}^{d_2-1}K_{p,d_2-1}(x_0,x)\sum_{i,k}   \Big[-\frac{N}{T\td{t}_{d_2}}\partial_x U_{0,p}^{(0)}(x)\cr
&&+\Delta_k\Phi_{i,p-1}\frac{1}{x-s_i}\frac{1}{(x-s_k)^2}+\Phi_{i,p-1}\mathcal{H}_{i,k}\frac{1}{(x-s_i)^2}\frac{1}{(x-s_k)^2}\Big]\cr
\eea
The term involving the derivatives is easy to compute. Indeed, it gives:
\beq \frac{N}{T\td{t}_{d_2}}\sum_i \sum_{k=0}^{d_2-1} K'(x_0,s_i)_{k,d_2-1}u_{k,i}\eeq
Now only the term with $k=d_2-1$ contributes from \eqref{assumption}. Moreover, $u_{d_2-1,i}=\frac{T\td{t}_{d_2}}{N}$ and from \eqref{K'} we have:
\beq K'(x_0,s_i)_{d_2-1,d_2-1}=\frac{N}{T(x_0-s_i)}+2\sum_{j\neq i}\frac{K(x_0,s_j)_{d_2-1,d_2-1}}{s_i-s_j} \eeq
so that the term involving the derivative is:

\beq \label{toutou2}\fbox{$
   \begin{array}{rcl}
-\frac{N}{T\td{t}_{d_2}}\underset{j}{\sum}\underset{x\to s_j}{\Res}\underset{p=0}{\overset{d_2-1}{\sum}}K(x_0,x)_{p,d_2-1}\partial_x U_{0,p}^{(0)}(x)&=&\frac{N}{T}\underset{i=1}{\overset{N}{\sum}} \frac{1}{x_0-s_i}\cr
&&+2\underset{i,j\neq i}{\sum} \frac{K(x_0,s_j)_{d_2-1,d_2-1}}{s_i-s_j} \end{array}
   $} 
\eeq

Note in particular that it contributes with a simple pole at $x_0=s_i$ with residue $\frac{N}{T}$.
where the kernel $K(x_0,x)$ satisfies
\beq \left[\mathcal{D}^t-\frac{T}{N}\frac{d}{dx}\right]K(x_0,x)=G(x_0,x)=\sum_i \frac{\alpha_i}{x-s_i}+\frac{Id}{x_0-x}\eeq
the matrices $\alpha_i$ and $Id$ are $d_2\times d_2$ matrices. We expand around $s_j$:
\bea &&\frac{1}{x-s_i}\frac{1}{(x-s_k)^2}=\frac{\delta_{i,j}\delta_{k,j}}{(x-s_j)^3}-\delta_{k,j}\sum_{q=0}^\infty \sum_{i\neq j} \frac{ (x-s_j)^{q-2}}{(s_i-s_j)^{q+1}} \cr
&&+\delta_{i,j}\sum_{q=0}^\infty\sum_{k\neq j}(q+1)\frac{ (x-s_j)^{q-1}}{(s_k-s_j)^{q+2}}\cr
&&-\sum_{q,r=0}^\infty \sum_{i,k\neq j} (q+1) \frac{ (x-s_j)^{q+r}}{(s_i-s_j)^{r+1}(s_k-s_j)^{q+2}}\eea
and also
\bea \frac{1}{(x-s_i)^2(x-s_k)^2}&=& \frac{\delta_{i,j}\delta_{k,j}}{(x-s_j)^4} \delta_{i,j}\sum_{q=0}^\infty \sum_{k\neq j} (q+1) \frac{(x-s_j)^{q-2}}{(s_k-s_j)^{q+2}}\cr
&&+\delta_{k,j}\sum_{r=0}^\infty\sum_{i\neq j} (r+1) \frac{(x-s_j)^{r-2}}{(s_i-s_j)^{r+2}}\cr
&&+\sum_{q,r=0}^\infty\sum_{i,k \neq j} (q+1)(r+1) \frac{ (x-s_j)^{q+r}}{(s_i-s_j)^{r+2}(s_k-s_j)^{q+2}}\eea
Consequently, from the expansion of $K(x_0,x)$ around $x=s_j$:
\beq U_{d_2-1}=\sum_j \sum_{p=1}^{d_2}U_{j,(p,d_2)}\eeq
with 
\bea U_{j,(p,d_2)}&=&\frac{1}{6} \Phi_{j,p-1}\mathcal{H}_{jj}k'''_{j,(p,d_2)}+\frac{1}{2}\Delta_j\Phi_{j,p-1}k''_{j,(p,d_2)}\cr
&&+\sum_{i\neq j}\left[\frac{ (\Phi_{i,p-1}+\Phi_{j,p-1})\mathcal{H}_{i,j}}{(s_i-s_j)^2} -\frac{\Delta_j \Phi_{i,p-1}}{s_i-s_j}\right]k'_{j,(p,d_2)}\cr
&&+\sum_{i\neq j}2\left[\frac{ (\Phi_{i,p-1}+\Phi_{j,p-1})\mathcal{H}_{i,j}}{(s_i-s_j)^3} +\frac{ (\Delta_i\Phi_{i,p-1}+\Delta_j\Phi_{i,p-1})}{(s_i-s_j)^2}\right]k_{j,(p,d_2)}\cr\eea

The expansion of $\mathcal{D}^t$ around $x=s_j$ is:

\bea \mathcal{D}^t&=& \frac{1}{x-s_j}\begin{pmatrix} \frac{T}{N}& & & & \cr
& \frac{T}{N} & & & &\cr
& & \dots & &\cr
& & & \frac{T}{N} & &\cr
\frac{\Phi_{j,0}}{\td{t}_{d_2}}& \frac{\Phi_{j,1}}{\td{t}_{d_2}}& \dots & \frac{\Phi_{j,d_2-2}}{\td{t}_{d_2}}& 2\end{pmatrix}\cr
&&+\begin{pmatrix} -B_{j,j} & 1 & 0 & \dots & 0\cr
0 &-B_{j,j} & 1 &\ddots &0\cr
\vdots & \ddots &\ddots &\ddots &0\cr
0& \dots & 0 &-B_{j,j}& 1\cr
D_0& D_1& \dots & D_{d_2-2}&-B_{j,j}-\frac{\td{t}_{d_2-1}}{\td{t}_{d_2}}\end{pmatrix}\cr
&&(x-s_j)\begin{pmatrix} -Q^j_{j,j} &  & & \cr
& -Q^j_{j,j} & &  \cr
& & \ddots & \cr
& & & -Q^j_{j,j} &  \cr
\frac{1}{\td{t}_{d_2}}\left[1-\underset{i\neq j}{\sum} \frac{\Phi_{i,0}}{(s_j-s_i)^2}\right]& -\frac{1}{\td{t}_{d_2}}\underset{i\neq j}{\sum} \frac{\Phi_{i,1}}{(s_j-s_i)^2}&\dots & \underset{i\neq j}{\sum} \frac{\Phi_{i,d_2-2}}{(s_j-s_i)^2}& -Q_{j,j}\end{pmatrix}\cr
&&+\frac{1}{2}(x-s_j)^2\begin{pmatrix} -R^j_{j,j,j} &  & & \cr
& -R^j_{j,j,j} & &  \cr
& & \dots & \cr
& & & -R^j_{j,j,j} &   \cr
\frac{2}{\td{t}_{d_2}}\underset{i\neq j}{\sum} \frac{\Phi_{i,0}}{(s_j-s_i)^3}& \frac{2}{\td{t}_{d_2}}\underset{i\neq j}{\sum} \frac{\Phi_{i,1}}{(s_j-s_i)^3}&\dots & \frac{2}{\td{t}_{d_2}}\underset{i\neq j}{\sum} \frac{\Phi_{i,d_2-2}}{(s_j-s_i)^3}& -R_{j,j,j}\end{pmatrix}\cr
&&\dots
\eea
where we have noted:
\bea D_0&=&\frac{1}{\td{t}_{d_2}}\left[\underset{i \neq j}{\sum} \frac{\Phi_{i,0}}{s_j-s_i}+s_j-\td{t}_0\right]\cr
D_p&=&\frac{1}{\td{t}_{d_2}}\left[\underset{i \neq j}{\sum} \frac{\Phi_{i,p}}{s_j-s_i}-\td{t}_p\right]
\eea
where
\beq Q_{i,j}=\sum_{p}Q_{i,j}^p=\frac{d B_i}{d s_j}\eeq
and 
\beq Q_{i,j}^p=\frac{d B_{i,p}}{d s_j}=\begin{pmatrix} V''_1(s_i)+\frac{T}{N}\sum_{k\neq i} \frac{1}{(s_i-s_k)^2} \,\, \text{if} \,i=j=p\cr
\frac{T}{N}\frac{1}{(s_i-s_p)^2}\,\, \text{if} \,\,i=j\neq p\cr
-\frac{T}{N}\frac{1}{(s_i-s_p)^2}\,\, \text{if} \,\,i=p\neq j \,\, \text{or} \,\, j=p\neq i\end{pmatrix}
\eeq
and where
\beq R_{i,j,k}\sum_{p}R_{i,j,k}^p=\frac{d Q_{i,j}}{d s_k} =\frac{d^2 B_i}{d s_j d s_k}\eeq
and
\beq R_{i,j,k}^p=\frac{d Q_{i,j}^p}{d s_k}=\frac{d^2 B_{i,p}}{d s_j ds_k}=\begin{pmatrix} V_1'''(s_i)\delta_{i,j}\delta_{i,k}-2\frac{T}{N}\sum_{q\neq i} \frac{ (\delta_{i,j}-\delta_{p,i})(\delta_{i,k}-\delta_{p,k})}{(s_i-s_q)^3} \,\,\, \text{if} \,\,i=p\cr
-2\frac{T}{N}\sum_{q\neq i} \frac{ (\delta_{i,j}-\delta_{p,i})(\delta_{i,k}-\delta_{p,k})}{(s_i-s_q)^3} \,\,\, \text{if} \,\,i\neq p\end{pmatrix}
\eeq

We also expand the kernels:
\beq G(x_0,x)=\frac{\alpha_j}{x-s_j}-\sum_{p=0}^\infty\sum_{i\neq j} \frac{(x-s_j)^p}{(s_i-s_j)^{p+1}}\alpha_i + \sum_{p=0}^\infty \frac{(x-s_j)^p}{(x_0-s_j)^{p+1}} Id\eeq
We define the matrices:
\beq M_j=\begin{pmatrix} 1 & & & &\cr
& 1 & & &\cr
& & \dots & &\cr
& & & 1 &\cr
\Phi_{j,0}& \Phi_{j,1}& \dots& \Phi_{j,d_2-2}& \Phi_{j,d_2-1}\end{pmatrix}
\eeq
and expand $M_j \left[ \mathcal{D}^t -\frac{T}{N}\frac{d}{dx}\right] K(x_0,x)=M_j G(x_0,x)$ and identify the powers in $(x-s_j)$:
\beq \frac{1}{x-s_j}\rightarrow \begin{pmatrix} 1 & & & &\cr
& \frac{T}{N} & & &\cr
& & \dots & &\cr
& & & \frac{T}{N} &\cr
2\Phi_{j,0}& 2\Phi_{j,1}& \dots& 2\Phi_{j,d_2-2}& 2\Phi_{j,d_2-1}\end{pmatrix}k_j=M_j \alpha_j
\eeq
and
\bea (x-s_j)^0&\rightarrow& \begin{pmatrix} 0 & & & &\cr
& 0 & & &\cr
& & \dots & &\cr
& & & 0 &\cr
\Phi_{j,0}& \Phi_{j,1}& \dots& \Phi_{j,d_2-2}& \Phi_{j,d_2-1}\end{pmatrix}k'_j\cr
&&+ \begin{pmatrix} -B_{j,j} &1 & & &\cr
& -B_{j,j} &1 & &\cr
& & \dots & &\cr
& & & -B_{j,j} &1\cr
0& 0&0& 0\end{pmatrix}k_j\cr
&&-\sum_{i \neq j} \frac{1}{s_i-s_j}M_j \alpha_i +\frac{M_j}{x_0-s_j}\eea
We note that
\beq M_j\alpha_i=\begin{pmatrix} 1 & & & &\cr
& \frac{T}{N} & & &\cr
& & \dots & &\cr
& & & \frac{T}{N} &\cr
\Phi_{j,0}+\Phi_{i,0}& \Phi_{j,1}+\Phi_{i,1}& \dots& \Phi_{j,d_2-2}+\Phi_{i,d_2-2}& \Phi_{j,d_2-1}+\Phi_{i,d_2-1}\end{pmatrix}k_i\eeq
The $d_2-1$ first equations tell that:
\beq k_{j,(1,d_2)}-\sum_r k_{r,(1,d_2)}\sum_{p=1}^{d_2}\sum_{i,q}[B^{p-1}]_{r,q}Q^i_{q,j}\Phi_{i,p-1}=\frac{\td{t}_{d_2}}{(x_0-s_j)^2}\eeq
We proved before that this is equivalent to
\beq k_{j,(1,d_2)}-\sum_{r} k_{r,(1,d_2)}\frac{d [V_2'(B)]_t}{d s_j}=\frac{\td{t}_{d_2}}{(x_0-s_j)^2}\eeq
so that:
\beq -\sum_r k_{r,(1,d_2)}Z_{r,j}=\frac{\td{t}_{d_2}}{(x_0-s_j)^2}\eeq
and
\bea k_{i,(1,d)}&=& -\sum_j \frac{\td{t}_{d_2}}{(x_0-s_j)^2}Z^{-1}_{j,i}\cr
k_{i,(p,d)}&=& -\sum_{j,r} \frac{\td{t}_{d_2}}{(x_0-s_j)^2}Z^{-1}_{j,r}[B^{p-1}]_{r,i}\eea
The other equations tell
\beq -\frac{1}{2}k''_{j,(p,d_2)}=B_{j,j}k'_{j,(p,d_2)}-k'_{j,(p+1,d_2)}+\sum_i k_{i,(p,d_2)}Q_{i,j}^j\eeq

Finally,
\bea &&(x-s_j)^2\rightarrow -\frac{1}{6} \begin{pmatrix} 2\frac{T}{N} & & & &\cr
& 2\frac{T}{N} & & &\cr
& & \dots & &\cr
& & & 2\frac{T}{N} &\cr
\Phi_{j,0}& \Phi_{j,1}& \dots& \Phi_{j,d_2-2}& \Phi_{j,d_2-1}\end{pmatrix}k_j'''\cr
&&+\frac{1}{2} \begin{pmatrix} -B_{j,j} &1 & & &\cr
& -B_{j,j} &1 & &\cr
& & \dots & &\cr
& & & -B_{j,j} &1\cr
0& 0&0& 0\end{pmatrix}k_j''\cr
&&+\begin{pmatrix} -Q_{j,j}^j & & & &\cr
& -Q_{j,j}^j & & &\cr
& & \dots & &\cr
& & & -Q_{j,j}^j &\cr
1-\underset{i}{\sum} Q_{j,j}^i\Phi_{i,0}& -\underset{i}{\sum} Q_{j,j}^i\Phi_{i,1}& \dots& -\underset{i}{\sum} Q_{j,j}^i\Phi_{i,d_2-2}& -\underset{i}{\sum} Q_{j,j}^i\Phi_{i,d_2-1}\end{pmatrix}k'_j\cr
&&+\frac{1}{2}\begin{pmatrix} -R_{j,j,j}^j & & & &\cr
& -R_{j,j,j}^j & & &\cr
& & \dots & &\cr
& & & -R_{j,j,j}^j &\cr
-\underset{i}{\sum} R_{j,j,j}^i\Phi_{i,0}& -\underset{i}{\sum} R_{j,j,j}^i\Phi_{i,1}& \dots& -\underset{i}{\sum} R_{j,j,j}^i\Phi_{i,d_2-2}& -\underset{i}{\sum} R_{j,j,j}^i\Phi_{i,d_2-1}\end{pmatrix}k_j\cr
&=&-\sum_{i\neq j}\frac{1}{(s_i-s_j)^3}M_j\alpha_i +\frac{M_j}{(x_0-s_i)^3}\eea
We need only to look at the last equation:
\bea &&\frac{1}{6}\sum_{p=1}^{d_2}\Phi_{j,p-1}k'''_{j,(p,d_2)}-k'_{j,(1,d_2)}+ \sum_{p=1}^{d_2}\sum_i Q^i_{j,j}\Phi_{i,p-1}k'_{j,(p,d)}+\frac{1}{2}\sum_{p=1}^{d_2}\sum_i R_{j,j,j}^i\Phi_{i,p-1}k_{j,(p,d)}\cr
&&=\sum_{i\neq j} \frac{1}{(s_i-s_j)^2}\sum_{p=1}^{d_2}(\Phi_{i,p-1}+\Phi_{j,p-1})k_{i,(p,d_2)} -\frac{\td{t}_{d_2}}{(x_0-s_i)^3}\eea
We now show that in $U_{d_2-1}$, if we replace the terms in $k'''_{i,(p,d_2)}$ and $k''_{i,(p,d_2)}$ by their value, then the terms in $k'_{i,(p,d_2)}$ disappear so that only the terms in $k_{i,(p,d_2)}$ which are known remain.

\underline{First Step}

\bea &&\frac{1}{2}\Delta_j\Phi_{j,p-1}k''_{j,(p,d_2)}-\sum_{i\neq j}\frac{\Delta_j \Phi_{i,p-1}}{s_i-s_j}k'_{j,(p,d_2)}\cr
&&=-\sum_i B_{j,i}\Delta_j\Phi_{i,p-1}k'_{j,(p,d_2)}+\Delta_j\Phi_{j,p-1}k'_{j,(p+1,d_2)}\cr
&&-\sum_ik_{i,(p,d_2)}Q_{i,j}^j\Delta_j \Phi_{i,p-1}\eea
or
\bea &&\frac{1}{2}\Delta_j \Phi_{j,p-1}k''_{j,(p,d_2)}-\sum_{i\neq j}\frac{ \Delta_j\Phi_{i,p-1}}{s_i-s_j}k'_{j,(p,d_2)}\cr
&&=\left[-\Delta_j \Phi_{j,p-2}-\Phi_{j,p-1}+\sum_{i,k} Q_{j,k}^i\Phi_{i,p-1}\mathcal{H}_{k,j}\right]k'_{j,(p,d_2)}\cr
&&+\Delta_j\Phi_{j,p-1}k'_{j,(p+1,d_2)}-\sum_ik_{i,(p,d_2)}Q_{i,j}^i\Delta_j\Phi_{j,p-1}\eea
where, for $p=1$,
\beq [\Delta_j \Phi_{j,p-2}]_{p-1}\rightarrow \Delta_j[V_2'(B)]_j=\Delta_j s_j=\mathcal{H}_{j,j}\eeq

\underline{Step 2}

\bea &&\sum_{p=1}^{d_2}\left[ \frac{\mathcal{H}_{j,j}}{6}\Phi_{j,p-1}k'''_{j,(p,d_2)}+\frac{1}{2}\delta_j \Phi_{j,p-1}k''_{j,(p,d_2)}-\sum_{i\neq j}\frac{\delta_j\Phi_{i,p-1}}{s_i-s_j} k'_{j,(p,d_2)}\right]\cr
 &&=\sum_{p=1}^{d_2}\sum_{i,k\neq j} Q_{j,k}^i\Phi_{i,p-1} \mathcal{H}_{k,j}k'_{j,(p,d_2)} -\frac{\mathcal{H}_{j,j}}{2} \sum_{p=1}^{d_2}\sum_i R_{j,j,j}^i \Phi_{i,p-1}k_{j,(p,d_2)}\cr
 &&+\mathcal{H}_{j,j}\left[\sum_{i\neq j}\frac{1}{(s_i-s_j)^3}\sum_{p=1}^{d_2}(\Phi_{i,p-1}+\Phi_{j,p-1})k_{i,(p,d_2)}-\frac{\td{t}_{d_2}}{(x_0-s_i)^3}\right]\cr
 &&-\frac{\td{t}_{d_2}}{x_0-s_j}+\sum_{i\neq j} \frac{1}{s_i-s_j}\sum_{p=1}^{d_2} (\Phi_{i,p-1}+\Phi_{j,p-1})k_{i,(p,d_2)}\cr
 &&-\sum_{i}\sum_{p=1}^{d_2}k_{i,(p,d_2)}Q_{i,j}^j\Delta_j\Phi_{j,p-1}\cr\eea

\underline{Step 3}

From the indices properties of $Q_{j,k}^i$ we obtain:

\bea \label{toutou} \sum_{p=1}^{d_2}U_{j,(p,d_2)}&=&-\td{t}_{d_2}\frac{\mathcal{H}_{j,j}}{(x_0-s_i)^3}- \td{t}_{d_2}\frac{T}{N(x_0-s_j)}-\frac{\mathcal{H}_{j,j}}{2}\sum_{p=1}^{d_2}\sum_i R_{j,j,j}^i\Phi_{i,p-1}k_{j,(p,d_2)}\cr
&&+\sum_{i\neq j}\sum_{p=1}^{d_2}\frac{ \Phi_{i,p-1}+\Phi_{j,p-1}}{(s_i-s_j)^3}(\mathcal{H}_{j,j}k_{i,(p,d_2)}+2\mathcal{H}_{i,j}k_{j,(p,d_2)})\cr
&&+\sum_{i\neq j}\sum_{p=1}^{d_2}\frac{ \Delta_i\Phi_{j,p-1}+\Delta_j\Phi_{i,p-1}}{(s_i-s_j)^2}k_{j,(p,d_2)}-\sum_{i,j}\sum_{p=1}^{d_2}k_{i,(p,d_2)}Q_{i,j}^j\Delta_j\Phi_{j,p-1}\cr
&&+\frac{ \Phi_{i,p-1}+\Phi_{j,p-1}}{s_i-s_j}k_{i,(p,d_2)}\eea

Consequently,

\beq \encadremath{ W_1^{(1)}(x)=\frac{U_{d_2-1}}{\td{t}_{d_2}}=-\sum_j \left[\frac{\mathcal{H}_{j,j}}{(x_0-s_j)^3}+\frac{B_j}{(x_0-s_j)^2}\right] } \eeq
Note in particular that the simple pole in \eqref{toutou} is canceled by the one coming from \eqref{toutou2} 

We now calculate $B_j$, we have
\bea &&\sum_j\sum_{i\neq j}\sum_{p=1}^{d_2}\frac{(\Phi_{i,p-1}+\Phi_{j,p-1})}{s_i-s_j}k_{i,(p,d_2)}\cr
&&=-\td{t}_{d_2}\sum_{i,j,k,n\neq i} \frac{1}{(x_0-s_j)^2}Z^{-1}_{j,k}\sum_{p=1}^{d_2}[B^{p-1}]_{k,i}\frac{(\Phi_{i,p-1}+\Phi_{n,p-1})}{s_i-s_n}\cr
&&=-\td{t}_{d_2}\sum_{i,j,k} \frac{1}{(x_0-s_j)^2}Z^{-1}_{j,k}\left[V_1'(s_i)V_{k,i}-B_{k,i}\sum_n V_{n,i}\right]\eea
We have also:
\bea&& \sum_j \Big( -\frac{\mathcal{H}_{j,j}}{2}\sum_{p=1}^{d_2}\sum_i R_{j,j,j}^i \Phi_{i,p-1}k_{j,(p,d_2)}\cr
&&+ \sum_{i \neq j}\sum_{p=1}^{d_2} \frac{(\Phi_{i,p-1}+\Phi_{j,p-1})}{(s_i-s_j)^3}(\mathcal{H}_{j,j}k_{i,(p,d_2)}+2\mathcal{H}_{i,j}k_{j,(p,d_2)})\Big)\cr
 &&=-\sum_{j,k,q}\frac{1}{(x_0-s_j)^2}Z^{-1}_{j,k}\sum_{p=1}^{d_2}[B^{p-1}]_{k,q}\Big( -\frac{\mathcal{H}_{q,q}}{2}\sum_i R_{q,q,q}^i \Phi_{i,p-1}\cr
 &&+\sum_{i\neq q}\frac{ \Phi_{q,p-1}+\Phi_{i,p-1}}{(s_q-s_i)^3}\mathcal{H}_{i,i}-2\sum_{i\neq q} \frac{ \Phi_{q,p-1}+\Phi_{i,p-1}}{(s_q-s_i)^3}\mathcal{H}_{i,q}\Big) \eea

The bracket is
 \beq \frac{\mathcal{H}_{q,q}}{2}V_1'''(s_i)\Phi_{q,p-1}-\sum_{i\neq q} \frac{ \Phi_{q,p-1}+\Phi_{i,p-1}}{(s_q-s_i)^3}(\mathcal{H}_{q,q}-2\mathcal{H}_{i,q}+\mathcal{H}_{i,i})=\frac{1}{2}\sum_{u,v,n}R_{q,u,v}^n\mathcal{H}_{u,v}\Phi_{n,p-1}\eeq
 Consequently,
 \bea && \sum_j \Big( -\frac{\mathcal{H}_{j,j}}{2}\sum_{p=1}^{d_2}\sum_i R_{j,j,j}^i \Phi_{i,p-1}k_{j,(p,d_2)}\cr
 &&+ \sum_{i \neq j}\sum_{p=1}^{d_2} \frac{(\Phi_{i,p-1}+\Phi_{j,p-1})}{(s_i-s_j)^3}(\mathcal{H}_{j,j}k_{i,(p,d_2)}+2\mathcal{H}_{i,j}k_{j,(p,d_2)})\Big)\cr
 &&=\frac{1}{2}\sum_{j,k,q} \frac{1}{(x_0-s_j)^2}Z^{-1}_{j,k}\sum_{p=1}^{d_2-1}[B^{p-1}]_{k,q}\sum_{u,v,n}\frac{d^2 [B]_{q,n}}{d s_u ds_v}\mathcal{H}_{u,v}\Phi_{n,p-1}\eea
We have:
\bea &&\sum_j\sum_{i\neq j}\sum_{p=1}^{d_2} \frac{ \Delta_i\Phi_{j,p-1}-\Delta_j\Phi_{i,p-1}}{(s_i-s_j)^2}k_{j,(p,d_2)}-\sum_{i,j}\sum_{p=1}^{d_2}k_{i,(p,d_2)}Q_{i,j}^i\Delta_j\Phi_{j,p-1}\cr
&&=-\sum_j \sum_{p=1}^{d_2}k_{j,(p,d_2)}\left[V_1''(s_j)\Delta_j\Phi_{j,p-1}+\sum_{i\neq j}\frac{1}{(s_i-s_j)^2}(\Delta_j-\Delta_i)(\Phi_{j,p-1}+\Phi_{i,p-1})\right]\cr
&&=\td{t}_{d_2}\sum_{j,k,q,n,m}\frac{1}{(x_0-s_j)^2}Z^{-1}_{j,k}\sum_{p=1}^{d_2}[B^{p-1}]_{k,q}Q^m_{q,n}\Delta_n\Phi_{m,p-1}\cr
&&=\td{t}_{d_2}\sum_{j,k,q,n,m}\frac{1}{(x_0-s_j)^2}Z^{-1}_{j,k}\sum_{p=1}^{d_2}[B^{p-1}]_{k,q}Q^m_{q,n}\left[-\frac{d \Phi_{m,p-1}}{d V_1'(s_n)}+\sum_u \frac{d \Phi_{m,p-1}}{d s_u} \mathcal{H}_{u,n}\right]\cr\eea
We note that
\bea &&\sum_{p=1}^{d_2}\sum_{q,u,v,n}\Big[ \frac{1}{2} [B^{p-1}]_{k,q}\frac{d^2[B]_{q,n}}{d s_u ds_v}\Phi_{n,p-1}+ [B^{p-1}]_{k,q}\frac{d B_{q,n}}{d s_v} \frac{d \Phi_{n, p-1}}{ds_u} \Big]\mathcal{H}_{u,v}\cr
&&=\frac{1}{2}\sum_{u,v} \frac{d^2[V_2'(B)]}{d s_u ds_v}\mathcal{H}_{u,v}\eea

Consequently, we find:
\bea \label{B_j} B_j&=&\sum_k Z^{-1}_{j,k}\Big[ -\frac{1}{2}\sum_{u,v} \frac{d^2[V_2'(B)]}{d s_u ds_v}\mathcal{H}_{u,v}+\sum_i V_1'(s_i)V_{k,i}-\sum_{i,n}B_{k,i}V_{i,n}\cr
&&+\sum_{p=1}^{d_2}\sum_{q,n,m} [B^{p-1}]_{k,q}Q_{q,n}^m \frac{d \Phi_{m,p-1}}{d V_1'(s_n)}\Big]\cr
&&+2\sum_{i,j\neq i,r}\frac{\td{t}_{d_2}}Z^{-1}_{j,r}[B^{d_2-1}]_{r,i}\eea
We have not been able to sum over $p$ the last term. This result is the same found from the differential operator $\hat{y}=-\frac{d}{dx}$ method.
We have
\beq \frac{d}{d V_1'(s_n)}\left[ \sum_{p=1}^{d_2}\sum_{q,n,m}[B^{p-1}]_{k,q}Q_{q,n}^m\Phi_{m,p-1}\right]=\frac{d}{d V_1'(s_n)}\frac{d[V_2'(B)]_k}{d s_n}=\frac{d}{d s_n}\frac{d[V_2'(B)]_k}{dV_1'(s_n)}=\frac{d V_{k,n}}{d s_n}\eeq
so that
\beq\sum_{p=1}^{d_2}\sum_{q,n,m}[B^{p-1}]_{k,q}Q_{q,n}^m\frac{d \Phi_{m,p-1}}{d V_1'(s_n)}=\sum_n \frac{d V_{k,n}}{d s_n}- \sum_{p=1}^{d_2}\sum_{q,n,m}\frac{d[B^{p-1}]_{k,q}}{d V_1'(s_n)} Q_{q,n}^m\Phi_{m,p-1}\eeq
or
\bea &&\sum_{p=1}^{d_2}\sum_{q,n,m}[B^{p-1}]_{k,q}Q_{q,n}^m\frac{d \Phi_{m,p-1}}{d V_1'(s_n)}=\frac{1}{2}\sum_{p=1}^{d_2}\sum_{q,n,m}\Big[ [B^{p-1}]_{k,q}Q_{q,n}^m\frac{d \Phi_{m,p-1}}{d V_1'(s_n)} \cr
&&-\frac{d[B^{p-1}]_{k,q}}{d V_1'(s_n)}Q_{q,n}^m\Phi_{m,p-1}\Big]+\frac{1}{2}\sum_n \frac{d V_{k,n}}{d s_n}\eea
We use the relation
\beq \frac{d V_{k,n}}{d s_n}=\sum_p \frac{d^2 [V_2'(B)]_k}{ds_n ds_p}\mathcal{H}_{p,n}+\sum_p Z_{k,p}\frac{\mathcal{H}_{p,n}}{d s_n}\eeq
and write:
\bea \label{B_j2} B_j&=&\frac{1}{2}\sum_n \frac{\mathcal{H}_{j,n}}{d s_n}+\sum_k Z_{j,k}^{-1}\Big[\frac{1}{2}\sum_{p=1}^{d_2}\sum_{q,n,m}\Big[ [B^{p-1}]_{k,q}Q_{q,n}^m\frac{d \Phi_{m,p-1}}{d V_1'(s_n)} \cr
&&-\frac{d[B^{p-1}]_{k,q}}{d V_1'(s_n)}Q_{q,n}^m\Phi_{m,p-1}\Big]+\sum_i V_1(s_i)V_{k,i}-\sum_{i,n}B_{k,j}V_{i,n}\Big]\cr
&&+2\sum_{i,j\neq i,r}\frac{\td{t}_{d_2}}{Z^{-1}_{j,r}}[B^{d_2-1}]_{r,i}\eea

\end{document}